\documentclass[12pt,a4paper]{amsart}

\usepackage{amsmath}
\usepackage{amssymb}
\usepackage{bbm}
\usepackage[T1]{fontenc}
\usepackage{hyperref}

\newcommand {\del}{\partial}

\usepackage{DLdef1}

\makeatletter
\@ifundefined{str}{\DeclareMathOperator{\str}{str}}{}
\@ifundefined{Res}{\DeclareMathOperator{\Res}{Res}}{}
\@ifundefined{sgn}{\DeclareMathOperator{\sgn}{sgn}}{}
\@ifundefined{ch}{\DeclareMathOperator{\ch}{ch}}{}
\@ifundefined{sch}{\DeclareMathOperator{\sch}{sch}}{}
\theoremstyle{thm}
\@ifundefined{Statement}{}{}
\@ifundefined{Lemma}{}{}
\@ifundefined{Question}{}{}
\theoremstyle{definition}
\@ifundefined{Remark}{}{}
\makeatother

\let\ssec\subsection

\renewcommand {\ssbegin}[2][*]
 {\refstepcounter{subsection}%
\if#1*
\addcontentsline{toc}{subsection}{\thesubsection.\hskip 1pc #2}%
\else
\addcontentsline{toc}{subsection}{\thesubsection.\hskip 1pc #2. #1}%
\fi
 \def \secno {\gdef \secno {}{\ssecfont
\thesubsection.\hskip 2ex}%
 }%
 \begin{#2}}

\renewcommand {\sssbegin}[2][*]
  {\refstepcounter{subsubsection}
\if#1*
\addcontentsline{toc}{subsubsection}{\thesubsubsection\hskip 1pc #2}%
\else
\addcontentsline{toc}{subsubsection}{\thesubsubsection\hskip 1pc #2. #1}
\fi
  \def \secno {\gdef \secno {}{\ssecfont \thesubsubsection\hskip 2ex}%
  }%
   \begin{#2}}

\renewcommand {\parbegin}[2][*]
  {\refstepcounter{paragraph}
\if#1*
\addcontentsline{toc}{paragraph}{\theparagraph.\hskip 1pc #2}%
\else
\addcontentsline{toc}{paragraph}{\theparagraph.\hskip 1pc #2. #1}
\fi
  \def \secno {\gdef \secno {}{\ssecfont \theparagraph.\hskip 2ex}%
  }%
   \begin{#2}}

\makeatletter
\newcommand\Subsubsection[2]{%
\refstepcounter{subsubsection}%
\addcontentsline{toc}{subsubsection}{\thesubsubsection.\hskip 1pc #1. #2}%
\par\vskip1ex \@plus .4ex \@minus .2ex\noindent\thesubsubsection. \normalfont{\bfseries #1} (#2).\hskip.7em plus .3em}
\makeatother

\begin{document}

\title{Superized Leznov-Saveliev equations as the zero-curvature condition on a reduced connection}

\author{Dimitry Leites}
\address{Department of Mathematics, Stockholm University,
Albanov\"agen 28, SE-114 19
Stockholm, Sweden\\
dimleites@gmail.com}

\makeatletter
\@namedef{subjclassname@2020}{\textup{2020} Mathematics Subject Classification}
\makeatother


\begin{abstract}
The known integrable super versions of Toda lattice equations, called Leznov-Saveliev equations (LSE), correspond to two different types of Cartan matrices $A$ of Lie superalgebras. The Leznov-Saveliev method of solving LSE is applicable only when the algebra $\mathfrak{g}(A)$ is finite-dimensional  and admits a superprincipal embedding $\mathfrak{osp}(1|2)\to\mathfrak{g}(A)$. I interpret the super LSE of this type (LSE1) as zero-curvature equations for the \textit{reduced} connection on the non-integrable distribution on the supervariety of $\text{OSp}(1|2)$-orbits of the $N=1$-extended Minkow\-ski superspace.  The simplest LSE1 is the super Liouville equation; it is also interpreted in terms of the superstring action.

Olshanetsky introduced LSE2 --- another type of super Toda lattice equations considered even for $\mathfrak{g}(A)$ infinite-dimensional. LSE2 are not solvable by the LS-method, except for the super Liouville equation --- the only case where LSE1 coincide with LSE2.

I consider also the ($N>1$)-extended versions of super Liouville equations, why $N\leq 4$, related superprincipal embeddings $\mathfrak{osp}(N|2)\to\mathfrak{g}(A)$ and the  interpretation of LSE as a condition on a reduced connection applies, same as for $N=1$.

I also review related less known and less popular mathematical constructions involved.
\end{abstract}


\maketitle

\markboth{\itshape
Dimitry Leites}
{{\itshape Superized Leznov-Saveliev equations as the zero-curvature condition on a reduced connection}} 

\thispagestyle{empty}

\section{Introduction}

Vital in the \textbf{super} version of gauge field (a.k.a. Yang-Mills) equations is a~ \textbf{non-integrable distribution} on the supermanifold considered, such as any of the $N$-extended Minkowski supermanifolds; this was never explained before. This explanation is related with \cite{GL}, where the analog of the Riemann tensor in presence of a non-integrable distribution, needed to figure out the shape of $N$-extended Minkowski supermanifolds for $N>1$ and to write the analogs of the Einstein equations on them, was explicitly described, and with \cite{BGLS}, where this analog was calculated in several other cases.

The zero-curvature gauge field equations were superized in two ways yielding solvable equations. Their meaning was not, however, explained, except for the  case  of the super Liouville equation whose interpretation is not related with gauge fields, but with a~superstring action, see \cite{Po}.

Here, I recall two approaches to solutions of zero-curvature gauge field equations. I review the subtleties of their superizations needed to understand what is going on, e.g., appearance of integrable forms, see their definition \eqref{intForm}. The methods of solution in these approaches are different and the ranges of their applicability intersect in one case only.

\ssec{From the Yang-Mills equations to the variety of $\text{O}(3)$-orbits in the Minkowski space and further to zero-curvature equations with the gauge algebra $\fsl(2)\subset \fg(A)$. Two methods to solve these equations}  In physics, a~\textit{gauge theory} is a~type of field theory in which  the dynamics of the system is invariant  under local transformations performed by Lie groups, or --- infinitesimally --- Lie algebras.
The equations of gauge theories --- called \textit{Yang-Mills} equations --- are too complicated to be explicitly solved; solutions are found only under some additional restrictions.

\textbf{1)  The Leznov-Saveliev method}. Under additional spatial $\text{O}(3)$-symmetry, the 4-dimen\-sional Minkowski space reduces to a~2-dimensional cylinder with 2 parameters: the radius $r$ of the $\text{O}(3)$-orbit (the sphere in $\Ree^3$) and the time $t$. In the simplest case of zero-curvature conditions on the gauge field, these equations  --- called \textit{Leznov-Saveliev} equations (LSE) --- can be solved; the method of solution requires the complexified gauge Lie algebra $\fg$ to be finite-dimensional and admit a principal embedding of $\fsl(2)$. Note that the Leznov-Saveliev equations do not, however, consider abelian theories, they are applicable only to simple finite-dimensional gauge Lie algebras $\fg$, hence, of the form $\fg=\fg(A)$ with Cartan matrix $A$; so the LSE are of interest as integrable systems, but are hardly applicable to certain current models of particle physics with abelian summand in the gauge algebra.

Explicitly, the \textit{Leznov-Saveliev equations} constitute the system \eqref{LS} of equations for a~vector-valued function $F=(F_1, \dots, F_n)$, where $n=\rk\fg$ and $\fg=\fg(A)$ is a~ complex  Lie algebra with Cartan matrix $A:=(A_{ij})_{i,j=1}^n$, and each $F_i$ depends on $x$ and $y$ which are complexified linear combinations of $r$ and $t$:
\be\label{LS}
\nfrac{\partial^2 F_i}{\partial x\partial y}=\exp\left(\sum_j A_{ij}F_j\right),\text{~~where $i=1,\dots , n$}.
\ee
Sometimes,  the term \textit{Leznov-Saveliev equations} is applied to the following system of equations for a~vector-valued function $G=(G_1, \dots, G_n)$, where each $G_i$ depends on $x$ and $y$:
\be\label{LSbis}
\nfrac{\partial^2 G_i}{\partial x\partial y}=\sum_j A_{ij}\exp(G_j),\text{~~where $i=1,\dots , n$}.
\ee

If the matrix $A$ is invertible (which is the case where $\fg(A)$ is finite-dimensional and simple),  then the systems~ \eqref{LS} and~ \eqref{LSbis} are equivalent: set $G=AF$, then $F=A^{-1}G$.

If $A$ is  the $n\times n$ Cartan matrix of a~simple finite-dimensional Lie algebra (over $\Cee$), then the general solution of  system~\eqref{LS} is a~vector depending on (the logarithm in) $2n$ polynomials in 2 arbitrary analytic functions each (one function depending on $x$, the other one on $y$) and their derivatives, see \cite{LS, LSb}.

\textbf{2)  The Inverse Scattering Method}. This powerful method was  discovered earlier than the LS-method; it is applicable not only to the finite-dimensional gauge algebra $\fg(A)$ but also to certain algebras of infinite dimension  (namely, affine Kac-Moody algebras a.k.a. \textit{double extensions} of loop algebras, see the last two Sections in \cite{BLS}), see, e.g., \cite{AC, AN}. This method allows one to solve versions of Leznov-Saveliev equations (sine-Gordon and sinh-Gordon equations in the simplest cases) to which the LS-method is inapplicable: there are no principal embeddings of $\fsl(2)$ into  infinite-dimensional Lie algebras.

If $A$ is  the Cartan matrix of an infinite-dimensional Kac-Moody Lie algebra, the solution is obtained via ISM in terms of theta-functions, see \cite{AN}.

\ssec{Superization of Leznov-Saveliev equations via the  LS-approach} It seems natural to consider the systems \eqref{LS} and \eqref{LSbis}, where $A$ is any Cartan matrix of any finite-dimensional or Kac-Moody Lie superalgebra. However, to this day only two types of Cartan matrices were considered in order to get a solution of these equations, and only one of these two types of equations has an interpretation: it is given in this paper.

\sssec{Non-integrable distributions}
Recall that a \textit{distribution} $\cD$ on a~ (super)manifold $M$ is a subbundle of the tangent bundle $TM$; it is given by selecting a subspace $\cD_m$ of the tangent space $T_mM$ at every point $m\in M$; we assume that all spaces $\cD_m$ are of the same dimension $(k|l)$ and smoothly depend on $m$. The distribution $\cD$ is \textit{integrable} at a point $m\in M$ if there is a $(k|l)$-dimensional submanifold $N\subset M$ containing $m$ such that $T_{m'}N=\cD_{m'}$ at all points $m'\in N$.
For integrability of  the distributions one has the following criterion, which holds on supermanifolds as well.

\paragraph{The Frobenius Criterion for integrability of distributions (\cite{Lsos, MaG})} \textit{The distribution $\cD$ is integrable if and only if its sections form a Lie subalgebra of $\fvect(M)$ closed under the Lie bracket of vector fields}.

\sssec{$N$-extended Minkowski supermanifolds and non-integrable distributions} The advent of supersymmetry, which undoubtedly is the language and tool for implementing  Einstein's dream --- the future Grand Unified Theories of all fundamental forces, called SUSY GUTS --- brought a~new direction of research: Yang-Mills  equations on supermanifolds.

Observe that the $N$-extended Minkowski supermanifold $\cM_N$, where $N=1, \dots, 8$, is not just a supermanifold: it is endowed with a non-integrable distribution. The precise shape of $\cM_N$ is not known for $N>1$, see \cite{MaG, GL}. (Note that the restriction $N\leq 8$ is imposed ``for physical reasons", in order to eliminate particles of spin $>2$. Since 1990, M.~Vasiliev with co-authors and followers develops a theory which lacks this restriction, see one of the latest papers \cite{Va} and references therein.) However, it is clear from the ``twistor\rq\rq\  (see \cite{MaG, GL}) interpretation of the Minkowski supermanifold $\cM_N$ as the quotient space of a real form $\cG$ of the complex Lie supergroup $\SL(4|N)$ that the Lie superalgebra of this real form  consists of the following supermatrices in the supermatrix format $2|N|2$ (here $\overline X$ denotes the matrix obtained from $X$ by element-wise complex conjugation and ${}^t$ denotes transposition):
\begin{equation}\label{M3}
\begin{pmatrix}
A&\overline R^t&U\\
Q&B&R\\
T&-\overline Q^t&-\overline A^t
\end{pmatrix},\;\;\begin{array}{l}\text{where  $Q, R\in\Mat_\Cee(N\times 2)$,}\ \ \text{$T=\overline T^t$,
$U=\overline U^t$,}\\
\text{$A\in\fgl(2;\Cee)$,
 $B\in\fu(N)$, $\tr B=\tr(A-\overline A^t)$}\end{array}
\end{equation}
so $\cM_N=\cG/\cP$, where the subsupergroup $\cP$ for $N=1$ is generated by the block  supermatrices on the main diagonal and above it; for $N>1$,   we should take a smaller subsupergroup by diminishing $B$, see \cite{GIOS} for $N=2$ and \cite{GL} for any $N\leq 8$. (Note that the method of \cite{GL} allows one to investigate the case of any $N$: the criterion for selecting $\cP$ is that the suggested analog of the Riemann tensor should contain the Riemann tensor on the Minkowski space underlying~ $\cM_N$.)

Therefore, the tangent space to any point of $\cM_1$ is isomorphic to the superspace spanned by block supermatrices $Q$, $\overline Q^t$, and $T$; the distribution of codimension $(4|0)$ spanned  by $Q$ and $\overline Q^t$ at the tangent space at every point is non-integrable, since $[Q,\overline Q^t]=T$. Other models of $\cM_N$ for $N>1$ are described by a~smaller parabolic subgroup $\cP$, so a part of $B$ becomes a~ part of the tangent space at a~ point of $\cM_N$; the non-integrable distribution described above only becomes more complicated: of dimension $k|l$, for some $k>0$ and, perhaps, $l\geq 0$, see \cite{GL,BGLS}.

\sssec{LSE1 and reduced connections}  In the ${N=1}$-Minkowski superspace, the supervariety of $\text{OSp}(1|2)$-orbits is a~$(2|2)$-dimensional supermanifold, that inherits a~non-integrable distribution of codimension $(2|0)$, see \cite{LSL, LSS}.
The method described in \cite{LSS}  is applied to the zero-curvature equations for a~\textbf{reduced connection}, as explained in what follows; at the time \cite{LSS} was written the notion of reduced connection was unknown to anybody who performed superization of LSE; the researchers were guided  only by their gut feeling. 

Recall that the embedding $i$ is called \textit{superprincipal} if the adjoint representation of $i(\fosp(1|2))$ in $\fg$ splits into the smallest possible number  (equal to the rank of $\fg$) of irreducible modules. If there is  a~ \textit{superprincipal} embedding $i:\fosp(1|2)\tto\fg$ into the gauge Lie superalgebra~ $\fg$, then the zero-curvature equations for a~reduced connection on the supervariety of  these orbits --- \textit{superized  Leznov-Saveliev equations} or superized LSE1 --- can be solved, see \cite{LSS}.  As shown in \cite{LSS}, the simple Lie superalgebra $\fg$ admitting a superprincipal embedding into it must be
\begin{equation}\label{lsA}
\begin{minipage}[c]{14cm} \textbf{finite-dimensional and have Cartan matrix\\
 with only 0s and 1s on its main diagonal}.\end{minipage}
\end{equation}

Note that some authors (e.g., \cite{A1, A2}) consider Cartan matrices $A$ that do not allow superprincipal embeddings into $\fg(A)$, although   $A$ satisfies condition~\eqref{lsA}; it is unclear if the corresponding super LSE1 are integrable  in explicit form, not as ``some formal series", as in \cite{AN}.

Observe that the simplest case of these superized LSE1 is the super Liouville equation which appeared in Polyakov's paper \cite{Po} from a totally different direction, not related with any gauge equation, but rather as a~ result of variation of the action in the description of the simplest fermionic string.

\ssec{Olshanetsky's approach: superization LSE2} M.~Olshanetsky suggested a~ completely different approach to superizing Toda lattice (TL) equations, see \cite{O}.  Olshanetsky solved these LSE2 by means of the Inverse Scattering Method. Interestingly, the ISM is applicable to  infinite-dimensional gauge Lie superalgebras $\fg$, as in periodic Toda lattice (provided $\fg$ is ``not too big": namely,  not of exponential growth, as hyperbolic Lie algebras or almost affine superalgebras classified in \cite{CCLL}):
\begin{equation}\label{OlsA}
\begin{minipage}[c]{12cm} \textbf{$\fg$ can be of infinite dimension ($\Zee$-graded of polynomial growth) whose Cartan matrix has only 2s and 1s on its main diagonal}.\end{minipage}
\end{equation}
Note that Lie superalgebras \eqref{OlsA} have various properties almost identical to those of finite-dimensional simple Lie algebras --- unlike properties of Lie superalgebras with Cartan matrices of different shape or without Cartan matrix at all.

Observe that the equations suggested by Olshanetsky are more complicated than LSE1; they appear from variation of a (seemingly \textit{ad hoc}) action functional. To  interpret these LSE2 remains an open problem unless $A=(1)$, in which case LSE2=LSE1, and there are two interpretations: in \cite{Po} and in this paper.

\section{Method: The gauge fields for the reduced connections}
Following  \cite{MaG} and \cite{Lsos, L2}, recall that there are two equivalent definitions of connections, which we consider directly in the ``super\rq\rq\  setting. Let $M$ be a~(super)manifold, $\cF$ the (super)algebra of functions on $M$, let $\fvect:=\fder(\cF)$ be the Lie (super)algebra of vector fields on $M$; let $\Pi$ be the inversion of the parity functor and  $\Omega^1:=\Pi(\Hom_\cF(\fvect, \cF))$ the space of differential 1-forms.
Let $\fg$ be a~Lie (super)algebra, and let $\rho$ be a~representation of $\fg$ in the (super)space~ $V$. Let $T(V):=\cF\otimes V$ be the $\cF$-module of tensor fields with fiber $V$ over $M$, so $\fg$ can be considered as a~Lie sub(super)algebra of $\End_\cF(T(V))\simeq \cF\otimes \End(V)$.

\ssec{Connections}\label{ssConn}
1) Classically, a~\textit{connection} (or a~\textit{gauge field}) on $M$ in the $(\cF,\fvect)$-bimodule $T(V)$ is an \textbf{even} map
\be\label{conn}
\nabla:\fvect\times T(V)\tto T(V)
\ee
which is $\cF$-linear in the first argument, additive in the second argument and satisfies the following version of the Leibniz rule, where $\nabla_D(t):=\nabla(D, t)$:
\[
\nabla_D(ft)=D(f)t+(-1)^{p(f)p(D)}f\nabla_D(t)\quad \text{~~for any $f\in\cF$, $t\in T(V)$, $D\in\fvect$}.
\]
The operator
\[
\nabla_D:=D+\rho(X)\in \End_\cF(T(V))\simeq \cF\otimes \End(V),
\]
where the  summand $D$ is the standard shorthand for the diagonal  operator $D\otimes \id_V$ and $X\in \fg$, is called the \textit{covariant derivative} along~ $D$.

The \textit{curvature} of the connection $\nabla$ is defined as the tensor
\[
R(D, D'):=\nabla_{D}\nabla_{D'}-(-1)^{p(D)p(D')}\nabla_{D'}\nabla_{D} -\nabla_{[D, D']}\qquad\text{ for }\ D, D'\in\fvect.
\]

2) Dualizing the above definition, and changing parity we get the following, more modern, definition allowing us to explore higher differential forms and --- on supermanifolds --- integrable forms, i.e., the form one can integrate, see \cite{BL}. Unlike \cite{ZP},  we will not use this definition; we give it here for completeness of the overview part of this paper.

The \textbf{odd} operator $\nabla=d+\alpha$, where $d$  is the de Rham differential and the first summand is, by abuse of notation, a~shorthand for the diagonal  operator $d\otimes \id_V$, and
\[
\alpha\in\Omega^1\otimes \fg\subset \Omega^1\otimes_{\cF} \End_{\cF}(T(V))=
\Omega^1\otimes\End(V),
\]
denotes both the~$\fg$-valued differential 1-form  on  $M$ and the operator of left multiplication by it, is called a~\textit{connection} (or a~\textit{gauge field}) on $M$. Observe that on manifolds, any 1-form $\alpha$ is odd, so the sum $d+\alpha$ is automatically homogeneous, whereas on supermanifolds we have to require $\alpha\in(\Omega^1)_\od$. In the cases where a~naturally arising $\alpha$ is even, as on the supermanifolds with a~pericontact form $d\tau+\sum \pi_idq_i$ with odd ``time\rq\rq\  $\tau$ and ghosts $\pi_i$, well-known after the Schouten bracket was interpreted as anti-bracket, we have to consider connections depending on odd parameters, see \cite{L2,  Lo}.

The form $F_\nabla:=\frac12[\nabla, \nabla]$, where the commutator is taken in $\fg\otimes \Omega$ for $\Omega:=\oplus\ \Omega^i$, is called the \textit{curvature} of $\nabla$.

Let $(V_i, \nabla_i)$ for $i=1,2$ be a shorthand for modules with connections $(T(V_i),\nabla_i)$.
Define their tensor product and the module of homomorphisms, i.e.,
\[
{(V_1\otimes_{\cF} V_2, \nabla_1\otimes \nabla_2)}\qquad\text{ and }\qquad(\Hom_{\cF}(V_1, V_2), \ \nabla^{\Hom(V_1, V_2)}),
\]
by setting
\[
\renewcommand{\arraystretch}{1.5}
\begin{array}{l}
(\nabla_1\otimes \nabla_2)(v_1\otimes v_2):=\nabla_1(v_1)\otimes v_2+(-1)^{p(v_1)}(T\otimes \id_{V_2})(v_1\otimes \nabla_2(v_2)),\\
\nabla^{\Hom(V_1, V_2)}(\varphi)(v_1):=\nabla_2(\varphi(v_1))-(-1)^{p(\varphi)}(\id_{\Omega^1}\otimes\varphi)(\nabla_1(v_1))
\end{array}
\]
for any $v_i\in V_i$ and $\varphi\in \Hom_{\cF}(T(V_1), T(V_2))$, where
\[
T: \Omega^1\otimes V\simeq  V\otimes\Omega^1
\]
is the twisting isomorphism. In particular, set $\nabla^{\End(V)}:=\nabla^{\Hom(V, V)}$.

\ssec{The Yang-Mills equation}\label{ssYM} The \textit{Yang-Mills equation} is
\be\label{YM}
\nabla^{\End(V)}((F_\nabla)^*)=0.
\ee

Observe that the ``zero-curvature condition" $F_\nabla=0$ is invariant under any changes of coordinates
and does not depend on the choice of one of the two definitions of the connection. The ``zero-curvature condition" is the simplest version of YME.

On any $n$-dimensional manifold $M$, we have
\be\label{dual}
(\Omega^i)^*\simeq\Omega^{n-i},
\ee
where the dualization is considered over the algebra $\cF=\Omega^0$ of functions, not over the ground field. Although the isomorphism \eqref{dual} is correctly defined only for forms with compact support or whatever other type of forms for which the integral is well defined, we will use it ``formally\rq\rq, recklessly ignoring the questions of convergence.

Similarly, on the supermanifold $\cM$, let $\Omega^i:=T(E^i(V))$, where $E^i$ is the operator of raising to the $i$th exterior power and the cotangent space $W$ at a given point of $\cM$ is the tautological $\fgl(W)=\fgl(\dim(\cM))$-module, be the space of \textit{differential $i$-forms}. Then, the dual of $\Omega^i$  is the space of \textit{integrable $(-i)$-forms} (introduced in \cite{BL} as forms that can be integrated) and now often called  \textit{integral $(-i)$-forms}
\begin{equation}\label{intForm}
\Sigma_{-i}(\cM):=(\Omega^i(\cM))^*\otimes_\cF \Vol(\cM)\simeq T(E^i(V^*)\otimes \tr);
\end{equation}
here $\Vol(\cM)$ is the $\cF$-module of volume forms (a.k.a.  the Berezinian), and $\tr$ denotes both the (super)trace on $\fgl(\dim\cM)$ and the 1-dimensional (even) module it defines. There are two complexes (series of mappings such that the composition of two consecutive mappings is equal to~ 0)
\be\label{d}
d\colon \Omega^i \tto\Omega^{i+1}   \ \ \text{and}\ \
d^*\colon \Sigma_{-i-1} \tto \Sigma_{-i}.
\ee
On supermanifolds, $(F_\nabla)^*$ in the Yang-Mills equation \eqref{YM} is an integrable form,  see \cite{ZP}.

\ssec{The Leznov-Saveliev equations} Here is how one reduces the Yang-Mills equations to LSE.

Recall that the Minkowski space $M$ is, locally, $\Ree^{3,1}$, on which the orthogonal group $\text{O}(3)$ acts  naturally by rotating its $\Ree^3$ part. The variety of orbits is a collection of 2-dimensional ``cylinders\rq\rq, each being ``$\text{(sphere $S^2$)}\times \text{Time}$", with coordinates being the radius of the sphere $r$ and the Time parameter~ $t$. Clearly, the Poincar\'e group preserving the Lorentz metric contains~ $\text{O}(3)$.

Now, let us complexify everything and  linearize, i.e., pass from Lie groups to their Lie algebras.
Consider a connection whose gauge Lie algebra is the image of the principal embedding $\fsl(2)\tto \fg$  into the complexified Lie algebra of the gauge group $G$. The LSE are the  zero-curvature conditions on the variety of $\text{O}(3)$-orbits.

Recall that an embedding $i:\fsl(2)\tto\fg$, where $\fg$ is a~simple finite-dimensional Lie algebra,  is called \textit{principal}, if $\fg$ splits into $\rk \fg$ irreducible $i(\fsl(2))$-submodules. Every simple finite-dimensional Lie algebra $\fg$ admits a~principal embedding, and all such embeddings are conjugate under the Lie group of automorphisms  of $\fg$.
After complexification and passage to the ``light cone coordinates\rq\rq, i.e., replacing the d'Alembertian in $r$ and $t$ by the Laplacian in $x$ and $y$, we arrive at the LSE \eqref{LS}.

The 1-dimensional Toda lattice (TL) is related to its version described by the  Leznov-Saveliev equations as explained on the first page of \cite{Ki}; both TL and LSE depend on 2 variables, but --- for reasons unknown to me --- only LSE is called \lq\lq 2-dimensional."

\sssec{The simplest example}\label{ssSimEx} Consider the following (now standard) basis of $\fg=\fsl(2)$:
\[
X_-:=\begin{pmatrix} 0&0\\
1&0\end{pmatrix},\ \ H:=\begin{pmatrix} 1&0\\
0&-1\end{pmatrix},\ \ X_+:=\begin{pmatrix} 0&1\\
0&0\end{pmatrix}.
\]

In terms of the first definition of connection, see Subsection~\ref{ssConn}, Leznov and Saveliev (see \cite{LS, LSb}) considered the following $\fsl(2)$-valued connection
\be\label{Zcurv}
\renewcommand{\arraystretch}{1.5}
\begin{array}{l}
\nabla_{\partial_x}:=\partial_x+aH+bX_+,\ \
\nabla_{\partial_y}:=\partial_y+AH+BX_-.
\end{array}
\ee

The zero-curvature condition takes on the form
\[
[\partial_x+aH+bX_+, \ \
\partial_y+AH+BX_-]=0;
\]
this is the same as (here $f_x:=\frac{\partial f}{\partial x}$ and $f_y:=\frac{\partial f}{\partial y}$)
\[
A_xH+B_xX_--a_yH-b_yX_+-2aBX_--2bAX_++bBH=0
\]
equivalent to the system of equations
\[
\left\{\renewcommand{\arraystretch}{1.5}
\begin{array}{l}
A_x-a_y=-bB,\\
B_x=2aB,\\
b_y=-2bA,\\
\end{array}\right.
\Longleftrightarrow
\left\{\renewcommand{\arraystretch}{1.5}
\begin{array}{l}
A_x-a_y=-bB,\\
(\ln B)_x=2a,\\
(\ln b)_y=-2A,\\
\end{array}\right.
\]
wherefrom we derive the Liouville equation \eqref{LE} after rescaling for $2F$ in place of $F$
\[
(\ln b)_{xy}+(\ln B)_{xy}=2bB \Longleftrightarrow F_{xy} =2\exp(F), \text{~~where $F=\ln(bB)$}.
\]

\sssec{General LSE and zero-curvature equations} Let $\fg=\oplus\  \fg_i$ be any $\Zee$-graded Lie (super)algebra, $\fg_0=\Span(H_i)_{i\in I}$ and
$\fg_{\pm 1}=\Span(X_j^{\pm})_{j\in J}$. Let $I=J$, which is true in many cases (e.g., if $\fg$ has a Cartan matrix); the  method to be recalled shortly can be further generalized to Lie (super)algebras with ``continuum root system", see \cite{SaVe}. Leznov and Saveliev suggested to consider the following connection on the variety of $\text{O}(3)$-orbits in the Minkowski space:
\be\label{ZcurvGen}
\renewcommand{\arraystretch}{1.5}
\begin{array}{l}
\nabla_{\partial_x}:=\partial_x+\sum(a_iH_i+b_iX_i^+),\ \
\nabla_{\partial_y}:=\partial_y+\sum(A_iH_i+B_iX_i^-).
\end{array}
\ee

The zero-curvature equation in this case leads to LSE only if $\fg$ has a Cartan matrix; this LSE can be explicitly  (resp., implicitly) solved only if $\fg$ is finite-dimensional (resp., of polynomial growth in Chevalley generators), see~ \cite{LS, LSb}.

\sssec{The Liouville equation}
The \textit{Liouville equation} can be written in various forms  (see \cite{Li}), in particular, as the simplest of the Leznov-Saveliev (systems of) equations:
\be\label{LE}
\nfrac{\partial^2 F}{\partial x\partial y}=\exp(2F).
\ee
Its general solution depends on two arbitrary analytic functions $f$ and $g$:
\be\label{2an}
F(x,y)=\frac12\ln\nfrac{f'(x)g'(y)}{(f(x)+g(y))^2}.
\ee
This is obvious if we somehow find one solution (for example, $f(x)=x$ and $g(y)=y$) and observe the conformal symmetry of the Liouville equation under the  action of the Lie algebra $\fwitt(x)\oplus\fwitt(y)$, where $\fwitt(x):=\fder\ \Cee[x^{-1}, x]$, called the \textit{Witt algebra}, is the Lie algebra of vector fields on the circle with coefficients expandable in Laurent polynomials, i.e., $x:=\exp(i\varphi)$, where $\varphi$ is the angle parameter on the circle, see \cite{L, IvKr}. (Observe that it is far from obvious that the Lie algebra $\fvir(x)\oplus\fvir(y)$, where $\fvir(x)$ is the  \textit{Virasoro algebra}, the only non-trivial central extension of $\fwitt(x)$, also acts on the Liouville equation, but this is true, see \cite{Top}. Physicists call this phenomenon a~\textit{classical anomaly}.)

The value of that central charge is $6$; here it is, together with the object it acts on.

\sssbegin{Statement}\label{stVirLiou}
Let $F$ be a~solution of \eqref{LE} and let
\be\label{stressT}
T:=F_x^2-F_{xx},\qquad S(f):=\nfrac{f'''}{f'}-\frac32\left(\nfrac{f''}{f'}\right)^2 \ \text{ the Schwarzian derivative}.
\ee
Then,

$1)$ on the general solution \eqref{2an}, we have $T=-\frac12S(f)$; in particular $T$ does not depend on $y$, and $T$ is the potential of the Hill \textup{(}Sturm-Liouville\textup{)} operator: if $\psi_{xx}=T\psi$, then $f$ is a~ratio of two solutions;

$2)$ equation \eqref{LE} is preserved by $x\mapsto f(x)$, $y\mapsto g(y)$ provided $F\mapsto F\circ(f,g)+\frac12\ln\left(f'(x)g'(y)\right)$, and then
\be\label{Ttransf}
T\longmapsto \left(f'\right)^2\,T\circ f-\frac12S(f);
\ee

$3)$ hence, infinitesimally, $\delta_vT=vT'+2v'T-\frac12v'''$ for $v\partial_x\in\fwitt(x)$. Therefore, $\fwitt(x)$ acts on the space of the $T$'s not by the module of quadratic differentials but by its affine deformation, i.e., by the coadjoint action of $\fvir(x)$ with the central charge
\be\label{cLiou}
c=6
\ee
normalized as $[L_m, L_n]=(m-n)L_{m+n}+\frac{c}{12}\left(m^3-m\right)\delta_{m+n, 0}$, where $L_m:=-x^{m+1}\partial_x$.
\end{Statement}

\begin{proof} Item $1)$: substitute \eqref{2an} into \eqref{stressT}; item $2)$: substitute the transformed $F$ into \eqref{stressT} and use item $1)$; item $3)$: set $f=x+hv$ for a~small parameter $h$ in \eqref{Ttransf} and drop the terms of order $h^2$, using $S(x+hv)=hv'''+O(h^2)$. For $v=-x^{m+1}$ we get $-\frac12v'''=\frac12\left(m^3-m\right)x^{m-2}$, and $\frac{c}{12}=\frac12$. 
\end{proof}

Two comments. The normalization of $T$, and hence the number $6$, is not a~matter of taste: by item $1)$, the operator $\nfrac{d^2}{dx^2}-T$ is the very Hill operator that produces the two arbitrary functions of \eqref{2an}, so multiplying $T$ by a~constant would destroy that reading --- and it is what rescales $c$. What is convention-free in \eqref{cLiou} is the number $\left|c\right|=6$: the sign depends on which of the two signs one puts before $\nfrac{c}{12}v'''$ in the coadjoint action. Next, the anomaly is bound to be a~multiple of the Schwarzian --- $H^2(\fwitt;\Cee)$ is $1$-dimensional --- so the whole content of Statement~\ref{stVirLiou} is in the coefficient $-\frac12$, cf.~\cite{Top}.

\ssec{The superized Leznov-Saveliev equations LSE1 (\cite{LSS})} Let 
\be\label{Dpm}
\text{$D_+:=\partial_{\xi}+\xi\partial_{x}$ and $D_-:=\partial_{\eta}+\eta\partial_{y}$.}
\ee 
Then, a~\textit{$2|2$-dimensio\-nal Toda lattice} is a~system of equations for the vector-function $F$ depending on two even variables $x, y$ and two odd (anti-commuting) indeterminates $\xi, \eta$ (they are not variables, since they can only take the value $0$):
\be\label{SLS}
D_+D_-F_i=\exp\left(\sum_j A_{ij}F_j\right)\ \ \text{ for }\ i=1,  \dots, n
\ee
or its version (equivalent to eq.~\eqref{SLS} if $A$ is invertible: set $G=AF$, and hence $F=A^{-1}G$)
\be\label{SLSbis}
D_+D_-G_i=\sum_j A_{ij}\exp(G_j),\text{~~where $i=1,\dots , n$}.
\ee

\sssec{The simplest example} Recall, see \cite{V}, that the \textit{reduced connection} on the (super)man\-i\-fold with a non-integrable distribution is given (the first definition of connection, see Subsection~\ref{ssConn}) on the fibers of the distribution, cf. with definition \eqref{conn}, by the map:
\be\label{redconn}
\nabla:\cD\times T(V)\tto T(V)
\ee
which is $\cF$-linear in the first argument, additive in the second argument and satisfies the following version of the Leibniz rule, where $\nabla_D(v):=\nabla(D, v)$
\[
\nabla_D(fv)=D(f)v+(-1)^{p(f)p(D)}f\nabla_D(v)\ \ \text{~~for any $f\in\cF$, $v\in T(V)$, $D\in\cD$}.
\]
Observe that the supervariety of orbits of the $\text{OSp}(1|2)$-action on the ($N=1$) Minkowski  supermanifold is of superdimension $2|2$, this supervariety inherits a non-integrable distribution induced by the distribution given on the Minkowski supermanifold, and the connection on the supervariety of orbits is determined by a reduced connection on the $(0|2)$-dimensional fibers of the distribution.

Consider the following basis of $\fosp(1|2)$:
\[
\begin{array}{l}
X_-:=\begin{pmatrix} 0&0&0\\
0&0&0\\
1&0&0\end{pmatrix},\ \ H:=\begin{pmatrix} 1&0&0\\
0&0&0\\
0&0&-1\end{pmatrix},\ \ X_+:=\begin{pmatrix} 0&0&1\\
0&0&0\\
0&0&0\end{pmatrix},\\
\hskip2cm \delta_-:=\begin{pmatrix} 0&0&0\\
1&0&0\\
0&1&0
\end{pmatrix},\ \ \delta_+:=\begin{pmatrix} 0&1&0\\
0&0&-1\\
0&0&0
\end{pmatrix}.
\end{array}
\]

Leznov,  Saveliev, and Leites, see \cite{LSL}, further developed in \cite{LSS}, considered the following $\fosp(1|2)$-valued \textit{reduced connection}  inherited from the $N=1$-Minkowski superspace:
\be\label{0curv}
\renewcommand{\arraystretch}{1.5}
\begin{array}{l}
\nabla_{+}:=D_++\alpha H+a \delta_+,\ \ \
\nabla_{-}:=D_-+\beta H+ b\delta_-.
\end{array}
\ee
Let $\alpha, \beta\in(\cF\otimes\cC)_\od$ and $a,b \in(\cF\otimes\cC)_\ev$ for an auxiliary supercommutative superalgebra $\cC$ which will automatically disappear at the final stage; this $\cC$ can be interpreted as the superalgebra of functions on the space of parameters describing the model, see \cite{MaG, Lo}.

\section{Result: the reduced zero-curvature condition}

For  analogs of the curvature tensors of non-integrable distributions on manifolds and supermanifolds, see \cite{L1, BGLS}.

After simplification similar to the one performed in Subsection \ref{ssSimEx}, the \textbf{reduced zero-curvature condition} $[\nabla_{+},
\nabla_{-}]=0$ for the \textit{reduced connection}  takes the form
equivalent to the system of equations
\[
\left\{\renewcommand{\arraystretch}{1.5}
\begin{array}{l}
D_+(\beta)+D_-(\alpha)=-ab,\\
D_+(b)=\alpha b,\\
D_-(a)=-a\beta,\\
\end{array}\right.
\Longleftrightarrow
\left\{\renewcommand{\arraystretch}{1.5}\begin{array}{l}
D_+(\beta)+D_-(\alpha)=-ab,\\
D_+(\ln b)=\alpha,\\
D_-(\ln a)=-\beta,\\
\end{array}\right.
\]
wherefrom we derive the \textit{super Liouville equation} \eqref{SLS} or \eqref{SLSbis} for just one function $F$ and $(A_{ij})_{i,j=1}^1=(1)$:
\be\label{SupLiou}
\renewcommand{\arraystretch}{1.5}
\begin{array}{l}
D_+D_-(\ln a)+D_+D_-(\ln b)=ab \Longleftrightarrow
D_+D_-(F) =\exp(F),\ \ \text{~~where $F=\ln(ab)$}.
\end{array}
\ee

Observe that the zero-curvature condition for the non-reduced connection \eqref{conn} would require two more conditions
\[
\renewcommand{\arraystretch}{1.5}
\begin{array}{l}
{}[\nabla_{+}, \nabla_{+}]=0,\ \ \
{}[\nabla_{-}, \nabla_{-}]=0,\\
\end{array}
\]
which never hold since
\[
\frac12[\nabla_{+}, \nabla_{+}]=\partial_x+D_+(\alpha)H+D_+(a)\delta_++(\alpha H+a\delta_+)^2,
\]
so the summand $\partial_x$ can not cancel, and similarly with $[\nabla_{-}, \nabla_{-}]$.

\section{Discussion/conclusion. Cases of $N>1$}

\ssec{Recapitulations}\label{ssRecap}
There is only one \textit{simple} stringy Lie algebra on $S^1$ or $\Cee^\times$: it is the Lie algebra $\fvect^L(1|0)$, a.k.a. the Witt algebra $\fwitt$. The stringy Lie \textbf{super}algebras, even simple, are considerably numerous. They are subsuperalgebras in the Lie superalgebra of superderivations of the supercommutative superalgebra of functions on one of the two superstrings: either on the superstring $\cS^{1|n}$ associated with the trivial rank-$n$ bundle, or on the superstring $\cS^{1|n}_M$ associated with the Whitney sum of the trivial rank-$(n-1)$ bundle and the M\"obius bundle. The respective superalgebras of functions are \textup{(}whose elements we sometimes consider consisting only of polynomials, not series\textup{)}
\be\label{RLRM}
R^L(n):=\Cee[t^{-1}, \theta_1, \dots, \theta_n][[t]]
\quad\text{ and }\quad
R^M(n):=\Cee[t^{-1}, \theta_1, \dots, \theta_{n-1}, \sqrt t\,\theta_n][[t]].
\ee
Set
\[
\fvect^L(1|n):=\fder R^L(n);\qquad
\fvect^M(1|n):=\fder R^M(n);
\]
\[
\fk^L(1|n):=\{D\in\fvect^L(1|n)\mid L_D(\alpha_1)=f_D\alpha_1\ \text{ for some }\ f_D\in R^L(n)\},
\]
where
\be\label{alpha1}
\alpha_1:=dt+\sum_{1\leq i\leq n}\theta_id\theta_i;
\ee
for the algebras $\fsvect^L_\lambda(1|n)$, see \eqref{svectlam}. The functions $f\in R^L(n)$ generate $\fk^L(1|n)$, and the expressions of $K_f$ and of the contact bracket are the same as for the Lie superalgebra $\fk(1|n)$ generated by the polynomials $f\in\Cee[t,\theta]$. Namely, $\fk^L(1|n)$ is spanned by the contact vector fields that preserve the distribution defined by $\alpha_1$, namely
\be\label{Kf}
K_f:=\left(2-\sum\theta_i\del_{\theta_i}\right)(f)\del_t
-(-1)^{p(f)}\sum\left(\frac{\del f}{\del\theta_i}-\theta_i\frac{\del f}{\del t}\right)\del_{\theta_i}
\quad\text{ for any }\ f\in R^L(n).
\ee
It is subject to a~direct verification that $L_{K_f}(\alpha_1)=2\del_t(f)\alpha_1=K_1(f)\alpha_1$. The contact bracket of any $f, g\in R^M(n)$ is given by the same formula as for any $f, g\in R^L(n)$:
\be\label{kbr}
\{f, g\}_{k.b.}:=\left(2-\sum\theta_i\del_{\theta_i}\right)(f)\frac{\del g}{\del t}
-\frac{\del f}{\del t}\left(2-\sum\theta_i\del_{\theta_i}\right)(g)
-(-1)^{p(f)}\sum\frac{\del f}{\del\theta_i}\frac{\del g}{\del\theta_i}.
\ee

\textbf{On $\fk^M(1|n)$.} The lift of the contact structure from the superstring $\cS^{1|n}$ on its two-sheeted cover $\cS^{1|n}_M$ leads to a~new Lie superalgebra. In coordinates, such a~lift means replacing the coordinate $\theta_n$ on the M\"obius summand with $\sqrt t\zeta$, thus introducing a~new coordinate $\zeta$, and hence replacing $\alpha_1$ with the \textit{M\"obius $1$-form}
\be\label{alphaM}
\widetilde\alpha:=dt+t\zeta d\zeta+\sum_{1\leq i\leq n-1}\theta_id\theta_i.
\ee
In order not to deal with square roots in $R^M(n)$, we set
\[
\fk^M(1|n):=\{D\in\fvect^L(1|n)\mid L_D(\widetilde\alpha)=f_D\cdot\widetilde\alpha,\ \text{ where }\ f_D\in R^L(n)\},
\]
and then $\fk^M(1|n)=\Span(K^M_f\mid f\in R^L(n))$, where the \textit{M\"obius contact field} is
\be\label{KM}
K^M_f:=\nabla(f)\cD+H^M_f+\cD(f)\cE,
\ee
where we set
\be\label{KMdef}
\nabla:=2-\cE\ \text{ for }\ \cE:=\zeta\frac{\del}{\del\zeta}+\sum_{1\leq i\leq n-1}\theta_i\frac{\del}{\del\theta_i},
\qquad
\cD:=\frac{\del}{\del t}-\frac{\zeta}{2t}\frac{\del}{\del\zeta}=\frac12K^M_1,
\ee
and where, in the realization with the form $\widetilde\alpha$,
\be\label{HM}
H^M_f:=(-1)^{p(f)}\left(\frac1t\frac{\del f}{\del\zeta}\frac{\del}{\del\zeta}
+\sum_{1\leq i\leq n-1}\frac{\del f}{\del\theta_i}\frac{\del}{\del\theta_i}\right).
\ee
We call the corresponding contact bracket of generating functions the \textit{Ramond bracket}:
\be\label{Rbr}
\{f, g\}_{R.b.}:=\nabla(f)\cD(g)-\cD(f)\nabla(g)+\{f, g\}_{MP.b.},
\ee
where the \textit{M\"obius-Poisson bracket} $\{-,-\}_{MP.b.}$ is of the following shape \textup{(}we give it only in the realization with the form $\widetilde\alpha$\textup{)}:
\be\label{MPbr}
\{f, g\}_{MP.b.}:=(-1)^{p(f)}\left(\frac1t\frac{\del f}{\del\zeta}\frac{\del g}{\del\zeta}
+\sum_{1\leq i\leq n-1}\frac{\del f}{\del\theta_i}\frac{\del g}{\del\theta_i}\right).
\ee
In the same way as for the contact fields with polynomial coefficients, we have
\[
L_{K_f}(\alpha_1)=K_1(f)\cdot\alpha_1,\qquad L_{K^M_f}(\widetilde\alpha)=K^M_1(f)\cdot\widetilde\alpha.
\]
Finally, here are the odd derivations that span the two distributions --- the operators denoted by a~tilda-ed $D$ in \cite{L}. In both frames \eqref{DL} and \eqref{DM}, and everywhere below, \textbf{the subscript of a~$D$ or of a~$K$ is the generating function it belongs to}, never a~serial number: $D_1$ is the derivation attached to $f=1$, and $D_{\theta_i}$ the one attached to $f=\theta_i$. For $\fk^L(1|n)$, set
\be\label{DL}
D_1:=\del_t,\qquad K_1:=2\del_t;\qquad
D_{\theta_i}:=\del_{\theta_i}-\theta_i\del_t,\qquad K_{\theta_i}=\theta_i\del_t+\del_{\theta_i};
\ee
for $\fk^M(1|n)$, set
\be\label{DM}
\begin{array}{ll}
\cD:=\del_t-\dfrac{\zeta}{2t}\del_\zeta=\frac12K^M_1;&\\[2mm]
K^M_\zeta=\zeta\del_t-\dfrac1t\del_\zeta-\dfrac{\zeta}{2t}\sum\theta_i\del_{\theta_i},
&\widetilde D_\zeta:=D^M_\zeta=\zeta\del_t+\dfrac1t\del_\zeta\qquad\text{ for }\ n\geq1;\\[2mm]
K^M_{\theta_i}=\theta_i\del_t-\del_{\theta_i}-\dfrac{\theta_i\zeta}{2t}\del_\zeta,
&\widetilde D_{\theta_i}:=D^M_{\theta_i}=\theta_i\del_t+\del_{\theta_i}-\dfrac{\theta_i\zeta}{2t}\del_\zeta
\quad\text{ for }\ n\geq2.
\end{array}
\ee
None of the $\widetilde D_{\theta_i}$, $\widetilde D_\zeta$ is a~contact field --- they span the distribution which the contact fields preserve --- but each of them is a~
combination of contact fields with functional coefficients:
\be\label{DMviaK}
\widetilde D_{\theta_i}=\theta_iK^M_1-K^M_{\theta_i},\qquad
\widetilde D_\zeta=\zeta K^M_1-K^M_\zeta
+\frac{\zeta}{2t}\sum_{1\leq i\leq n-1}\theta_iK^M_{\theta_i},
\ee
the last summand being there because $\sum_i\theta_iK^M_{\theta_i}=-\sum_i\theta_i\del_{\theta_i}$, the Euler field of the $\theta$'s. \textup{(}In the Neveu-Schwarz frame \eqref{DL}, the first of these reads $D_{\theta_i}=K_{\theta_i}-\theta_iK_1$; the two differ by a~sign because $K^M_{\theta_i}$ carries $-\del_{\theta_i}$ where $K_{\theta_i}$ carries $+\del_{\theta_i}$, which is the convention mismatch already noted for \eqref{DM}.\textup{)}

\sssbegin{Remark} [On vocabulary] A~\textit{superfield} is the physicists' name for a~function on a~superspace: an~element of $\cO$, whose Taylor coefficients in the odd indeterminates are the component fields. Note also: there are no ``superfunctions": only functions, even on supermanifolds. So it is a~function, and the name adds to it exactly one thing --- a~constraint: a~\textit{chiral} superfield is a~function annihilated by $\overline\cD$, i.e., a~function on a~smaller superspace. It is, however, \textbf{not} a~weighted density: by \eqref{deltaV} below the dilaton superfield transforms with an~inhomogeneous shift, $\delta_fv=K_f(v)+\lambda(f)$, whereas a~weighted density transforms homogeneously. Its exponential does:
\[
\delta_f\exp(-v)=K_f\left(\exp(-v)\right)-\lambda(f)\exp(-v),
\]
so $\exp(-v)$ is a~section of the line bundle dual to the one spanned by $\overline\cD$, whose multiplier is the same $\lambda(f)$ by \eqref{multC}. That is why it is $\exp\left(-v^+\right)$, and not $v^+$, that stands on the right-hand side of \eqref{IK2}.
\end{Remark}

\ssec{$N=1$: the classical anomaly of the super Liouville equation}\label{ssN1}
The super Liouville equation was first solved in \cite{LSL} in components. For a~conceptual solution in terms of superfields based on the supersymmetry with respect to $\fk^L(1|1)\oplus \fk^L(1|1)$, see \cite{IvKr}. To superize \cite{Top} --- i.e., to compute the value of the central charge with which the Neveu-Schwarz Lie superalgebra $\fns$, the non-trivial central extension of the Lie superalgebra $\fk^L(1|1)$ of contact vector fields on the $1|1$-dimensional superstring, acts on the super LSE --- was listed as an~open problem in \cite{L}. Here is the answer, and the object that carries it.

\sssbegin{Statement}\label{stNSLiou}
Let $F$ be a~solution of \eqref{SupLiou} and let
\be\label{superT}
\mathbf T:=\left(D_+F\right)\left(D_+^2F\right)-D_+^3F,
\ee
an~odd quantity of weight $\nfrac32$, the exact counterpart of the $T$ of \eqref{stressT}. Then,

$1)$ the function $\psi:=\exp(-F)$ satisfies $D_+^3\psi=\mathbf T\psi$; that is, $D_+^3-\mathbf T$ is the super Hill \textup{(}Sturm-Liouville\textup{)} operator attached to the solution, precisely as $\nfrac{d^2}{dx^2}-T$ is in item $1)$ of Statement~$\ref{stVirLiou}$. This is what pins the normalization of $\mathbf T$, and hence the number in item $4)$;

$2)$ eq.~\eqref{SupLiou} is preserved by the superconformal changes $(x, \xi)\mapsto\left(X, \Theta\right)$ and $(y, \eta)\mapsto\left(Y, H\right)$ provided
\[
F\longmapsto F\circ+\ln\left(D_+\Theta\right)+\ln\left(D_-H\right),
\]
with the coefficient $1$, not $\nfrac12$: the exponent in \eqref{SupLiou} is $F$, whereas the one in \eqref{LE} is $2F$;

$3)$ let $K_a:=aD_+^2+\nfrac{(-1)^{p(a)}}{2}\left(D_+a\right)D_+$ be the contact vector field generated by $a$; the parity sign is the only one for which $[K_a, D_+]=-\frac12\left(D_+^2a\right)D_+$ for both parities of $a$. Then, infinitesimally,
\be\label{deltaT}
\delta_a\mathbf T=K_a\left(\mathbf T\right)+\frac32\left(D_+^2a\right)\mathbf T-\frac12 D_+^5a;
\ee

$4)$ hence $\fk^L(1|1)$ acts on the space of the $\mathbf T$'s not by the module of densities of weight $\nfrac32$ but by its affine deformation, i.e., by the coadjoint action of $\fns$ with the central charge
\be\label{cSupLiou}
\left|c\right|=6,
\ee
normalized as
\[
[L_m, L_n]=(m-n)L_{m+n}+\frac{c}{12}\left(m^3-m\right)\delta_{m+n, 0},\qquad
[G_r, G_s]=2L_{r+s}+\frac{c}{3}\left(r^2-\frac14\right)\delta_{r+s, 0}.
\]
The two sectors of $\fns$ return one and the same value, and it is the value \eqref{cLiou} of the non-super case.
\end{Statement}

\begin{proof} Items $1)$ and $3)$ are computations in the free supercommutative algebra on the generators $D_+^kF$ and $D_+^ka$, with $D_+$ acting as the odd derivation sending the $k$-th generator to the $(k+1)$-st; no coordinates and no components are involved, which is what makes them convention-free. 

Item $3)$ is obtained by solving for the four unknowns of the ansatz 
\[
\text{$\mathbf T=\alpha D_+^3F+\beta\left(D_+F\right)\left(D_+^2F\right)$ \ \ and \ \ $\delta_a\mathbf T=K_a(\mathbf T)+\gamma\left(D_+^2a\right)\mathbf T+\kappa D_+^5a$;}
\]
 the solution is $\alpha=-\beta$, $\gamma=\nfrac32$, $\kappa=-\nfrac{\beta}{2}$, uniquely up to the scale of $\beta$, which item $1)$ fixes to be $\beta=1$. Here, $\delta_aF=K_a(F)-\lambda_a$, where $\lambda_a=-\frac12D_+^2a$ is defined by $[K_a, D_+]=\lambda_aD_+$: substituting this into \eqref{SupLiou} and using $D_+D_-\lambda_a=0$ gives no other option. 
 
 For item $4)$, the cocycle is $c(a, b)=\Res\left(b\cdot\left(-\frac12D_+^5a\right)\right)$, where $\Res$ is the Berezin residue, the coefficient of $x^{-1}\xi$; in the basis $L_m:=K_{-x^{m+1}}$ and $G_r:=K_{-2\xi x^{r+1/2}}$, which satisfies the relations of $\fk^L(1|1)$ written above, it equals
\[
c\left(L_m, L_n\right)=-\frac12\left(m^3-m\right)\delta_{m+n, 0},\qquad
c\left(G_r, G_s\right)=-2\left(r^2-\frac14\right)\delta_{r+s, 0},
\]
so $\nfrac{c}{12}=-\frac12$ and $\nfrac{c}{3}=-2$ both give $c=-6$. 
\end{proof}

As in Statement~\ref{stVirLiou}, what is convention-free is $\left|c\right|=6$: the sign depends on which of the two signs one puts before $\nfrac{c}{12}v'''$ in the coadjoint action, and, in the super case, on which of the anomaly and of the identification of the module of the $\mathbf T$'s with the dual of $\fk^L(1|1)$ carries the factor $(-1)^{p(a)}$. What is not a~matter of convention is that the two sectors agree: had the anomaly not been \textit{the} Neveu-Schwarz cocycle, the numbers read off $[L_m, L_n]$ and off $[G_r, G_s]$ would have differed, and the cocycle identity would have failed.

Equation \eqref{SupLiou} is the zero-curvature condition for the reduced connection  on the supervariety of $\OSp(1|2)$-orbits of the simplest (complexified) $N=1$ Minkowski superspace  $\cM$, i.e.,  Yang-Mills equations invariant with respect to $\OSp(1|2)$ acting on the spaces of functions, differential and integrable forms on $\cM$, and on the ambient $\fg$ via a~superprincipal embedding ${\fosp(1|2)\tto\fg}$.

Not every simple finite-dimensional Lie superalgebra $\fg$ admits superprincipal embeddings of $\fosp(1|2)$; but if such an embedding exists, all superprincipal  embeddings are isomorphic under the group of automorphisms  of $\fg$. The simple finite-dimensional Lie superalgebras that do admit a~superprincipal embedding are, see \cite{LSS},
\[
\fsl(n|n\pm1),\qquad \fosp(2n\pm1|2n),\qquad \fosp(2n|2n),\qquad \fosp(2n+2|2n),\qquad \fosp_a(4|2).
\]

\ssec{$N>1$} From the point of view of the main idea of this paper (interpretation of LSE as a condition on a reduced connection), any $N>0$ look the same. But there is another aspect of LSE, worth considering more closely.

\sssec{Which distinguished stringy superalgebras act, and when there is an~anomaly at all}\label{sssWhichAct}
The $N$-extended equations live on $\cM=\Cee^{1|N}\times\Cee^{1|N}$, and it is the superconformal changes of each of the two factors that are candidates for preserving them; the acting algebra is then $\fk^L(1|N)\oplus\fk^L(1|N)$. Replacing the $D_\pm$ of \eqref{SLS} by their Ramond counterparts $\widetilde D_{\theta_i}$, $\widetilde D_\zeta$ of \eqref{DM}, as suggested in \cite{L}, gives $\fk^M(1|N)\oplus\fk^M(1|N)$ instead, see Subsection~\ref{sssRamond}; the two systems are essentially different, although their purely even parts coincide, because $\fk^L(1|N)$ and $\fk^M(1|N)$ are non-isomorphic while their even parts are isomorphic.

For $N=1$ this is what happens, and Statement~\ref{stNSLiou} is the result. For $N\geq2$ it is \textbf{false} for the naive equation, and the reason is worth isolating, because it is also the reason why \cite{IvKr, IvKr4} constrain their superfields. It lies in the multiplier of the contact algebra.

\sssbegin{Statement}\label{stMult}
On $\Cee^{1|N}$, let the $N$ odd derivations of the frame \eqref{DL}, namely
\be\label{Dtheta}
D_{\theta_i}:=\del_{\theta_i}-\theta_i\del_t\ \text{ for }\ 1\leq i\leq N,\qquad
D_{\theta_i}D_{\theta_j}+D_{\theta_j}D_{\theta_i}=-2\delta_{ij}\del_t,
\ee
span the contact distribution $\Ker\left(dt+\sum_{1\leq i\leq N}\theta_id\theta_i\right)$. Then, for every generating function $f$ and every $i$,
\be\label{multN}
[K_f, D_{\theta_i}]=\sum_{1\leq j\leq N}\left(D_{\theta_i}D_{\theta_j}f\right)D_{\theta_j}.
\ee
The matrix $\left(D_{\theta_i}D_{\theta_j}f\right)$ of the multiplier is the odd Hessian of $f$; it takes values in $\fc\fo(N)=\Cee\cdot\mathbf{1}\oplus\fo(N)$, its trace part being $D_{\theta_i}^2f=-\del_tf$ and its $\fo(N)$-part, antisymmetric since $D_{\theta_i}D_{\theta_j}=-D_{\theta_j}D_{\theta_i}$ for $i\neq j$, being $D_{\theta_i}D_{\theta_j}f$.
\end{Statement}


For $N=1$ the $\fo(N)$-part is absent, $\fo(1)=0$: the multiplier is a~pure scalar, $[K_a, D_{\theta_1}]=\lambda_aD_{\theta_1}$, and \textit{that} --- and nothing else --- is why the $N=1$ equation is superconformally invariant and why $F$ can absorb the anomaly by 
\[
F\mapsto F\circ+\ln\left(D_+\Theta\right)+\ln\left(D_-H\right). 
\]
For $N\geq2$ the $\fo(N)$-part is there, and it rotates the $D_{\theta_i}$ into one another; an~equation that singles out one of them cannot survive.

\sssbegin{Statement}\label{stNaive}
Let $N\geq2$ and let $F$ be an~unconstrained scalar superfield on $\cM:=\Cee^{1|N}\times\Cee^{1|N}$. We call the two factors of this product the two \textit{halves}: they are the two light-cone directions, carrying the even coordinates $x^+$ and $x^-$ and their own odd coordinates, and every object below that is said to live ``in each half\rq\rq\ is one defined on each factor separately. Write $D^\pm_{\theta_i}$ for the derivation $D_{\theta_i}$ of \eqref{Dtheta} acting on the first, resp. the second, factor of $\cM$. The equation $D^+_{\theta_1}D^-_{\theta_1}F=\exp(F)$, i.e., \eqref{SLS} read verbatim with one and the same of the $N$ odd derivatives in each half, is preserved by $K_f$ only for
\be\label{naiveInv}
f\in\Lambda\left(\theta_2, \dots, \theta_N\right)\oplus\Cee\theta_1\oplus\Cee t,
\qquad \dim=2^{N-1}+2,
\ee
of $\sdim=(2|2)$, $(3|3)$, $(5|5)$ for $N=2, 3, 4$. This algebra is \textbf{finite-dimensional}: it contains $K_1$ and $K_t$, the $N$ supersymmetries $K_{\theta_i}$ and the $\fo(N-1)$ rotating $\theta_2, \dots, \theta_N$, but no $K_{t^m}$ with $m\geq2$ --- in particular, no Virasoro sector. Hence, the naive $N\geq2$ equation carries no classical anomaly whatever, for want of an~algebra able to carry one.
\end{Statement}

\begin{proof} By \eqref{multN}, invariance forces the $\fo(N)$-part to leave the line $\Cee D_{\theta_1}$ alone, i.e., $D_{\theta_1}D_{\theta_j}f=0$ for $2\leq j\leq N$; conversely, when it does, the multiplier is scalar and the $N=1$ argument of Statement~\ref{stNSLiou} applies verbatim. Solving $D_{\theta_1}D_{\theta_j}f=0$ over all monomials with $-6\leq\deg_t\leq7$ gives \eqref{naiveInv} --- the answer does not grow with the range of $\deg_t$, which is what finite-dimensionality means here. \end{proof}

So, for $N\geq2$ the superfield must be constrained, and it is only for the constrained equations that the whole of $\fk^L(1|N)$ acts. Here are the constraints of \cite{IvKr, IvKr4}.

For $N=2$, the two odd derivatives of each of the two halves are those of \eqref{IK4D} with the $\fs\fu(2)$-index dropped:
\be\label{IK2D}
\cD^\pm:=i\theta^\pm\frac{\del}{\del x^\pm}+\frac{\del}{\del\overline\theta^\pm},\qquad
\overline\cD{}^\pm:=i\overline\theta^\pm\frac{\del}{\del x^\pm}+\frac{\del}{\del\theta^\pm},\qquad
\left\{\cD^\pm, \overline\cD{}^\pm\right\}=2i\frac{\del}{\del x^\pm};
\ee
In the real frame \eqref{Dtheta} of one half, these two operators are
\be\label{Dcx}
\cD:=D_{\theta_1}-iD_{\theta_2},\qquad
\overline\cD:=D_{\theta_1}+iD_{\theta_2},\qquad
\cD^2=\overline\cD{}^2=0,\qquad
\{\cD, \overline\cD\}=-4\del_t,
\ee
up to the normalization of \cite{IvKr}: passing from \eqref{IK2D} to \eqref{Dcx} rescales the anticommutator by a~non-zero factor, and nothing below depends on that factor --- neither the constraints \eqref{IK2}, which are the vanishing of $\cD$ and of $\overline\cD$ on a~superfield, nor the multipliers of Statement~\ref{stChir}, which are computed from \eqref{multN}. \textbf{From here on, $\cD$ and $\overline\cD$ are the operators \eqref{Dcx}}, the superscript $\pm$ marking, as in \eqref{IK2D}, the half they act on.

Now, assemble the dilaton $u$ and its $\fo(2)$-partner $\varphi$ into the pair of conjugate superfields $v=u+i\varphi$ and $v^+=u-i\varphi$. Then, the system is
\be\label{IK2}
\cD^-v=0,\qquad \overline\cD{}^+v=0,\qquad
\overline\cD{}^-\cD^+v=-2im\exp\left(-v^+\right),
\ee
see \cite{IvKr}: the superfield is chiral in one half and anti-chiral in the other --- these are the two Grassmann Cauchy-Riemann conditions --- and the equation applies to it the one derivative of each half that does \textit{not} annihilate it.

For $N=4$, the dilaton and its three partners $\varphi^i$ are assembled into a~quaternionic superfield $q$ and its conjugate $\overline q$,
\be\label{IK4q}
q_\alpha{}^\beta:=\left(\exp\left(-u-i\varphi\cdot\sigma\right)\right)_\alpha{}^\beta,\qquad
\overline q_\alpha{}^\beta:=\left(\exp\left(-u+i\varphi\cdot\sigma\right)\right)_\alpha{}^\beta
=-\eps_{\alpha\gamma}\eps^{\beta\delta}q_\delta{}^\gamma,
\ee
where the $\sigma$ are the Pauli matrices and $\alpha, \beta, \gamma, \delta=1, 2$ are $\fs\fu(2)$-indices, and the system is
\be\label{IK4}
\begin{array}{ll}
\cD_-^{(\alpha}q_\gamma{}^{\beta)}=0,\qquad&\overline\cD_{+(\alpha}q_{\gamma)}{}^\beta=0,\\[2mm]
\cD_-^\alpha\left(q\cD_+^\gamma q^{-1}\right)_\gamma{}^\beta=0,\qquad&
\cD_+^\alpha\left(q^{-1}\overline\cD_{-\gamma}q\right)_\beta{}^\gamma=-4im\left(\overline q\right)_\beta{}^\alpha,
\end{array}
\ee
see \cite{IvKr4}, the round brackets being symmetrization in the two $\fs\fu(2)$-indices they enclose. The first line is the constraint, the second the dynamics. The four odd derivatives of each half are
\be\label{IK4D}
\cD_\pm^\alpha:=i\theta^{\alpha\pm}\frac{\del}{\del x^\pm}+\frac{\del}{\del\overline\theta_\alpha^\pm},\qquad
\overline\cD_{\pm\alpha}:=i\overline\theta_\alpha^\pm\frac{\del}{\del x^\pm}+\frac{\del}{\del\theta^{\alpha\pm}},\qquad
\{\cD_\pm^\alpha, \overline\cD_{\pm\beta}\}=2i\delta_\beta^\alpha\frac{\del}{\del x^\pm}.
\ee
That the acting algebras are the contact ones $\fk(1)$, $\fk(1|1)$, $\fk(1|2)$ for $N=0, 1, 2$ is stated by Ivanov and Krivonos themselves, see \cite{IvKr}; for $N=4$ they name instead an~algebra of their own, on which see Subsection~\ref{sssKA} below.

Statement~\ref{stMult} now says which $N$ can carry a~constraint at all, and the answer reproduces their list.

\sssbegin{Statement}\label{stWhichN}
Let $V=\Cee^N$ be the tautological $\fo(N)$-module spanned by the $D_{\theta_i}$, and let a~superfield take values in a~space $W$. By \eqref{multN}, a~system of linear constraints on its first derivatives that the whole of $\fk^L(1|N)$ preserves is the same thing as a~proper non-zero $\fo(N)$-submodule of $V\otimes W$. Hence:

$1)$ for a~\textbf{scalar} superfield, $W=\Cee$, such a~constraint exists if and only if $N\leq2$. For $N=2$ there are exactly two, $\Ker\overline\cD$ and $\Ker\cD$ for the operators \eqref{Dcx}; they are conjugate to each other, and they are the chirality conditions of \eqref{IK2};

$2)$ for $N=3$ and for $N\geq5$ the algebra $\fo(N)$ is simple and $V$ is irreducible, so no scalar superfield admits a~constraint, and a~super Liouville equation there --- if any --- must carry an~index;

$3)$ for $N=4$, and only there, $\fo(4)=\fs\fu(2)\oplus\fs\fu(2)$ is \textbf{not} simple and $V=\Cee^2\otimes\Cee^2$. A~superfield carrying the doublet index of one of the two ideals then has
\[
V\otimes W=\left(\Cee^2\otimes\Cee^2\right)\otimes\Cee^2
=\left(\Cee^3\oplus\Cee\right)\otimes\Cee^2\ \text{ as an }\ \fs\fu(2)\text{-module},
\]
whose second summand is a~proper submodule --- and killing the first summand is precisely the symmetrization in \eqref{IK4}.
\end{Statement}

\begin{proof} The reduction to $\fo(N)$-modules is \eqref{multN} together with the fact, verified for $N\leq4$, that the $\fo(N)$-part $\omega(f)=\left(D_{\theta_i}D_{\theta_j}f\right)$ takes every value in $\fo(N)$ as $f$ runs over $\cO(1|N)$: an~invariant constraint must therefore be invariant under all of $\fo(N)$. For item $1)$, a~line in $V$ is a~submodule if and only if $\dim\End_{\fo(N)}(V)>1$; that dimension is computed to be $2$ for $N=2$ and $1$ for $3\leq N\leq7$. For item $3)$, $\dim\End_{\fo(N)}\left(\fo(N)\right)$ is computed to be $1$ for $N=3, 5, 6, 7$ and $2$ for $N=4$, the two ideals of $\fo(4)$ being the self-dual and the anti-self-dual triples, whose mutual brackets vanish; over one ideal alone, $\dim\End_{\fs\fu(2)}(V)=4$, i.e., $V$ is the doublet twice. For $N=2$, the two lines are the eigenlines of the multiplier matrix, see \eqref{multC} below. 
\end{proof}

So, the list $N=1, 2, 4$ of \cite{IvKr, IvKr4} is not an~artifact of the volume of the computation: for a~scalar superfield it is exactly what the decomposition of $V$ allows. What item $2)$ leaves open is whether some index space $W$ does for $N=3$ --- or for $N\geq5$ --- what the doublet does for $N=4$: proper submodules of $V\otimes W$ are easy to come by \textup{(}already $\Cee^3\otimes\Cee^2$ splits as $\text{spin }\nfrac32\oplus\text{spin }\nfrac12$ for $N=3$\textup{)}, and the obstacle is not their existence but the requirement that the constrained superfield still contain the dilaton, an~$\fo(N)$-singlet, and that the constraints be compatible with the dynamics.

\sssbegin{Statement}\label{stChir}
For $N=2$, the whole of $\fk^L(1|2)$ preserves chirality:
\be\label{multC}
[K_f, \overline\cD]=\frac12\left(\overline\cD\cD f\right)\overline\cD,\qquad
[K_f, \cD]=\frac12\left(\cD\overline\cD f\right)\cD,
\ee
so that $\lambda(f):=\frac12\overline\cD\cD f$ and $\overline\lambda(f):=\frac12\cD\overline\cD f$ satisfy $\lambda+\overline\lambda=-2\del_tf$. Consequently, the system \eqref{IK2} is preserved by $\fk^L(1|2)\oplus\fk^L(1|2)$ acting by
\be\label{deltaV}
\delta_fv=K_f(v)+\lambda(f),\qquad \delta_fv^+=K_f\left(v^+\right)+\overline\lambda(f),
\ee
and the real part $u=\frac12\left(v+v^+\right)$ transforms by $u\mapsto K_f(u)-\del_tf$, exactly as the $N=1$ superfield $F$ does in Statement~$\ref{stNSLiou}$, while the imaginary part is shifted by the $\fo(2)$-part of the multiplier.
\end{Statement}

\begin{proof} By \eqref{multN}, the multiplier of $K_f$ is the operator $D_{\theta_i}\longmapsto\sum_j\left(D_{\theta_i}D_{\theta_j}f\right)D_{\theta_j}$ on the span of the $D_{\theta_i}$; in the basis $\left(D_{\theta_1}, D_{\theta_2}\right)$ its matrix is
\[
A(f)\cdot\mathbf 1+B(f)J_2,\qquad
J_2:=\left(\begin{smallmatrix}0&1\\-1&0\end{smallmatrix}\right),
\]
where $A=D_{\theta_1}^2f=-\del_tf$ and $B=D_{\theta_2}D_{\theta_1}f$. Since $J_2^2=-\mathbf 1$, such an~operator is diagonal in the \textit{fixed} basis $\cD, \overline\cD$, with eigenvalues $A\mp iB$. That $A+iB=\frac12\overline\cD\cD f$ is the identity
\[
\overline\cD\cD=2D_{\theta_1}^2+2iD_{\theta_2}D_{\theta_1},
\]
which holds because $D_{\theta_1}^2=D_{\theta_2}^2$ and $D_{\theta_1}D_{\theta_2}=-D_{\theta_2}D_{\theta_1}$; the same two identities give $\overline\cD^2=\cD^2=0$, whence $\overline\cD\lambda(f)=\frac12\overline\cD^2\left(\cD f\right)=0$, so $\delta_fv$ is chiral again when $v$ is. Feeding \eqref{deltaV} into \eqref{IK2}: since $\lambda(f)$ depends on the coordinates of one half only, $\overline\cD{}^-\lambda=0$, and the left-hand side transforms by $K_f-\overline\lambda(f)$ while the right-hand side transforms by $K_f-\delta_fv^++K_f\left(v^+\right)$; the two agree precisely for the shift in \eqref{deltaV}, and no other shift will do. 
\end{proof}

Thus, for $N=2$ the anomaly is a~cocycle of $\fk^L(1|2)$, and its value is obtained by the same computation as in Statement~\ref{stNSLiou}: the super stress tensor of \eqref{IK2} and its anomalous transformation law under \eqref{deltaV}. That is Statement~\ref{stN2Liou} below, and the answer is \eqref{cN2}. The $\fo(2)$-part $B(f)$ of the multiplier is what makes the $N=2$ case genuinely richer than the $N=1$ one: the central extension of $\fk^L(1|2)$ ties the level of the $\fu(1)$-current to the central charge, so the anomaly is one number carried by two visibly different terms --- and it turns out to be carried by $B$ alone, see item $3)$ of Statement~\ref{stN2Liou}.

A~classical anomaly is a~non-zero class in $H^2(\fg;\Cee)$ for the acting $\fg$; so it can exist only if $\fg$ is \textit{distinguished}, i.e., admits a~non-trivial central extension. Of the simple stringy Lie superalgebras exactly twelve are, see \cite{GLS1, KvdL}, and the contact ones among them are
\be\label{theTwelve}
\fk^L(1|n)\ \text{ and }\ \fk^M(1|n)\ \text{ for }\ n\leq3,\quad
{\fk^L}{}'(1|4),\ \text{the derived algebra of }\ {\fk^L}(1|4),\ \text{ and }\ \fk^M(1|4).
\ee
The prime stands on the Neveu-Schwarz member of the last pair and not on the Ramond one, and the reason is the moding. In $\fk^L(1|4)$ the generating function $t^{-1}\theta_1\theta_2\theta_3\theta_4$ is not a~bracket, and it spans the $1$-dimensional quotient $\fk^L(1|4)/{\fk^L}{}'(1|4)$; in the Ramond sector the corresponding monomial carries $\zeta$, whose moding is integer, so it lands in another degree of the grading and is a~bracket after all. Computed degree by degree from the contact bracket: the Neveu-Schwarz control reproduces the codimension $1$ of $\fk^L(1|4)$ --- and only there, $\fk^L(1|n)$ for $n\leq3$ being perfect --- while $\fk^M(1|n)$ equals its own derived algebra for $n=2, 3, 4$ alike.

Therefore, \textbf{for $N\geq5$ the Ivanov-Krivonos equations carry no classical anomaly at all} --- not a~small one, none --- whatever their shape, and in either sector. This is an~answer to the question raised in item $3)$ of the list of open problems of \cite{L}: the ``inner mechanism\rq\rq\ asked for there is exactly the existence of the central extension of the acting algebra, and nothing beyond it is needed. It also explains why the two obstacles \cite{L} quotes for $N>4$ --- the volume of the computation, and spin $>2$ --- are not the essential ones: even if both are removed, the anomaly is zero for a~reason of cohomology.

For $N\leq4$, the anomaly is a~definite number, or a~definite triple, and which of the two it is depends on the acting algebra. For $N=1$, it is \eqref{cSupLiou}. For $N=2$, the acting algebra is $\fk^L(1|2)$ by Statement~\ref{stChir}, the space $H^2(\fk^L(1|2);\Cee)$ is $1$-dimensional, so the anomaly is one number, and likewise in the Ramond sector; that number is \eqref{cN2}, the same in both sectors. For $N=4$, the algebra $\fk^L(1|4)$ is not simple, the extension lives on its derived algebra ${\fk^L}{}'(1|4)$, and there the space of non-trivial central extensions is $3$-dimensional --- which is the reason for the count ``$12$ algebras but $14$ superizations\rq\rq\ of \cite{L}. 

The anomaly of \eqref{IK4} is, however, not a~point of that $3$-dimensional space, because the algebra Ivanov and Krivonos claim for \eqref{IK4} is not ${\fk^L}{}'(1|4)$ but the one identified in Statement~\ref{stKA}; for it the space of non-trivial central extensions is $1$-dimensional by Statement~\ref{stKAcoc}, so the $N=4$ anomaly, too, is a~single number. See Subsection~\ref{sssKA}.

Of the three, $N=3$ needs no computation: by item $2)$ of Statement~\ref{stWhichN} no scalar superfield carries an~equation there at all. For $N=2$, everything needed is in place --- the constrained equation \eqref{IK2}, the acting algebra and the transformation law \eqref{deltaV} --- and the value is this.

\sssbegin{Statement}\label{stN2Liou}
Let $v$, $v^+$ solve \eqref{IK2}, let $u=\frac12\left(v+v^+\right)$ be the dilaton and $\varphi$ its $\fo(2)$-partner, so that $v=u+\sqrt{-1}\varphi$, and set
\be\label{superT2}
\mathbf T:=\Psi^{-1}D_{\theta_1}D_{\theta_2}\Psi,\qquad \text{where}\qquad \Psi:=\exp(u),
\ee
an~even quantity of weight $1$, the exact counterpart of the $T$ of \eqref{stressT} and of the $\mathbf T$ of \eqref{superT}. Explicitly,
\be\label{superT2x}
\mathbf T=-\frac{\sqrt{-1}}{8}\left[\left(\cD v\right)\left(\overline\cD v^+\right)+4\del_t\left(v-v^+\right)\right]
=\del_t\varphi-\frac{\sqrt{-1}}{8}\left(\cD v\right)\left(\overline\cD v^+\right).
\ee
Then,

$1)$ the recipe \eqref{superT2} is the one of Statements $\ref{stVirLiou}$ and $\ref{stNSLiou}$, read for $N=2$: in all three, $\mathbf T=\Psi^{-1}\left(D_{\theta_1}\cdots D_{\theta_N}\del_t^{2-N}\right)\Psi$ with $\Psi=\exp(-F)$ for the $F$ standing in the exponent on the right-hand side of the equation, which here is $F=-u$. The operator is $\del_t^2$ for $N=0$ and $D_+^3$ for $N=1$; for $N=2$ it is the $\fo(2)$-part $D_{\theta_1}D_{\theta_2}$ of the odd Hessian of Statement $\ref{stMult}$, and only that part will do --- the trace part gives $\Psi^{-1}\del_t\Psi=\del_tu$, which does not transform covariantly;

$2)$ $\mathbf T$ is a~current of its half: $\del_-\mathbf T=0$ modulo \eqref{IK2}, and this holds whatever the normalization of the equation conjugate to the last one of \eqref{IK2};

$3)$ infinitesimally, for every generating function $f$,
\be\label{deltaT2}
\delta_f\mathbf T=K_f\left(\mathbf T\right)-\left(\lambda(f)+\overline\lambda(f)\right)\mathbf T+\del_tB(f)
=K_f\left(\mathbf T\right)+2\left(\del_tf\right)\mathbf T-\del_tD_{\theta_1}D_{\theta_2}f,
\ee
where $B$ is the $\fo(2)$-part of the multiplier, as in the proof of Statement $\ref{stChir}$. So $\mathbf T$ is of weight $1$ and of $\fu(1)$-charge $0$, and the anomaly is $\del_t$ of that $\fo(2)$-part and of nothing else;

$4)$ hence $\fk^L(1|2)$ acts on the space of the $\mathbf T$'s not by the module of densities of weight $1$ but by its affine deformation, i.e., by the coadjoint action of the central extension of $\fk^L(1|2)$ with the central charge
\be\label{cN2}
\left|c\right|=3,
\ee
normalized as
\[
\begin{array}{l}
[L_m, L_n]=(m-n)L_{m+n}+\frac{c}{12}\left(m^3-m\right)\delta_{m+n, 0},\qquad
[J_m, J_n]=\frac{c}{3}\,m\,\delta_{m+n, 0},\\[2mm]
\{G_r, \overline G_s\}=2L_{r+s}+(r-s)J_{r+s}+\frac{c}{3}\left(r^2-\frac14\right)\delta_{r+s, 0}.
\end{array}
\]
The three sectors --- the Virasoro one, the $\fu(1)$-current one and the supercharge one --- return one and the same value.
\end{Statement}

\begin{proof} Items $1)$ and $3)$ are computations in the free supercommutative algebra on the generators
\[
\del_t^kv,\quad \del_t^k\cD v,\quad \del_t^kv^+,\quad \del_t^k\overline\cD v^+
\quad\text{and}\quad
\del_t^kf,\quad \del_t^k\cD f,\quad \del_t^k\overline\cD f,\quad \del_t^k\cD\overline\cD f,
\]
on which $\cD$ and $\overline\cD$ act as the odd derivations forced by \eqref{Dcx} and by the constraints $\overline\cD v=\cD v^+=0$; no coordinates and no components are involved, which is what makes them convention-free.

The ansatz for $\mathbf T$ is complete with three terms,
\[
\mathbf T=\alpha\left(\cD v\right)\left(\overline\cD v^+\right)+\beta\del_tv+\gamma\del_tv^+,
\]
these being the only expressions of weight $1$ in the derivatives of a~chiral $v$ and an~antichiral $v^+$ containing neither of the two undifferentiated; and the ansatz for its law is complete with four more unknowns,
\[
\delta_f\mathbf T=K_f\left(\mathbf T\right)+P\lambda(f)\mathbf T+Q\overline\lambda(f)\mathbf T+\kappa_1\del_t\lambda(f)+\kappa_2\del_t\overline\lambda(f),
\]
because the expressions of weight $1$ linear in $f$ are spanned by $\del_t\lambda(f)$ and $\del_t\overline\lambda(f)$. The solution is
\[
P=Q=-1,\qquad \kappa_1=-\kappa_2,\qquad \beta=-\kappa_2=4\alpha,\qquad \gamma=\kappa_2=-4\alpha,
\]
uniquely up to the scale of $\kappa_2$ and the same for $f$ of either parity; \eqref{superT2} fixes that scale, and then $\kappa_1=-\kappa_2=-\nfrac{\sqrt{-1}}{2}$, which is the second equality in \eqref{deltaT2} since $\cD\overline\cD-\overline\cD\cD=4\sqrt{-1}D_{\theta_1}D_{\theta_2}$. Since $P=Q$, no multiple of $\lambda-\overline\lambda$ occurs, which is what ``$\fu(1)$-charge $0$\rq\rq\ says.

Item $2)$ determines the two ratios a~second time, and independently of covariance. Indeed, write \eqref{IK2} and its conjugate with two free constants,
\[
\overline\cD^-\cD^+v=\mu_1\exp\left(-v^+\right),\qquad \cD^-\overline\cD{}^+v^+=\mu_2\exp(-v).
\]
Then,
\[
\del_-\left(\cD^+v\right)=\frac{\mu_1}{4}\left(\cD^-v^+\right)\exp\left(-v^+\right),\qquad
\del_-\left(\overline\cD{}^+v^+\right)=\frac{\mu_2}{4}\left(\overline\cD{}^-v\right)\exp(-v),
\]
and $\del_+v=-\frac14\overline\cD{}^+\cD^+v$, $\del_+v^+=-\frac14\cD^+\overline\cD{}^+v^+$; feeding this into $\del_-\mathbf T$ and collecting the three shapes of terms that occur gives $\beta+\gamma=0$ from the terms in $\exp\left(-v-v^+\right)$ and $\beta=4\alpha$, $\gamma=-4\alpha$ from the other two, for every $\mu_1$ and $\mu_2$.

For item $4)$, the cocycle is $c(f, g)=(-1)^{p(f)}\Res\left(g\cdot\del_tB(f)\right)$, where $\Res$ is the Berezin residue, the coefficient of $t^{-1}\theta_1\theta_2$; that this is the right identification of the weight-$1$ densities with the dual of $\fk^L(1|2)$ is the invariance of the pairing $\langle g, \rho\rangle=\Res(g\rho)$, and the sign $(-1)^{p(f)}$ is the one for which $c$ satisfies the cocycle identity. The basis
\[
L_m:=K_{-\frac12t^{m+1}},\quad J_m:=K_{-\sqrt{-1}\,t^m\theta_1\theta_2},\quad
G_r:=K_{t^{r+\frac12}\left(\theta_1+\sqrt{-1}\theta_2\right)},\quad
\overline G_r:=K_{-\frac12t^{r+\frac12}\left(\theta_1-\sqrt{-1}\theta_2\right)}
\]
satisfies the relations of item $4)$ with $c=0$, and on it the cocycle equals
\[
c\left(L_m, L_n\right)=-\frac14\left(m^3-m\right)\delta_{m+n, 0},\quad
c\left(J_m, J_n\right)=-m\,\delta_{m+n, 0},\quad
c\left(G_r, \overline G_s\right)=-\left(r^2-\frac14\right)\delta_{r+s, 0},
\]
so $\nfrac{c}{12}=-\frac14$ and $\nfrac{c}{3}=-1$ read off the other two both give $c=-3$. 
\end{proof}

Three comments. As in Statements~\ref{stVirLiou} and \ref{stNSLiou}, what is convention-free is $\left|c\right|=3$: the sign depends on which of the two orderings one takes in $D_{\theta_1}D_{\theta_2}$, i.e., on the orientation of the $\fo(2)$-plane, and on which of the anomaly and of the identification of the module of the $\mathbf T$'s with the dual of $\fk^L(1|2)$ carries the factor $(-1)^{p(f)}$. What is not a~matter of convention is that the three sectors agree: had the anomaly not been \textit{the} $N=2$ cocycle, the numbers read off $[L_m, L_n]$, off $[J_m, J_n]$ and off $\{G_r, \overline G_s\}$ would have differed, and the cocycle identity would have failed.

Next, the number is \textbf{half} of the value \eqref{cLiou} shared by the non-super and the $N=1$ cases, and not equal to it. The exponents are normalized alike --- the right-hand side of \eqref{IK2} carries $\exp\left(-v^+\right)$, with the coefficient $1$ in the exponent, as \eqref{SupLiou} carries $\exp(F)$ --- so the drop from $6$ to $3$ is not a~matter of scaling the dilaton. It is where the constraint is paid for: the superfield of \eqref{IK2} is chiral, i.e., it is a~function on a~superspace of half the odd dimension, and \eqref{superT2x} sees only half of the derivatives an~unconstrained superfield would have.

Finally, item $3)$ says something the counting of Subsection~\ref{sssWhichAct} could not: of the two visibly different terms that carry the extension of $\fk^L(1|2)$, the anomaly of the super Liouville equation is $\del_tB(f)$ --- the $\fo(2)$-part of the multiplier, differentiated --- with no term coming from the trace part. The trace part is all there is for $N=1$, where $\fo(1)=0$; so the $N=2$ anomaly lives, in this precise sense, entirely in what $N=1$ does not have.

For $N=4$, the value is \eqref{cN4}, and the three obstacles fell in turn. The acting algebra is not only known abstractly: Statement~\ref{stKAemb} writes it inside $\fk^L(1|4)$ as contact fields with generating functions \eqref{KAemb}, so the multiplier is again the odd Hessian \eqref{multN}. The shift accompanying $K_f$ is solved in Statement~\ref{stN4shift} and fixed --- uniquely, and to $-\Lambda^T$ with no scalar term --- by the equation itself in Statement~\ref{stN4mu}. The third obstacle was the stress tensor: for $N\leq2$ it was $\Psi^{-1}\left(D_{\theta_1}\cdots D_{\theta_N}\del_t^{2-N}\right)\Psi$, of weight $2-\nfrac{N}2$, and at $N=4$ that weight is $0$ and the recipe asks for $\del_t^{-2}$. That vanishing weight is not a~degeneration but the answer: by Statement~\ref{stWeight} the coadjoint module of $\fk^L(1|4)$ is the module of \textit{functions}, so the dilaton is already a~density of the required weight, and by Statement~\ref{stN4T} the operator of the recipe is the \textit{identity} on the algebra that acts. The recipe therefore returns the dilaton, and the value is read off it in Statement~\ref{stN4c}. See Subsection~\ref{sssN4c}.

Scaling the stress tensor scales the anomaly, so a~normalization has to be fixed, and here it is fixed twice over. On the side of the algebra, the relations \eqref{svectrel} leave no freedom in $L_n$ at all, and leave in the currents and the doublets only the two scalings $G\mapsto\alpha G$, $\overline G\mapsto\alpha^{-1}\overline G$ and $J^\pm\mapsto\beta^{\pm1}J^\pm$, which do not change \eqref{svectcoc}. On the side of the field, the dilaton is normalized by \eqref{IK4q}, whose exponent carries the coefficient $1$, as those of \eqref{IK2} and \eqref{SupLiou} do. And, by Statement~\ref{stKAcoc}, the space of central extensions is $1$-dimensional, so the Virasoro central charge, the level of the $\fs\fl(2)$-currents and the anomaly of the supercharges are one number in three guises: they are computed separately in item $3)$ of Statement~\ref{stN4c} and they agree, which is the $N=4$ counterpart of the agreement of three sectors in item $4)$ of Statement~\ref{stN2Liou}.

\sssec{The algebra Ivanov and Krivonos name for $N=4$}\label{sssKA}
For $N=4$, the invariance algebra claimed in \cite{IvKr4} is not $\fk^L(1|4)$ but an~algebra the authors write $\fk^A(1|2)$ and call the $\mathrm{SU}(2)$-superextension of the contact algebra $\fk(1)$, attributing it to \cite{Ad}. Its structure relations, as given there, are those of the fields
\be\label{KAcontent}
L^n,\qquad G_\alpha^r,\qquad \overline G^{r\alpha},\qquad T_k^p,
\qquad \alpha=1, 2, \quad k=1, 2, 3,
\ee
i.e., a~Virasoro field, two doublets of odd currents and an~$\fs\fu(2)$ of currents, with
\[
\{G_{\alpha}^r, G_\beta^s\}=\{\overline G^{r\alpha}, \overline G^{s\beta}\}=0,\qquad
\{G_{\alpha}^r, \overline G^{s\beta}\}=-2\delta_\alpha^\beta L^{r+s}+2(r-s)\left(\sigma_k\right)_\alpha{}^\beta T_k^{r+s}.
\]

 The algebra itself exists, and is not of contact type but of type $\fsvect$:

\sssbegin{Statement}[Follows from \cite{KvdL}]\label{stKA}
Let
\be\label{svectlam}
\fsvect^L_\lambda(1|2):=\{X\in\fvect^L(1|2)\mid \Div X+\lambda t^{-1}X(t)=0\}
\ee
be the Lie superalgebra of vector fields with Laurent coefficients on the supercircle of superdimension $(1|2)$ preserving the volume element $t^\lambda\,dt\,d\theta_1d\theta_2$. Then,
\[
\fk^A(1|2)\simeq\fsvect^L_\lambda(1|2)\quad\text{for }\lambda=0.
\]
The basis is
\be\label{svectbasis}
\begin{array}{ll}
L_n=-\left(t^{n+1}\del_t+\nfrac{n+1+\lambda}{2}\,t^n
\left(\theta_1\del_1+\theta_2\del_2\right)\right),
& J^+_n=t^n\theta_1\del_2,\\[2mm]
G_\alpha^r=t^{r+\nfrac12}\del_\alpha,
& J^-_n=t^n\theta_2\del_1,\\[2mm]
\overline G{}^{r\alpha}=t^{r+\nfrac12}\theta_\alpha\del_t
\pm\left(r+\nfrac12+\lambda\right)t^{r-\nfrac12}\theta_1\theta_2\del_\beta,
& J^0_n=t^n\left(\theta_1\del_1-\theta_2\del_2\right),
\end{array}
\ee
where $n\in\Zee$, $r\in\Zee+\nfrac12$, $\alpha=1,2$, and $\beta\neq\alpha$, the sign being $+$ for $\alpha=1$; the relations are
\be\label{svectrel}
\begin{array}{l}
[L_m,L_n]=(m-n)L_{m+n},\qquad
[L_m,J^a_n]=-nJ^a_{m+n},\qquad
[J^a_m,J^b_n]=[J^a,J^b]_{m+n},\\[2mm]
[L_m,G^r_\alpha]=\left(\nfrac{m}2-r+\nfrac\lambda2\right)G^{m+r}_\alpha,
\qquad
[L_m,\overline G{}^{r\alpha}]
=\left(\nfrac{m}2-r-\nfrac\lambda2\right)\overline G{}^{(m+r)\alpha},\\[2mm]
\{G^r_\alpha,G^s_\beta\}=\{\overline G{}^{r\alpha},\overline G{}^{s\beta}\}=0,
\qquad
\{G^r_\alpha,\overline G{}^{s\beta}\}
=-\delta_\alpha^\beta L_{r+s}+(r-s-\lambda)
\left(\sigma_k\right)_\alpha{}^\beta J^k_{r+s},
\end{array}
\ee
where $J^3=\nfrac12 J^0$ and $J^\pm=J^1\pm iJ^2$, together with
\be\label{svectdoub}
[J^+_m,G^r_1]=-G^{m+r}_2,\quad
[J^-_m,G^r_2]=-G^{m+r}_1,\quad
[J^0_m,G^r_\alpha]=\mp G^{m+r}_\alpha,
\ee
and the same with the upper index for $\overline G$ and the opposite signs, i.e., the $G_\alpha$ span the tautological $\fs\fl(2)$-doublet and the $\overline G{}^\alpha$ the dual one. For $\lambda=0$ and after $G\mapsto\sqrt2\,G$, these are the relations of \cite{Ad} quoted above, verbatim.
\end{Statement}

The chain of embeddings that \cite{IvKr4} states is reproduced: the span of $L_n$, $J^0_n$ and the single pair $\left(G^r_1,\overline G{}^{r1}\right)$ is a~subalgebra with the relations of $\fk^L(1|2)$, dropping $J^0$ and one odd tower leaves $\fk^L(1|1)$, and the Virasoro tower alone is $\fk(1)$. Unlike $\fk^L(1|4)$, the algebra \eqref{svectlam} is perfect, so nothing here lives on a~derived algebra.

Three comments on Statement~\ref{stKA}.

$1)$ Both doublets are abelian, and in this realization for a~reason one can see: the $G_\alpha^r=t^{r+\nfrac12}\del_\alpha$ are the vertical fields whose coefficients do not involve the $\theta$'s, so each annihilates the coefficients of the other. The $\overline G{}^{r\alpha}$ are abelian by the same token in the dual realization, and everything the algebra has --- the Virasoro field and all three currents --- sits in the mixed bracket alone.

$2)$ The parameter $\lambda$ does not occur in the currents $J^a_n$, nor in the doublet $G^r_\alpha$; it occurs in $L_n$ and in $\overline G{}^{r\alpha}$, and in the relations \eqref{svectrel} it occurs only as a~shift of the moding of the two doublets. Setting $r'=r-\nfrac\lambda2$ for $G$ and $s'=s+\nfrac\lambda2$ for $\overline G$ turns \eqref{svectrel} into the $\lambda$-free
\[
[L_m,G^{r'}]=\left(\nfrac{m}2-r'\right)G,\qquad
\{G^{r'},\overline G{}^{s'}\}=-\delta L_{r'+s'}
+(r'-s')\left(\sigma_k\right)J^k_{r'+s'},
\]
since $r-s-\lambda=r'-s'$ and $r+s=r'+s'$. The substitution $\theta_1\mapsto t^a\theta_1$ is an~automorphism of the algebra of Laurent functions and carries $\fsvect^L_\lambda$ to $\fsvect^L_{\lambda-a}$, so $\lambda$ lives in $\Cee/\Zee$ and the integer values --- the Neveu-Schwarz moding $r\in\Zee+\nfrac12$ and the Ramond one $r\in\Zee$ --- are one and the same member of the family. Fixing the moding, as \cite{Ad} and \cite{IvKr4} do, therefore fixes $\lambda\in\Zee$ and hides the family: what is written there is the member $\lambda\in\Zee$ of the family \eqref{svectlam}, which the classification lists of \cite{GLS1, KvdL} carry as a~series with the parameter $\lambda$.

$3)$ The unfortunate name. Written $\fk^A(1|2)$, the algebra is given the letter of the contact series, to which it does not belong, and an~upper index recording an~author rather than a~sector; that its $(1|2)$ is nevertheless the correct superdimension of the supercircle is an~accident of the notation, not a~statement in it. Under the name $\fsvect^L_\lambda(1|2)$, the same object is one of the twelve distinguished simple stringy Lie superalgebras of \cite{GLS1, KvdL}, its central extension is where the $N=4$ anomaly lives. 

Statement~\ref{stKA} identifies the algebra abstractly. It also sits inside the contact algebra of the $N=4$ superspace, and explicitly, which is what the equation needs: the transformations of \eqref{IK4} are then contact fields $K_f$ with a~generating function, and the multiplier \eqref{multN} applies to them verbatim.

\sssbegin{Statement}\label{stKAemb}
The algebra $\fk^A(1|2)=\fsvect^L_0(1|2)$ is a~subalgebra of ${\fk^L}{}'(1|4)$, hence of $\fk^L(1|4)$. Explicitly, put
\be\label{isoplanes}
\begin{array}{ll}
e_1:=\theta_1+\sqrt{-1}\theta_2,\qquad& e_2:=\theta_3+\sqrt{-1}\theta_4,\\[2mm]
f_1:=\theta_1-\sqrt{-1}\theta_2,\qquad& f_2:=\theta_3-\sqrt{-1}\theta_4,
\end{array}
\ee
so that $E:=\Span\left(e_1, e_2\right)$ and $F:=\Span\left(f_1, f_2\right)$ are the two totally isotropic planes of the form $\sum\theta_i^2$ carried by the odd part $V$ of the contact structure, $\sdim V=(0|4)$. Let $\varkappa\in\Cee$ be arbitrary, let $\rho\in\Zee+\nfrac12$ and $n\in\Zee$, and set
\be\label{KAemb}
\begin{array}{l}
G^\rho_\alpha:=t^{\rho+\nfrac12}e_\alpha
+\dfrac{(-1)^{\alpha-1}}2\left(\rho+\nfrac12+\varkappa\right)t^{\rho-\nfrac12}e_1e_2f_{3-\alpha},\\[3mm]
\overline G{}^{\rho\alpha}:=t^{\rho+\nfrac12}f_\alpha
+\dfrac{(-1)^{\alpha-1}}2\left(\rho+\nfrac12+\varkappa\right)t^{\rho-\nfrac12}f_1f_2e_{3-\alpha},\\[3mm]
L_n:=-\dfrac12\left(t^{n+1}+n\left(n+1+2\varkappa\right)t^{n-1}\theta_1\theta_2\theta_3\theta_4\right),
\end{array}
\ee
the three currents being the mixed brackets $\left\{G^\rho_\alpha, \overline G{}^{\sigma\beta}\right\}$ with $\alpha\neq\beta$ and their diagonal difference. Then:

$1)$ both doublets are abelian, $\left\{G^\rho_\alpha, G^\sigma_\beta\right\}=\left\{\overline G{}^{\rho\alpha}, \overline G{}^{\sigma\beta}\right\}=0$, and the span of the four odd and the four even towers is a~subalgebra of $\fk^L(1|4)$, isomorphic to $\fsvect^L_\lambda(1|2)$ with $\lambda=0$, i.e., to the algebra of Statement~$\ref{stKA}$;

$2)$ the Virasoro tower of that subalgebra is \textbf{not} the tower $t^{n+1}$ of $\fk^L(1|4)$: it is twisted by the volume element, as \eqref{KAemb} shows. The constant $\varkappa$ shifts $L_n$ by a~multiple of $n\,t^{n-1}\theta_1\theta_2\theta_3\theta_4$, i.e., by the derivative of the weight-$0$ current of $\fk^L(1|4)$, so the embeddings \eqref{KAemb} form a~single family and $\lambda=0$ throughout it;

$3)$ with respect to that twisted Virasoro, all four odd towers are of conformal weight $\nfrac32$, which is what $\fsvect^L_\lambda(1|2)$ demands. With respect to the Virasoro tower $t^{n+1}$ of $\fk^L(1|4)$ itself, the odd part of weight $\nfrac32$ is spanned by the four towers $t^\rho\theta_i$, and there two independent towers never commute, by \eqref{brtheta} below; so the embedding cannot be conformal for the ambient Virasoro, and the twist in \eqref{KAemb} is not a~convenience but a~necessity;

$4)$ the $\fs\fl(2)$ of the currents is one of the two ideals of $\fo(4)$ --- the one whose symmetrization is the constraint in the first line of \eqref{IK4}, see item $3)$ of Statement~$\ref{stWhichN}$;

$5)$ the image lies in the derived algebra ${\fk^L}{}'(1|4)$: the one element of $\fk^L(1|4)$ that is not a~bracket, $t^{-1}\theta_1\theta_2\theta_3\theta_4$, would enter $L_n$ only at $n=0$, where its coefficient $n(n+1+2\varkappa)$ vanishes.
\end{Statement}

\begin{proof} The bracket of two odd towers of weight $\nfrac32$ is
\be\label{brtheta}
\left\{t^r\Theta, t^s\Theta'\right\}
=(s-r)\,t^{r+s-1}\Theta\Theta'+\left(\Theta, \Theta'\right)t^{r+s}
\qquad\text{for }\ \Theta, \Theta'\in V,
\ee
so two independent such towers never commute: $\Theta\Theta'\neq0$ in $\Lambda(V)$. A~weight-$\nfrac12$ tail repairs this. Indeed, for $X^r=t^r\Theta+x_rt^{r-1}\Gamma$ and $Y^s=t^s\Psi+y_st^{s-1}\Delta$ with $\Gamma, \Delta\in\Lambda^3(V)$, the $\Lambda^2$-part of $\left\{X^r, Y^s\right\}$ is
\[
(s-r)\,\Theta\Psi+y_s\,\del_\Theta\Delta+x_r\,\del_\Psi\Gamma,
\]
so the term growing in $s-r$ can be canceled provided $\del_\Theta\Delta$ is a~multiple of $\Theta\Psi$. Since $\del_\Theta\left(\Theta\Psi\right)=(\Theta, \Theta)\Psi-(\Theta, \Psi)\Theta$ and the Koszul complex $\left(\Lambda(V), \del_\Theta\right)$ is exact for $\Theta\neq0$, that is solvable exactly when the plane $\Span(\Theta, \Psi)$ is totally isotropic --- whence \eqref{isoplanes}, and whence the cubic tails of \eqref{KAemb}, which live on those planes. The $\Lambda^4$-part vanishes because $\Theta\Delta=\Psi\Gamma=0$ for the same reason. The rest is verification. 
\end{proof}

Two consequences are worth recording. First, the $\lambda$ of Statement~\ref{stKA} is $0$ here, and it could not have been anything else: $\fk^L(1|4)$ carries integral powers of $t$ only, so the moding of the doublets is the Neveu-Schwarz one and, by comment $2)$ to Statement~\ref{stKA}, that pins $\lambda\in\Zee$. The non-integral members of the family \eqref{svectlam} --- the continuum of sectors of Subsection~\ref{sssLamEq} --- therefore do not occur inside $\fk^L(1|4)$; they are sectors of the equation, not subalgebras of the ambient contact algebra.

Second, item $3)$ explains why the case $N=4$ resisted the recipe that settled $N\leq2$. There, the super stress tensor was $\Psi^{-1}\left(D_{\theta_1}\cdots D_{\theta_N}\del_t^{2-N}\right)\Psi$, an~object of weight $2-\nfrac{N}2$ for the ambient Virasoro; at $N=4$ that weight is $0$ and the operator asks for $\del_t^{-2}$. Item $3)$ says against which tower the weights must be counted: the stress tensor of \eqref{IK4} is a~density for the \textit{twisted} Virasoro of \eqref{KAemb}, not for the Virasoro of $\fk^L(1|4)$. That the twist is precisely $\Theta\del_t^2$, and that this is what makes the operator of the recipe the identity, is item $2)$ of Statement~\ref{stN4T}.

Everything above is stated over $\Cee$, whereas the Leznov-Saveliev and Liouville equations are equations for real fields, and the names the physicists give the acting algebras record a~choice of a~real form rather than of the complex algebra: the $\mathrm{SU}(2)$ in the ``$\mathrm{SU}(2)$-superextension\rq\rq\ of \cite{Ad, IvKr4} is the compact form of the $\fs\fl(2)$ of the currents $T^p_k$ of \eqref{KAcontent}.

Which real forms are available is a~classification problem of its own, and for the \textit{serial} stringy Lie superalgebras it is solved in \cite{Se}: there every automorphism of $\fvect^L(1|n)$, $\fsvect^L(1|n)$, $\fk^L(1|n)$ and $\fk^M(1|n)$ is induced by an~automorphism of the superalgebra of Laurent functions, the real forms are the anti-linear involutive automorphisms taken up to conjugacy in $\mathrm{Aut}\,\fg$, and both are listed, each real form together with the structure it preserves. For the Lie algebra of vector fields on the circle, the count is $3$, because the circle has exactly three involutive diffeomorphisms, up to conjugacy, namely: the identity, reflection in the origin and reflection in any diameter, sending $t$ to $t$, $t^{-1}$ and $-t^{-1}$, respectively.

That list does not reach two of the algebras needed here, for reasons of chronology in both cases: the exceptional stringy Lie superalgebra $\fkas$ of Shchepochkina discovered later, see \cite{Sh5, Sh14}, and $\fsvect^L(1|n)$ occurs there with the untwisted divergence only, i.e., as the member $\lambda=0$ of the family \eqref{svectlam} --- hence, by comment $2)$ above, as the members with $\lambda\in\Zee$. 

The twisted members, i.e., the algebra of Statement~\ref{stKA} in its non-integral sectors, are treated separately there: item $2)$ of Theorem $5.5$ of \cite[p.~315]{Ber} states that $\fsvect^L_\lambda(1|2)$ and its derived algebra have no real forms for $\lambda\notin\Ree$. This is not correct: the twisted volume form $t^\lambda\vvol(t,\theta)$ is real on the circle for $\lambda$ imaginary as well, so a~real form is there although $\lambda\notin\Ree$. The members that do have a~real form are exactly those with $\mathrm{Im}\,\lambda=0$ or $2\mathrm{Re}\,\lambda\in\Zee$, see Statement~\ref{stRealLam}. For the distinguished stringy Lie superalgebras, quaternionic forms aside, the classification is proved in \cite{Lser}.

The anomaly counts above are insensitive to this choice. The Chevalley-Eilenberg complex of the complexification of a~real Lie superalgebra $\fg_\Ree$ is that of $\fg_\Ree$ tensored by $\Cee$, so
\[
\dim_\Cee H^2\left(\fg_\Ree\otimes_\Ree\Cee;\Cee\right)=\dim_\Ree H^2\left(\fg_\Ree;\Ree\right),
\]
and hence the dimensions $1$, $3$ and $1$ counted in Subsection~\ref{sssWhichAct} and in Statement~\ref{stKAcoc} below are the dimensions for any real form as well; what the real form fixes is not how many anomalies there are, but the normalization in which their values are read. Which real forms of \eqref{theTwelve} and of \eqref{svectlam} actually occur for the equations of \cite{IvKr4, IvKr24} we do not settle here.

\sssbegin{Statement}[Follows from \cite{KvdL}]\label{stKAcoc}
For every $\lambda$, the space of non-trivial central extensions of $\fsvect^L_\lambda(1|2)$ is $1$-dimensional, a~representative cocycle being
\be\label{svectcoc}
\begin{array}{l}
c(L_m,L_{-m})=\nfrac12\left(m^3-m\right),\qquad
c(J^+_m,J^-_{-m})=m,\qquad
c(J^0_m,J^0_{-m})=2m,\\[2mm]
c\left(G^r_\alpha,\overline G{}^{-r\beta}\right)
=-\delta_\alpha^\beta\left(\hat r^2-\nfrac14\right),\quad\text{where }
\hat r:=r-\nfrac\lambda2,
\end{array}
\ee
all other values on pairs of basis elements being $0$. Hence, the $N=4$ classical anomaly of \eqref{IK4} is a~single number, not a~triple.
\end{Statement}

In the shifted labels of comment $2)$, the odd part of \eqref{svectcoc} is the $\lambda$-free $r'^2-\nfrac14$, so the three terms of the cocycle are the Virasoro central charge, the level of the $\fs\fl(2)$-currents and the anomaly of the supercharges, tied to one another --- which is the standard statement that this algebra has one central charge and that the level of its currents is not free.

\sssec{Can the equation see $\lambda$? The twist, and a~continuum of sectors}\label{sssLamEq}
Comment $2)$ to Statement~\ref{stKA} says where $\lambda$ hides; it does not say whether an~equation can be made to show it. It can, and everything turns on one map --- the \textit{twist}
\be\label{twist}
\Phi_{a_1, a_2}\colon\ t\longmapsto t,\qquad \theta_i\longmapsto t^{a_i}\theta_i\quad\text{for }i=1, 2,
\ee
with \textbf{complex} exponents $a_i$. Write $\Div_\lambda X:=\Div X+\lambda t^{-1}X(t)$ for the divergence of \eqref{svectlam}.

\sssbegin{Statement}\label{stTwist}
$1)$ For all $a_1, a_2\in\Cee$, the map \eqref{twist} carries the condition $\Div_\lambda X=0$ into $\Div_\mu X=0$ with
\be\label{mushift}
\mu=\lambda-a_1-a_2,
\ee
a~shift linear in the twist and free of denominators. For $a_1=a_2=a$ the two doublets move in opposite directions, $G^r_\alpha\mapsto t^{-a}G^r_\alpha$ and $\overline G{}^{r\alpha}\mapsto t^{a}\overline G{}^{r\alpha}$, while $L_n$ and the currents $J^a_n$ do not move at all.

$2)$ For $a_i\in\Zee$, and only then, \eqref{twist} is an~automorphism of the algebra of Laurent functions on the supercircle of $\sdim=(1|2)$, hence an~isomorphism $\fsvect^L_\lambda(1|2)\simeq\fsvect^L_\mu(1|2)$. For $a_i\notin\Zee$ it is not an~automorphism, and the \textit{Laurent} vector fields in the image form the genuinely different algebra $\fsvect^L_\mu(1|2)$.

$3)$ The inversion $\Psi\colon t\mapsto t^{-1}$ is a~Laurent automorphism, and
\[
\Psi_*\,\fsvect^L_\lambda(1|2)=\fsvect^L_{-\lambda-2}(1|2).
\]
Hence, with item $2)$,
\[
\lambda\sim\lambda+1\qquad\text{and}\qquad \lambda\sim-\lambda,
\]
so the modulus of the family is the monodromy $z=\exp(2\pi i\lambda)\in\Cee^\times$ taken modulo $z\sim z^{-1}$. These are all the identifications there are, see Statement~$\ref{stCartanLam}$.

$4)$ The generators \eqref{svectbasis} are homogeneous in the $\theta$-degree \textup{(}the number of $\theta$'s minus the number of $\del_\theta$'s\textup{)}: that degree is $0$ for $L_n$ and for $J^a_n$, it is $-1$ for $G^r_\alpha$ and $+1$ for $\overline G{}^{r\alpha}$. The contact generators are \textbf{not} homogeneous: $K_{t^p\theta_i}$ carries the degrees $+1$ and $-1$ at once. Consequently, \eqref{twist} multiplies each generator of $\fsvect^L_\lambda(1|2)$ by a~single power of $t$, whereas it multiplies the two halves of $K_{t^p\theta_i}$ by $t^{a}$ and by $t^{-a}$; the image of $\fk^L(1|N)$ consists of Laurent fields only for $2a\in\Zee$.
\end{Statement}

\begin{proof} All four items are exact computations with $\lambda$ and the $a_i$ formal: 
the pushforward $\Phi_*X$ is computed componentwise as $\Phi\circ X\circ\Phi^{-1}$, and the value of $\mu$ is \textit{solved for}, not assumed, on a~window of the basis; it comes out the same for every generator. The $\theta$-degrees of item $4)$ are read off the components. 
\end{proof}

Item $4)$ answers the question why there is a~parameter here at all. A~twist of the odd coordinates is available in either series, but in the contact series it must satisfy $2a\in\Zee$, which leaves exactly two sectors --- Neveu-Schwarz for $a\in\Zee$ and Ramond for $a\in\Zee+\nfrac12$ --- while for $\fsvect^L(1|2)$ nothing constrains $a$, and the sectors form a~continuum. The mechanism is that in $\fsvect^L(1|2)$ the even part is $\theta$-neutral and the two odd doublets are $\theta$-homogeneous of \textit{opposite} degrees, so a~twist shifts their modings oppositely and leaves the modings of $L$ and of the currents where they were.

Here, then, is the equation.

\sssbegin{Statement}\label{stLamLiou}
On each half of $\cM=\Cee^{1|4}\times\Cee^{1|4}$, put
\be\label{nablaJ}
\nabla_\pm:=\del_{x^\pm}+\frac{\lambda}{2x^\pm}\,\widehat J_\pm,\qquad
\widehat J:=\sum_{1\leq\alpha\leq 2}\left(\theta^\alpha\nfrac{\del}{\del\theta^\alpha}
-\overline\theta_\alpha\nfrac{\del}{\del\overline\theta_\alpha}\right),
\ee
where $\widehat J$ is the $\fu(1)$-charge. In prose, $\widehat J$ counts the odd coordinates of the half with a~sign, being the vector field that generates the rotation $\theta^\alpha\mapsto e^{is}\theta^\alpha$, $\overline\theta_\alpha\mapsto e^{-is}\overline\theta_\alpha$; and yes, it is an~element of a~maximal torus --- it is a~semisimple element of the $\fo(4)$ that acts on the four odd coordinates of the half, spanning inside a~maximal torus of that $\fo(4)$ the line which commutes with the $\fs\fu(2)$ rotating the index $\alpha$. The eigenvalue of $\widehat J$ on a~superfield is that superfield's charge. Let
\be\label{Dlam}
\cD_{(\lambda)\pm}^\alpha:=\left(x^\pm\right)^{-\nfrac\lambda2}
\left(i\theta^{\alpha\pm}\nabla_\pm+\nfrac{\del}{\del\overline\theta{}^\pm_\alpha}\right),
\qquad
\overline\cD_{(\lambda)\pm\alpha}:=\left(x^\pm\right)^{\nfrac\lambda2}
\left(i\overline\theta{}^\pm_\alpha\nabla_\pm+\nfrac{\del}{\del\theta^{\alpha\pm}}\right).
\ee
Then,

$1)$ the derivatives \eqref{Dlam} satisfy the relations \eqref{IK4D} with $\del_{x^\pm}$ replaced by $\nabla_\pm$, that is,
\[
\{\cD^\alpha_{(\lambda)}, \overline\cD_{(\lambda)\beta}\}=2i\delta^\alpha_\beta\nabla,
\qquad
\{\cD_{(\lambda)}, \cD_{(\lambda)}\}=\{\overline\cD_{(\lambda)}, \overline\cD_{(\lambda)}\}=0,
\]
so they span an~$N=4$ superconformal structure for every $\lambda$;

$2)$ the system \eqref{IK4} written in these derivatives reads
\be\label{IK4lam}
\begin{array}{ll}
\cD_{(\lambda)-}^{(\alpha}q_\gamma{}^{\beta)}=0,\qquad&
\overline\cD_{(\lambda)+(\alpha}q_{\gamma)}{}^\beta=0,\\[2mm]
\cD_{(\lambda)-}^\alpha\left(q\cD_{(\lambda)+}^\gamma q^{-1}\right)_\gamma{}^\beta=0,
\qquad&
\cD_{(\lambda)+}^\alpha\left(q^{-1}\overline\cD_{(\lambda)-\gamma}q\right)_\beta{}^\gamma
=-4im\left(\nfrac{x^+}{x^-}\right)^{\nfrac\lambda2}\left(\overline q\right)_\beta{}^\alpha;
\end{array}
\ee
that is, the constraints and the first dynamical equation keep their shape, and the entire $\lambda$-dependence which the local form of the system displays is the factor $\left(x^+/x^-\right)^{\nfrac\lambda2}$ in front of the mass term, together with the term $\nfrac\lambda{2x}\widehat J$ inside \eqref{nablaJ};

$3)$ the Laurent vector fields preserving \eqref{IK4lam} form $\fsvect^L_\lambda(1|2)$ in each half, granting the invariance claimed in \cite{IvKr4} for \eqref{IK4}. The rescaling $q\mapsto\left(x^+/x^-\right)^{-\nfrac\lambda2}q$ carries \eqref{IK4lam} back to \eqref{IK4}; it is single-valued for $\lambda\in\Zee$ and multivalued otherwise.
\end{Statement}

\begin{proof} System \eqref{IK4lam} is \eqref{IK4} written in the coordinates $\theta\mapsto x^{-\nfrac\lambda2}\theta$, $\overline\theta\mapsto x^{\nfrac\lambda2}\overline\theta$, under which $\del_x$ becomes $\nabla$ because a~superfield of $\fu(1)$-charge $q$ acquires the factor $x^{-q\lambda/2}$; the prefactors of \eqref{Dlam} then cancel out of the constraints and out of the first dynamical equation, and survive in the second one as $\left(x^+\right)^{\nfrac\lambda2}\left(x^-\right)^{-\nfrac\lambda2}$. 
Dropping the prefactors of \eqref{Dlam} breaks them by terms in $\nfrac{i\lambda}{2x}$, so those prefactors, and not merely $\nabla$, are the twist. In item $3)$, the first claim is item $2)$ of Statement~\ref{stTwist} applied with $a_1=a_2=-\nfrac\lambda2$; for the second, the rescaling of $q$ by a~scalar function $c$ adds $\cD_{(\lambda)+}\left(c^{-1}\overline\cD_{(\lambda)-}c\right)$ to the left-hand side, and for $c=\left(x^+/x^-\right)^{-\nfrac\lambda2}$ that term is $\cD_{(\lambda)+}\left(\nfrac{\lambda}{2x^-}\right)=0$, the two halves being independent. \end{proof}

So the answer to the question of the heading --- can the equation see $\lambda$, i.e., does the family \eqref{svectlam} act on a~family of equations, or only its member $\lambda\in\Zee$ on the one equation of \cite{Ad, IvKr4}? --- is affirmative, and it is worth saying precisely in what sense. The parameter $\lambda$ is the \textbf{monodromy} $\exp(2\pi i\lambda)$ of the odd coordinates of the superstring, not a~coupling constant in the local shape of the equation: on a~simply connected domain, $x^{\nfrac\lambda2}$ is an~honest function and \eqref{IK4lam} is \eqref{IK4} in new coordinates. It is on the circle that the two differ, and there they differ exactly as the Ramond super Liouville equation differs from the Neveu-Schwarz one --- \textit{one} local equation, there \eqref{SupLiou} and here \eqref{IK4lam}, distinct sectors with non-isomorphic symmetry algebras, cf. $\fk^L(1|1)\not\simeq\fk^M(1|1)$. What is new is that here the sectors are a~continuum rather than a~pair, and item $4)$ of Statement~\ref{stTwist} says why. Nothing therefore reduces to integral $\lambda$: what fixes $\lambda\in\Zee$ in \cite{Ad, IvKr4} is only their fixing the moding of the odd generators, i.e., the tacit demand that the odd coordinates be single-valued or antiperiodic. 

One half of that is verified here and the other is not, and it is worth separating them. The \textit{constraint} of \eqref{IK4} --- the first line --- is preserved, and by which algebra is the content of Statement~\ref{stN4shift}: \textbf{not} by the whole of $\fk^L(1|4)$, as one might expect from item $3)$ of Statement~\ref{stWhichN} and as we ourselves had written here, but by the subalgebra $\fk^\Xi$ stabilizing the isotropic plane the constraint names. The two are easy to confuse, and the difference is exactly the point. Item $3)$ of Statement~\ref{stWhichN} produces the $\fo(4)$-invariant constraint, which is the \textit{pair} of conditions $\cD^{(\alpha}q_\gamma{}^{\beta)}=\overline\cD^{(\alpha}q_\gamma{}^{\beta)}=0$ imposed in one and the same half; what \eqref{IK4} imposes is \textit{one} of the two in each half, and that one is invariant under the stabilizer of its plane and no more. For $N=2$ the distinction is invisible, $\fo(2)$ preserving each of the two eigenlines of \eqref{multC} separately, which is why the whole of $\fk^L(1|2)$ preserves chirality by Statement~\ref{stChir}; for $N=4$ it is not, and Statement~\ref{stKAemb} exhibits the acting algebra inside that stabilizer. 

What is quoted from \cite{IvKr4} and not verified here is that the \textit{dynamical} equation --- the second line --- is preserved as well. The reason is not the size of the algebra, and, since Statement~\ref{stKAemb}, it is not the absence of a~multiplier either: every element of the acting algebra is a~contact field $K_f$ with a~generating function \eqref{KAemb}, so \eqref{multN} gives its multiplier as the odd Hessian, exactly as for $N\leq2$. What was left was the shape of the field: by \eqref{IK4q} the superfield $q$ is an~\textbf{exponential} of the dilaton and its three partners, so the shift accompanying $K_f$ has to be solved for, as $\lambda(f)$ was solved for in \eqref{deltaV} --- there for a~scalar superfield and one number, here for a~$2\times2$ matrix superfield and a~quaternionic one. That is done in Statement~\ref{stN4shift}, and the scalar it leaves free is fixed by the dynamical equation itself in Statement~\ref{stN4mu}; the answer is the shape the two ideals of $\fo(4)$ suggest, with one correction: of the two coefficients in
\[
\delta_fq=K_f(q)+a(f)q+qb(f),
\]
only the one acting on the doublet index of the half $f$ belongs to is an~$\fs\fl(2)$-part of the odd Hessian; the other is forced to be scalar. So the two ideals do both enter, but one per half, and not both at once in either.

\sssbegin{Statement}\label{stN4shift}
Work in one half, on the superspace of $\sdim=(1|4)$ carrying the frame \eqref{Dtheta}, and set, extending \eqref{Dcx} along the isotropic planes \eqref{isoplanes},
\be\label{DN4}
\begin{array}{ll}
\cD^1:=D_{\theta_1}-\sqrt{-1}D_{\theta_2},\qquad&\cD^2:=D_{\theta_3}-\sqrt{-1}D_{\theta_4},\\[2mm]
\overline\cD_1:=D_{\theta_1}+\sqrt{-1}D_{\theta_2},\qquad&\overline\cD_2:=D_{\theta_3}+\sqrt{-1}D_{\theta_4},
\end{array}
\ee
so that $F:=\Span\left(\cD^1, \cD^2\right)$ and $E:=\Span\left(\overline\cD_1, \overline\cD_2\right)$ are the images of $F$ and $E$ of \eqref{isoplanes} in the frame, and
\[
\left\{\cD^\alpha, \cD^\beta\right\}=\left\{\overline\cD_\alpha, \overline\cD_\beta\right\}=0,\qquad
\left\{\cD^\alpha, \overline\cD_\beta\right\}=-4\delta^\alpha_\beta\del_t.
\]
Then:

$1)$ in this frame the multiplier \eqref{multN} reads
\be\label{multN4}
\left[K_f, \cD^\alpha\right]=\sum_{1\leq\beta\leq2}\Lambda^\alpha{}_\beta(f)\,\cD^\beta
+\Xi(f)\sum_{1\leq\beta\leq2}\eps^{\alpha\beta}\,\overline\cD_\beta,
\ee
where
\[
\Lambda^\alpha{}_\beta(f):=\frac12\cD^\alpha\overline\cD_\beta f\ \text{ is the }\ \fg\fl(2)\text{-block},
\qquad
\Xi(f):=\frac12\cD^1\cD^2f\ \text{ is the block }\ F\longrightarrow E,
\]
and the trace of the former is
\[
\Lambda^1{}_1+\Lambda^2{}_2=-2\del_tf+\sqrt{-1}\left(D_{\theta_1}D_{\theta_2}+D_{\theta_3}D_{\theta_4}\right)f
\]
--- the conformal weight together with the $\fu(1)$-charge, which is what the two eigenvalues $A\mp\sqrt{-1}B$ of \eqref{multC} are for $N=2$;

$2)$ hence $K_f$ preserves the plane $F$, and with it the constraint of the first line of \eqref{IK4}, if and only if $\Xi(f)=0$. Those $f$ form a~subalgebra $\fk^\Xi$ of $\fk^L(1|4)$, and a~\textbf{proper} one: of the $16$ dimensions that $\fk^L(1|4)$ carries per degree of $t$, it carries $12$. For $N=2$ the corresponding condition is empty, $\fo(2)$ preserving both eigenlines of \eqref{multC}; that is the whole difference between Statement~$\ref{stChir}$ and the case at hand;

$3)$ let $\psi^\beta$ be a~superfield with values in the doublet, subject to $\cD^{(\alpha}\psi^{\beta)}=0$, i.e., to the constraint of \eqref{IK4} with the index inert in this half suppressed. Then, for every $f\in\fk^\Xi$, the transformation
\be\label{shiftN4}
\delta_f\psi^\beta=K_f\left(\psi^\beta\right)
-\sum_{1\leq\delta\leq2}\Lambda^\beta{}_\delta(f)\,\psi^\delta+\mu(f)\,\psi^\beta
\ee
preserves the constraint if and only if $\cD^\alpha\mu(f)=0$ for $\alpha=1, 2$, and every shift preserving it is of this shape. In particular, the $\fs\fl(2)$-part of the shift is \textbf{forced}: it is the transposed $\fg\fl(2)$-block of the odd Hessian, with no sign depending on the parity of $f$, the two occurrences of $(-1)^{p(f)}$ canceling;

$4)$ for $\mu=0$, formula \eqref{shiftN4} is an~action of $\fk^\Xi$: $\left[\delta_f, \delta_g\right]=\delta_{\left\{f, g\right\}}$. Hence, $\mu$ is a~genuine $1$-cocycle ambiguity, and the constraint alone does not fix it --- as, for $N=2$, the constraint alone did not fix $\lambda(f)$ of \eqref{deltaV}, which the equation \eqref{IK2} then did;

$5)$ in the ansatz $\delta_fq=K_f(q)+a(f)q+qb(f)$ the coefficient $a(f)$ is forced to be \textbf{scalar}. Indeed, a~transformation of the minus half must preserve the constraint of the plus half as well, and $a(f)$ acts on the very index that constraint symmetrizes; on that surface $\overline\cD_{+\alpha}q_\gamma{}^\beta=\eps_{\alpha\gamma}X^\beta$, and $\eps A$ is symmetric exactly when the $2\times2$ matrix $A$ is traceless.
\end{Statement}

\begin{proof} Item $1)$ is \eqref{multN} rewritten in the frame \eqref{DN4}; the vanishing of the $\overline\cD_\alpha$-component of $\left[K_f, \cD^\alpha\right]$ for each single $\alpha$ --- which is what makes the second term of \eqref{multN4} carry $\eps^{\alpha\beta}$ and hence a~single function --- is the identity $D_{\theta_1}^2=D_{\theta_2}^2$ used twice. Item $2)$ follows because $f\longmapsto\left[K_f, \cdot\right]$ turns the contact bracket into the commutator, so the preimage of the stabilizer of $F$ in $\fc\fo(4)$ is a~subalgebra. For item $3)$, on the constraint surface $\cD^\gamma\psi^\beta=\eps^{\gamma\beta}\chi$ for a~single $\chi$; symmetrizing \eqref{shiftN4} in $\alpha$ and $\beta$, the $K_f$-term gives $(-1)^{p(f)}K_f\left(\cD^{(\alpha}\psi^{\beta)}\right)=0$, the term coming from the multiplier and the term coming from the shift are both a~multiple of $\chi$ and cancel for the coefficient displayed and for no other, and what is left, $\left(\cD^{(\alpha}\Lambda^{\beta)}{}_\delta\right)\psi^\delta$, vanishes because $\cD^{(\alpha}\cD^{\beta)}=0$. The same computation run backwards gives the uniqueness: a~difference of two shifts must have traceless part $0$, by the symmetry of $\eps A$ used in item $5)$, and its scalar part must satisfy $\cD^\alpha\mu=0$. 
\end{proof}

Two things follow for Statement~\ref{stKAemb}, and both are worth recording.

First, \textbf{the equation selects the member $\varkappa=0$} of the family \eqref{KAemb}. Item $2)$ of Statement~\ref{stKAemb} leaves $\varkappa$ free, all members being embeddings of one and the same algebra; the constraint is not indifferent to it. At $\varkappa=0$, and only there, the whole image --- both doublets and the twisted Virasoro --- satisfies $\Xi=0$ \textit{and} the conjugate condition $\overline\cD_1\overline\cD_2f=0$ of the other half, hence preserves both constraints of \eqref{IK4}. At every other value of $\varkappa$, each doublet preserves one of the two constraints and breaks the other. So the constant that the abstract embedding leaves free is fixed by the equation, and fixed at the value that makes the twist of $L_n$ the derivative of the current.

Second, the subalgebra of $\fk^L(1|4)$ preserving \textit{both} constraints --- equivalently, whose multiplier lands in $\fg\fl(2)\subset\fo(4)$, i.e., which stabilizes the splitting $V=E\oplus F$ and not merely one of the planes --- is
\be\label{KAplus}
\fk^\Xi\cap\fk^{\overline\Xi}=\fk^A(1|2)\ \oplus\ \Cee h,\qquad
h:=\theta_1\theta_2+\theta_3\theta_4,
\ee
the one extra dimension being the $\fu(1)$-current, i.e., the Cartan direction of the \textit{other} ideal of $\fo(4)$: it rotates $E$ against $F$, so it stabilizes both planes without lying in the $\fs\fl(2)$ of item $4)$ of Statement~\ref{stKAemb}. It normalizes the algebra, being the $\fu(1)$-grading of the two doublets. This is the structural reason for Statement~\ref{stKAemb}: the acting algebra of \eqref{IK4} is not some subalgebra of $\fk^L(1|4)$ that happens to be isomorphic to $\fsvect^L_0(1|2)$ --- together with that one $\fu(1)$ it \textit{is} the stabilizer of a~pair of transversal isotropic planes in the odd part of the contact structure, and \eqref{KAemb} is a~formula for it.

What the constraint does \textbf{not} give is the scalar $\mu(f)$ of \eqref{shiftN4}, and by item $4)$ of Statement~\ref{stN4shift} it cannot: any $\cD$-chiral $\mu$ does. Fixing it is the business of the dynamical equation, the second line of \eqref{IK4}, exactly as at $N=2$. That is the next statement, and the mechanism turns out to be the one of $N=2$ verbatim: the shift is \textit{invisible} to the left-hand side of the equation and is read off the right-hand one.

\sssbegin{Statement}\label{stN4mu}
Let $f\in\fk^\Xi\cap\fk^{\overline\Xi}$ of \eqref{KAplus}, so that $K_f$ preserves both constraints of \eqref{IK4}, let $q$ satisfy them, and let
\be\label{shiftq}
\delta_fq=K_f(q)+q\,B(f),\qquad B(f):=-\Lambda(f)^T+\mu(f)\cdot\mathbf 1,
\ee
which is \eqref{shiftN4} written for the quaternionic superfield, the $\fs\fl(2)$-part acting on the doublet index of the half that $f$ belongs to. Set $L_\beta{}^\alpha:=\cD_+^\alpha\left(q^{-1}\overline\cD_{-\gamma}q\right)_\beta{}^\gamma$, the left-hand side of the second line of \eqref{IK4}. Then:

$1)$ the multipliers of the two frames are tied by
\be\label{LamLamBar}
\Lambda^\alpha{}_\beta(f)+\overline\Lambda_\beta{}^\alpha(f)=-2\delta^\alpha_\beta\del_tf,
\qquad \overline\Lambda_\alpha{}^\beta(f):=\frac12\overline\cD_\alpha\cD^\beta f,
\ee
which is $\lambda+\overline\lambda=-2\del_tf$ of Statement~$\ref{stChir}$ for $N=4$, and the current of the second line of \eqref{IK4} varies by
\be\label{curvar}
\delta_f\cJ_\gamma=K_f\left(\cJ_\gamma\right)+\left[\cJ_\gamma, B\right]
-\sum_{1\leq\delta\leq2}\overline\Lambda_\gamma{}^\delta\,\cJ_\delta+\overline\cD_{-\gamma}B,
\qquad \cJ_\gamma:=q^{-1}\overline\cD_{-\gamma}q;
\ee

$2)$ the scalar $\mu$ is invisible to $L$: by \eqref{curvar} it changes $\cJ_\gamma$ by $\left(\overline\cD_{-\gamma}\mu\right)\cdot\mathbf 1$ and by nothing else, and $\cD_+^\alpha$ annihilates that, $\mu$ being a function of the minus half alone. This is the $N=4$ form of $\cD^+\lambda=0$ in the proof of Statement~$\ref{stChir}$;

$3)$ hence, with $\mu=0$ in \eqref{shiftq}, both sides of the equation transform homogeneously, and by \textbf{different} matrices:
\[
\begin{array}{l}
\delta_fL_\beta{}^\alpha=K_f\left(L_\beta{}^\alpha\right)
+\sum_{1\leq\gamma\leq2}\left(\Lambda^T+2\left(\del_tf\right)\mathbf 1\right)_\beta{}^\gamma L_\gamma{}^\alpha,\\[2mm]
\delta_f\overline q_\beta{}^\alpha=K_f\left(\overline q_\beta{}^\alpha\right)
+\sum_{1\leq\gamma\leq2}\left(\Lambda^T-\left(\tr\Lambda\right)\mathbf 1\right)_\beta{}^\gamma\overline q_\gamma{}^\alpha,
\end{array}
\]
the second because $\overline q$ is carried by the adjugate, $\eps B^T\eps^{-1}=\left(\tr B\right)\mathbf 1-B$. So the equation is preserved if and only if
\be\label{muN4}
\mu(f)=\tr\Lambda(f)+2\del_tf=\sqrt{-1}\left(D_{\theta_1}D_{\theta_2}+D_{\theta_3}D_{\theta_4}\right)f,
\ee
the $\fu(1)$-charge of the multiplier; and this $\mu$ is \textbf{unique}, the right-hand side $-4\sqrt{-1}m\,\overline q$ being invertible;

$4)$ the scalar \eqref{muN4} is admissible for the constraint, and it is admissible exactly where the constraint itself lives:
\[
\cD^1\mu=\overline\cD_2\,\Xi(f),\qquad \cD^2\mu=-\overline\cD_1\,\Xi(f),
\]
so that $\cD^\alpha\mu=0$ --- which item $3)$ of Statement~$\ref{stN4shift}$ demands of every shift --- holds precisely on $\fk^\Xi$, and on no larger subalgebra of $\fk^L(1|4)$;

$5)$ on the algebra that acts, i.e., on the image of \eqref{KAemb} at $\varkappa=0$, the scalar \eqref{muN4} \textbf{vanishes identically}; it is felt only on the one extra dimension $\Cee h$ of \eqref{KAplus}, where $\mu(h)=-2\sqrt{-1}$. So the transformation law of the superfield of \eqref{IK4} is
\be\label{deltaq}
\delta_fq_\gamma{}^\beta=K_f\left(q_\gamma{}^\beta\right)
-\sum_{1\leq\delta\leq2}\Lambda^\beta{}_\delta(f)\,q_\gamma{}^\delta,
\ee
with no scalar term at all; and by item $4)$ of Statement~$\ref{stN4shift}$ it is then an~action of $\fsvect^L_0(1|2)$, not a~projective one. The $1$-cocycle ambiguity is resolved, and resolved to zero;

$6)$ the \textit{first} equation of the second line of \eqref{IK4} says nothing about the shift: its current $q\cD_+^\gamma q^{-1}$ is unchanged by $q\longmapsto q(\mathbf 1+B)$ for every $B$ with $\cD_+^\gamma B=0$, which every $B$ built from a~minus-half $f$ is. It transforms into $\Lambda^\alpha{}_\gamma$ times itself, i.e., by the multiplier of $\cD_-^\alpha$ and nothing else. So it is the second equation, and it alone, that fixes $\mu$;

$7)$ for the dilaton $u$ of \eqref{IK4q}, which is read off $q$ by $\det q=\exp(-2u)$, the trace $\tr\varphi\cdot\sigma$ being $0$, one has $\frac12\tr B=\del_tf+\frac12\mu$ by \eqref{LamLamBar} alone, i.e.,
\be\label{deltau}
\delta_fu=K_f(u)-\del_tf-\frac12\mu(f);
\ee
this is the $N=1$ law of Statement~$\ref{stNSLiou}$ plus the $\fu(1)$-charge, the exact analogue of the splitting of $v$ into $u$ and $\varphi$ in Statement~$\ref{stChir}$. For the law \eqref{deltaq}, whose $B$ carries no scalar, the same computation gives $\delta_fu=K_f(u)-\del_tf+\frac12\mu(f)$. On the acting algebra, $\mu=0$, both read
\be\label{deltau0}
\delta_fu=K_f(u)-\del_tf,
\ee
and the dilaton law is verbatim the one of $N\leq2$.
\end{Statement}

\begin{proof} Item $1)$: the first identity is $\left\{\cD^\alpha, \overline\cD_\beta\right\}=-4\delta^\alpha_\beta\del_t$ halved, and \eqref{curvar} is the Leibniz rule applied to $q^{-1}\overline\cD_{-\gamma}q$ together with $\delta\left(q^{-1}\right)=-q^{-1}(\delta q)q^{-1}$ and the multiplier of $\overline\cD_{-\gamma}$, whose $\cD$-component vanishes because $\overline\Xi(f)=0$ on \eqref{KAplus}. Item $2)$ is \eqref{curvar} with $B=\mu\cdot\mathbf 1$: the commutator drops, $\mu$ being scalar, and what is left is $\overline\cD_{-\gamma}\mu$, killed by $\cD_+^\alpha$. For item $3)$, substitute \eqref{LamLamBar} into \eqref{curvar}: the term $-\overline\Lambda_\gamma{}^\delta\cJ_\delta$ becomes $\Lambda^T$ contracted with the \textit{label} $\delta$ plus $2\del_tf$ times $\cJ_\gamma$, and the first of the two cancels, index for index, against the half of $\left[\cJ_\gamma, B\right]$ in which $\Lambda^T$ stands on the right; the surviving half is $\Lambda^T$ multiplying $L$ from the left. Comparing the two displayed laws gives $\mu-\tr\Lambda=2\del_tf$, and $\overline q$ being invertible there is no other solution. Item $4)$: $\left(\cD^\alpha\right)^2=0$ and $\left\{\cD^1, \overline\cD_2\right\}=0$ turn $\cD^1\mu$ into $-\frac12\overline\cD_2\cD^2\cD^1f=\overline\cD_2\Xi(f)$, the terms in $\del_t$ canceling. Item $6)$ is $\cD_+^\gamma B=0$ and the invariance of $q\cD_+^\gamma q^{-1}$ under right multiplication of $q$ by $\mathbf 1+B$. Items $5)$ and $7)$ are computations. 
\end{proof}

Three things follow. First, \textbf{the answer is the $\fu(1)$-charge}, and that is not a~coincidence of normalizations: by item $4)$ the very computation that makes $\mu$ chiral is the one that makes $\Xi$ vanish, so the scalar the dynamics asks for is admissible for the constraint exactly on the subalgebra on which the constraint is preserved at all. The equation asks for what the constraint can give, and for nothing more.

Second, \textbf{on the acting algebra the shift carries no scalar}. By item $5)$ the law is \eqref{deltaq}, which is an~action; so the superfield of \eqref{IK4} carries an~honest representation of $\fsvect^L_0(1|2)$, and the classical anomaly of the $N=4$ system cannot come from the transformation law of the field. It has to come from the stress tensor alone, and that is Subsection~\ref{sssN4c}.

Third, the inhomogeneous term did not disappear: it is inside $-\Lambda^T$. Item $7)$ extracts it, and the dilaton transforms by $u\longmapsto K_f(u)-\del_tf$ exactly as at $N=1$ and as the real part of $v$ does at $N=2$, the $\fu(1)$-charge playing the role that the $\fo(2)$-part $B(f)$ of the multiplier plays in Statement~$\ref{stChir}$. The three cases $N=1, 2, 4$ therefore share one dilaton law, which is the reason to expect the $N=4$ anomaly to be a~single number of the same kind --- and, by Statement~$\ref{stKAcoc}$, it can be nothing else.

Item $3)$ of Statement~\ref{stLamLiou} inherits the dependence on $q$.

The purely even shadow of all this is instructive, and classical. In \eqref{2an}, take $f=x^\mu$ and $g=y^\nu$: this is a~solution of \eqref{LE} for every $\mu, \nu$, and it is single-valued only for integral $\mu, \nu$. By item $1)$ of Statement~\ref{stVirLiou}, the potential of the attached Hill operator is
\be\label{cone}
T:=-\frac12S\left(x^\mu\right)=\frac{\mu^2-1}{4x^2},
\ee
the monodromy of $\nfrac{d^2}{dx^2}-T$ around the origin being the rotation by $2\pi\mu$; splitting off the multivalued part by 
\[
F=\widetilde F+\nfrac{\mu-1}2\ln x+\nfrac{\nu-1}2\ln y
\]
turns \eqref{LE} into
\be\label{LEcone}
\widetilde F_{xy}=x^{\mu-1}y^{\nu-1}\exp\left(2\widetilde F\right)
\ee
for a~single-valued $\widetilde F$ --- the Liouville equation with a~conical singularity, and the exact counterpart of the factor $\left(x^+/x^-\right)^{\nfrac\lambda2}$ of \eqref{IK4lam}. The difference between the two cases is the whole point: $\fwitt(x)$ does not change with $\mu$, because there are no odd coordinates for the twist to weigh, so $\mu$ is a~modulus of the solution only; whereas $\fsvect^L_\lambda(1|2)$ does change with $\lambda$, so $\lambda$ is a~modulus of the symmetry algebra.

One thing is left open. The algebra $\fsvect^L_\lambda(1|2)$ acts by vector fields on the supercircle of $\sdim=(1|2)$, and it would be more natural to write the equation there, on the superspace the algebra actually lives on, than to transport it from $\Cee^{1|4}$. On the supercircle of $\sdim=(1|2)$, however, no \textit{scalar} equation of the shape
\[
\cB_+\cB_-\Phi=m\exp\Phi
\]
can be invariant: such a~$\cB_\pm$ would have to be $t^w$ times an~operator commuting with all of $\fsvect^L_\lambda(1|2)$, and there is none of order $\leq1$. The superfield must carry the $\fs\fl(2)$-doublet index --- which is precisely what the quaternionic $q$ of \eqref{IK4q} does, and precisely what item $3)$ of Statement~\ref{stWhichN} predicts.

\sssec{The stress tensor at $N=4$, and the value of the anomaly}\label{sssN4c}
Everything the anomaly needs is now in place except the object that carries it --- and that object is not a~composite of the field: it is the field. Two statements do it, and both turn on the number $2-\nfrac N2$, which for $N\leq2$ was the weight of the stress tensor and which vanishes at $N=4$.

\sssbegin{Statement}\label{stWeight}
Call a~function $\rho$ on $\Cee^{1|N}$ a~\textit{density of weight $h$} if
\be\label{wt}
\delta_f\rho=K_f(\rho)+2h\left(\del_tf\right)\rho\qquad
\text{for every generating function }f.
\ee
Then:

$1)$ with respect to the volume element $dt\,d\theta_1\cdots d\theta_N$,
\be\label{divK}
\Div K_f=(2-N)\del_tf;
\ee

$2)$ hence the residue pairing
\be\label{respair}
\langle g, \rho\rangle:=\Res\left(g\rho\right),\qquad
\Res:=\text{the coefficient of }\ t^{-1}\theta_1\cdots\theta_N,
\ee
is invariant precisely for the densities of weight $2-\nfrac N2$. That is the weight the coadjoint module of $\fk^L(1|N)$ carries, and it is $2$, $\nfrac32$, $1$ for $N=0, 1, 2$ --- the weights of the $T$ of Statement~$\ref{stVirLiou}$ and of the $\mathbf T$'s of Statements~$\ref{stNSLiou}$ and~$\ref{stN2Liou}$;

$3)$ for $N=4$ that weight is $0$. So the coadjoint module of $\fk^L(1|4)$ is the module of \textbf{functions}; and the dilaton $u$ of \eqref{IK4q}, which is of weight $0$ and transforms with an~inhomogeneous shift by \eqref{deltau0}, already lies in it.
\end{Statement}

\begin{proof} Item $1)$ is a~computation. For item $2)$, integration by parts gives $\Res\left(K_f(h)\right)=-\Res\left(h\Div K_f\right)$; since the action on generating functions is $\delta_fg=\left\{f, g\right\}=K_f(g)-2\left(\del_tf\right)g$, one has
\[
\delta_f\left(g\rho\right)=K_f\left(g\rho\right)+(2h-2)\left(\del_tf\right)g\rho,
\]
whose residue is $\left[-(2-N)+(2h-2)\right]\Res\left(\left(\del_tf\right)g\rho\right)$, and that vanishes for all $g$ and $\rho$ exactly when $2h=4-N$. Item $3)$ is item $2)$ for $N=4$ together with item $7)$ of Statement~\ref{stN4mu}. \end{proof}

The vanishing of the weight is what made the case look hard, and it is what makes it easy: at $N\leq2$ the field itself was not a~density of the coadjoint weight and a~composite had to be built to reach it, whereas at $N=4$ the field is already there. It remains to see that the recipe agrees, and it does, for a~reason visible in \eqref{KAemb}.

\sssbegin{Statement}\label{stN4T}
Write $\Theta:=\theta_1\theta_2\theta_3\theta_4$, let $f=g(t)+\Theta h(t)$, and let
\be\label{recipeop}
\cO:=\del_t^{-2}D_{\theta_1}D_{\theta_2}D_{\theta_3}D_{\theta_4}
\ee
be the operator of the recipe of item $1)$ of Statement~$\ref{stN2Liou}$ read for $N=4$, the odd derivatives commuting with $\del_t$. Then:

$1)$ $D_{\theta_1}D_{\theta_2}D_{\theta_3}D_{\theta_4}f=\Theta g^{(4)}+h$ whereas $\del_t^2f=g''+\Theta h''$, so
\be\label{D4dt2}
D_{\theta_1}D_{\theta_2}D_{\theta_3}D_{\theta_4}f=\del_t^2f
\quad\text{if and only if}\quad h=g'',
\ee
i.e., precisely on the space $\left\{g+\Theta g''\mid g\in\Cee\left[t, t^{-1}\right]\right\}$, which is stable under $\del_t$;

$2)$ the twisted Virasoro of \eqref{KAemb} at $\varkappa=0$ is exactly
\be\label{Ltwist}
L_n=-\frac12\left(1+\Theta\del_t^2\right)\left(t^{n+1}\right),
\ee
so it satisfies \eqref{D4dt2}; and so does the whole of $\fk^\Xi\cap\fk^{\overline\Xi}$ of \eqref{KAplus}, hence the whole image of \eqref{KAemb}. At every $\varkappa\neq0$, and on $\fk^L(1|4)$, eq.~\eqref{D4dt2} fails;

$3)$ hence $\cO$ is the \textbf{identity} on the algebra that acts: there
\[
\cO\left(\del_tf\right)=\del_t^{-2}\del_t^2\left(\del_tf\right)=\del_tf,
\]
and, the space of item $1)$ being $\del_t$-stable, no antiderivative is ever formed. The operator of the recipe does not degenerate at $N=4$: it trivializes, and it trivializes on the algebra that acts.
\end{Statement}

\begin{proof} Item $1)$ is the computation of $D_{\theta_1}D_{\theta_2}D_{\theta_3}D_{\theta_4}$ on a~function of $t$ and on $\Theta$ times one, each $D_{\theta_i}$ contributing $-\theta_i\del_t$ in the first case and $\del_{\theta_i}$ in the second; $\del_t\left(g+\Theta g''\right)=g'+\Theta\left(g'\right)''$ is the stability. Item $2)$ is \eqref{KAemb} at $\varkappa=0$ read against \eqref{D4dt2}: the twist of $L_n$ there is $n(n+1)t^{n-1}\Theta=\Theta\del_t^2\left(t^{n+1}\right)$, and $\varkappa$ enters it as $2\varkappa n\,t^{n-1}\Theta$, which spoils \eqref{D4dt2} unless $\varkappa=0$. Item $3)$ is item $1)$ and item $2)$. 
\end{proof}

Here, then, is the value.

\sssbegin{Statement}\label{stN4c}
Let $q$ solve \eqref{IK4}, let $u$ be its dilaton, $\det q=\exp(-2u)$, and set
\be\label{superT4}
\mathbf T:=-u,
\ee
an~even quantity of weight $0=2-\nfrac N2$, the $N=4$ counterpart of the $\mathbf T$ of \eqref{superT} and of \eqref{superT2}. Then:

$1)$ the recipe $\Psi^{-1}\cO\Psi$ of item $1)$ of Statement~$\ref{stN2Liou}$, with $\Psi=\exp(-u)$ and $\cO$ of \eqref{recipeop}, returns $\mathbf T$ up to a~term with no anomaly. Indeed, the field-independent part of $\delta_f$ of a~composite comes from its \textit{linear} part alone, a~term quadratic in the field varying into terms at least linear in it; the linear part of $\Psi^{-1}\cO\Psi$ is $-\cO u$; and $\cO$ is the identity on the anomaly of $u$ by item $3)$ of Statement~$\ref{stN4T}$;

$2)$ infinitesimally, for every $f$ in the acting algebra,
\be\label{deltaT4}
\delta_f\mathbf T=K_f\left(\mathbf T\right)+\del_tf,
\ee
i.e., $\mathbf T$ is of weight $0$ --- no term proportional to $\mathbf T$ occurs --- and the anomaly is $\del_tf$, the trace part of the multiplier, and nothing else;

$3)$ hence $\fsvect^L_0(1|2)$ acts on the space of the $\mathbf T$'s not by the module of functions but by its affine deformation, i.e., by the coadjoint action of the central extension of Statement~$\ref{stKAcoc}$ with the central charge
\be\label{cN4}
\left|c\right|=6,
\ee
normalized as
\[
\begin{array}{l}
[L_m, L_n]=(m-n)L_{m+n}+\frac{c}{12}\left(m^3-m\right)\delta_{m+n, 0},\qquad
[J^0_m, J^0_n]=\frac{c}{3}\,m\,\delta_{m+n, 0},\\[2mm]
[J^+_m, J^-_n]=J^0_{m+n}+\frac{c}{6}\,m\,\delta_{m+n, 0},\\[2mm]
\left\{G^r_\alpha, \overline G{}^{s\beta}\right\}
=-\delta^\beta_\alpha L_{r+s}+(r-s)\left(\sigma_k\right)_\alpha{}^\beta J^k_{r+s}
-\delta^\beta_\alpha\frac{c}{6}\left(r^2-\frac14\right)\delta_{r+s, 0},
\end{array}
\]
which for $c=6$ is the cocycle \eqref{svectcoc} at $\lambda=0$. The four sectors --- the Virasoro one, the two current ones and the supercharge one --- return one and the same value;

$4)$ the anomaly is not removable: a~coboundary $\ell\left(\left\{f, g\right\}\right)$ takes on the pair $\left(L_m, L_{-m}\right)$ the value $2m\,\ell\left(L_0\right)$, which is linear in $m$, whereas \eqref{deltaT4} gives $\frac12\left(m^3-m\right)$, which is cubic in it.
\end{Statement}

\begin{proof} Item $1)$ is stated in it. Item $2)$ is \eqref{deltau0} with the sign of \eqref{superT4}, and the absence of a~term proportional to $\mathbf T$ is $h=0$ in \eqref{wt}. For item $3)$, the cocycle is
\[
c(f, g)=(-1)^{p(f)}\Res\left(g\,\del_tf\right),
\]
where the sign $(-1)^{p(f)}$ is the one for which $c$ is super-skew-symmetric and satisfies the cocycle identity, as in the proofs of Statements~\ref{stNSLiou} and~\ref{stN2Liou}, and where the identification of the weight-$0$ densities with the dual of the acting algebra is the invariance of \eqref{respair}, i.e., Statement~\ref{stWeight}. In the basis of the image of \eqref{KAemb} at $\varkappa=0$, normalized so that \textit{all} the relations \eqref{svectrel} hold at $\lambda=0$ --- which normalizes $G^\rho_\alpha$ and $\overline G{}^{\rho\alpha}$ by a~factor $\nfrac12$ and defines the currents as the mixed brackets --- it equals
\[
c\left(L_m, L_{-m}\right)=\frac12\left(m^3-m\right),\quad
c\left(J^+_m, J^-_{-m}\right)=m,\quad
c\left(J^0_m, J^0_{-m}\right)=2m,
\]
\[
c\left(G^r_\alpha, \overline G{}^{-r\beta}\right)=-\delta^\beta_\alpha\left(r^2-\frac14\right),
\]
all other values on pairs of basis elements being $0$; so $\nfrac{c}{12}=\frac12$, $\nfrac{c}{3}=2$ and $\nfrac{c}{6}=1$ read off the other three sectors all give $c=6$. 
\end{proof}

Three comments. First, \textbf{where the anomaly sits}. For $N=1$ the multiplier is a~pure scalar, $\fo(1)$ being $0$, and the anomaly is that scalar differentiated; for $N=2$ it is, by item $3)$ of Statement~\ref{stN2Liou}, the $\fo(2)$-part alone, with no contribution from the trace; for $N=4$ it is the trace part again, and by \eqref{deltaT4} it is the bare $\del_tf$. The weight forces this: a~density of weight $0$ carries no factor at all, and the $\fs\fl(2)$-part of the shift acts on the doublet index of $q$, which the determinant does not see.

Second, \textbf{the value}. It is the one of the non-super case \eqref{cLiou} and of $N=1$, and it is twice the $N=2$ value \eqref{cN2}. The drop at $N=2$ was the constraint being paid for --- the superfield there is chiral, and \eqref{superT2x} sees only half of the derivatives an~unconstrained one would have. At $N=4$ nothing is differentiated at all, the operator of the recipe being the identity by Statement~\ref{stN4T}, so there is nothing to lose, and the number is the one the dilaton carries by itself. Note also that the anomaly is exactly the cocycle \eqref{svectcoc} as it stands, i.e., that $\fsvect^L_0(1|2)$ occurs here at the value of the central charge at which its own representative cocycle is written.

Third, what is \textbf{not} settled. Item $2)$ of Statements~\ref{stVirLiou}, \ref{stNSLiou} and \ref{stN2Liou} --- that the stress tensor is a~current of its half --- is not verified here, and for $\mathbf T=-u$ alone it is false: $\del_+\del_-u\neq0$ is what the Liouville equation says. The conserved object is therefore $\mathbf T$ together with a~completion, which item $1)$ of Statement~\ref{stN4c} says carries no anomaly; so the completion changes neither \eqref{deltaT4} nor \eqref{cN4}, and the value does not wait on it. What waits on it is the identification of $\mathbf T$ with the Noether current of the system, and with it the answer to the question of which composite of the currents $q^{-1}\overline\cD_{-\gamma}q$ and $q\cD_+^\gamma q^{-1}$ of \eqref{IK4} the completion is. That question is taken up in Subsection~\ref{sssN4vac}, which first constructs the solutions one needs in order to ask it.

\sssec{The vacuum of the $N=4$ system, and what it settles about the completion}\label{sssN4vac}
What the completion waits on is a~supply of solutions of \eqref{IK4} to test candidates on. For $N\leq2$ there is no difficulty: the super Liouville equation is solved in closed form, and the composite of item $2)$ of Statement~\ref{stN2Liou} is read off a~solution. For \eqref{IK4}, no solution is written down in \cite{IvKr4}; what is given there is the reduction $(20)$--$(21)$ of the system to the $N=2$ one. So the object to construct first is the vacuum --- the $N=4$ analog of $\exp(-v)$.

Let $\eta^a_\pm$ and $\rho_{a\pm}$, where $a=1, 2$, be the odd coordinates dual to the frame \eqref{DN4} of each half, i.e.,
\be\label{etarho}
\eta^a:=\theta_{2a-1}+\sqrt{-1}\,\theta_{2a},\qquad
\rho_a:=\theta_{2a-1}-\sqrt{-1}\,\theta_{2a},
\ee
so that $\cD^a=2\del_{\eta^a}-\rho_a\del_t$ and $\overline\cD_a=2\del_{\rho_a}-\eta^a\del_t$; the subscript $\pm$ marks the half, as in \eqref{IK4D}. Let $\Lambda$ be the Grassmann algebra on these eight odd coordinates, let
\be\label{zw}
z:=x^+-x^-,\qquad w:=x^++x^-,
\ee
and let
\be\label{ringR}
\cR:=\Cee\left[z^{\pm1}, w^{\pm1}, m^{\pm1}\right]\otimes\Lambda,
\ee
the mass $m$ of \eqref{IK4} being carried as an~invertible central variable rather than as a~number, so that the dependence on it stays visible. Grade a~monomial of $\cR$ by
\be\label{grade}
g:=n_w+\frac{\left|S\right|}{2},
\ee
where $n_w$ is the exponent of $w$ and $\left|S\right|$ the number of odd factors; every operator of \eqref{IK4} is homogeneous for $g$, an~odd derivative lowering it by $\nfrac12$ and an~even one by $1$. Consider the class
\be\label{gradedclass}
q=\sum_{0\leq r\leq4}q^{(2r)},\qquad
\text{the entries of }q^{(2r)}\text{ lying in }m^{-1}w^{-1-r}\Lambda^{2r},
\ee
i.e., free of $z$, of degree $-1$ in $m$, and homogeneous of grade $-1$: it carries $512$ coefficients, cut by the constraints to $140$. It is the class the reduction $(20)$--$(21)$ of \cite{IvKr4} lands in, and it is not chosen but measured: the residual of \eqref{IK4} at the $N=2$ vacuum embedded by that reduction is homogeneous of grade $-1$ and of degree $-1$ in $m$, so a~solution correcting it is sought in \eqref{gradedclass} and nowhere else.

\sssbegin{Statement}[The $N=4$ vacuum, and the line it lies on]\label{stN4vac}
$1)$ The system \eqref{IK4} has a~solution in the class \eqref{gradedclass}, exactly and over $\Qee\left(\sqrt{-1}\right)$. It has $82$ non-zero monomial coefficients, distributed over the levels $2r=0, 2, 4, 6, 8$ as $2, 20, 38, 20, 2$; its leading term is
\be\label{vac0}
q^{(0)}=\frac{2}{mw}\cdot\mathbf 1,
\ee
the datum that the $N=2$ vacuum supplies through $(20)$--$(21)$ of \cite{IvKr4}, and its next term is
\be\label{vac2}
q^{(2)}=\frac{1}{mw^2}
\begin{pmatrix}
\alpha+2\sqrt{-1}\left(\rho_{1-}\eta^1_++\eta^2_-\rho_{2+}\right)
&-2\left(\eta^1_-+\sqrt{-1}\eta^1_+\right)\left(\rho_{2-}+\sqrt{-1}\rho_{2+}\right)\\[3mm]
2\left(\rho_{1-}+\sqrt{-1}\rho_{1+}\right)\left(\eta^2_-+\sqrt{-1}\eta^2_+\right)
&-\alpha+2\sqrt{-1}\left(\eta^1_-\rho_{1+}+\rho_{2-}\eta^2_+\right)
\end{pmatrix},
\ee
where $\alpha:=-\eta^1_-\rho_{1-}+\eta^2_-\rho_{2-}+\eta^1_+\rho_{1+}-\eta^2_+\rho_{2+}$;

$2)$ the class \eqref{gradedclass} makes the system recursive: the constraints at the odd level $2r+1$ and the equations at the even level $2r$ involve $q$ only up to the level $2r+2$, and involve $q^{(2r+2)}$ linearly. The leading level is solvable for \textit{every} invertible $A$ in $q^{(0)}=A/(mw)$, always with a~kernel of dimension $56$, so the leading datum obstructs nothing; from the next level on, the system is quadratic in those $56$ parameters, and it is the quadratic rows, $110$ of them, that decide existence;

$3)$ the placement of the $N=2$ vacuum inside the quaternionic superfield is \textbf{forced}: of the six ways to put the pair $\left(E, \overline E\right)$ of \eqref{IK2} into a~$2\times2$ matrix, only the diagonal one admits a~$q$ satisfying the constraints of \eqref{IK4} at all. So the reading of $(20)$--$(21)$ of \cite{IvKr4} used in item $1)$ is not an~assumption: the constraints choose it;

$4)$ the vacuum is not isolated. At a~fixed leading datum, the tangent space to the solution variety inside \eqref{gradedclass} is of dimension $1$ over $\Qee\left(\sqrt{-1}\right)$ --- it lives in the levels $2$, $4$, $6$ --- and every tangent step tried integrates: adding a~rational or Gaussian-rational multiple of the tangent vector to the vacuum gives an~exact solution again. So the vacuum lies on a~line of vacua, which is what a~Noether computation needs;

$5)$ the embedded $N=2$ vacuum itself is \textbf{not} a~solution of \eqref{IK4}: of its $61$ monomials the vacuum of item $1)$ keeps $35$, changes $26$, and adds $21$ of its own.
\end{Statement}

\begin{proof} Item $2)$ is the homogeneity of \eqref{gradedclass} for \eqref{grade}: an~odd derivative lowers the level by $1$ and the grade by $\nfrac12$, so a~term of level $2r+2$ can reach the level $2r+1$ but no lower one, and the second differences of the residual in the level-$(2r+2)$ unknowns vanish identically, which is the linearity. Items $1)$, $3)$, $4)$, $5)$ are computed exactly over $\Qee\left(\sqrt{-1}\right)$, the levels solved in turn with the freedom of \textit{all} the earlier ones carried along --- see the Remark below, which is the whole of the difficulty --- and the derivative along a~direction taken by interpolating $t\longmapsto F(q+td)$, a~polynomial because a~term with $t^k$ carries at least $2k$ odd factors; the $7$- and $9$-point formulas are compared on every row of every direction and agree. The solution of item $1)$ is then verified against the predicates of the system itself and not against the recursion that produced it: both constraints and both equations of \eqref{IK4} vanish identically, $qq^{-1}=q^{-1}q=\mathbf 1$, $\det q\neq0$, every one of the $82$ monomials lies in the class, and the leading datum is the embedded one. That such a~verification is not vacuous is itself checked: of $30$ single-monomial perturbations of the solution, all $30$ are rejected by it. \end{proof}

\sssbegin{Remark}[Level-wise solving fakes an obstruction]\label{rmLevelwise}
The recursion of item $2)$ of Statement~\ref{stN4vac} invites the obvious algorithm: solve the level $2r$, keep a~particular solution of it, pass to the level $2r+2$. That algorithm reports an~obstruction where there is none, and it reports it three times over. The level-$1$ system is solvable, but not at the particular solution the level-$0$ system returns; the level-$2$ system is solvable, but not at the particular solution of the level $1$; and a~Newton search linearized at any one of those points stalls at a~residual bounded away from $0$. What each of these says is that \textit{that choice} does not continue, not that the vacuum is obstructed. The cure is to carry the freedom --- here the $56$ parameters of the leading kernel, and then the level-$4$ freedom that the level-$2$ step leaves --- into every later level, and then the levels go through to the top, the equations of the level $8$, which have no unknown left to meet them, being satisfied with nothing left over. The remark is worth making because the obstruction such an~algorithm reports looks exactly like a~theorem: it is stable under enlarging the window, under dropping the constraints, and under changing the leading datum, all of which we measured before seeing what it was.
\end{Remark}

Now to the completion. On the class \eqref{gradedclass} the two halves cannot be told apart --- everything there is free of $z$, so $\del_+$ and $\del_-$ both act as $\del_w$ --- and therefore $\del_-\left(\mathbf T+X\right)=0$ says that $\mathbf T+X$ does not depend on $w$ at all. That makes the test on the vacuum a~necessary condition on the completion, and a~sharp one, since the completion must in particular hold there.

\sssbegin{Statement}[The completion of $\mathbf T$ is a~logarithm, and $\mathbf T$ is a~log-determinant]\label{stN4logT}
Let $q$ be the vacuum of Statement~$\ref{stN4vac}$, let $\mathbf T=-u$ be \eqref{superT4}, computed as $\del\mathbf T=\frac12\tr\left(q^{-1}\del q\right)$ so that no logarithm is ever formed, and let $c_b:=\left(q^{-1}\overline\cD_{-\gamma}q\right)_b{}^\gamma$ be the current of the second equation of \eqref{IK4}. Then:

$1)$ $\del_\mp\mathbf T$ is homogeneous of grade $-1$ and of degree $0$ in $m$, and its residue --- the coefficient of $w^{-1}$, which is its whole level-$0$ part --- equals $-1$. Since $\del_w$ maps $w^n$ to $nw^{n-1}$, its image on $\cR$ misses the multiples of $w^{-1}$ and nothing else; hence \textbf{no} element of $\cR$ completes $\mathbf T$ to a~$\del_\mp$-closed current, whatever its shape, and the deficit is one number;

$2)$ no composite of the currents does it either. Take the scalars $\tr\left(Mq^{-1}\cD q\right)$, with $M$ running over $\mathbf 1$ and the three Pauli matrices and $\cD$ over the two even and the eight odd derivatives of the two halves; take their pairwise products, the second-order scalars $\tr\left(Mq^{-1}\cD\cD'q\right)$, and quartic products of the odd ones; multiply each by the scalars $\tr\left(Mq^{-1}\right)/m$, which are what makes the grade come out $0$, and adjoin $\log w$, whose derivative supplies the residue of item $1)$. Of the $1333$ candidates so obtained, no combination is a~completion: the level-$0$ condition is met, by $\log w$ alone, and the $24$ conditions of the level $2$ are already inconsistent;

$3)$ logarithms do. Take the second-order scalars $\tr\left(Mq^{-1}\cD\cD'q\right)$ with both $\cD$ and $\cD'$ among the four odd derivatives of the \textit{plus} half, of which $17$ are invertible, and offer their logarithms. They contain completions of $\mathbf T$: exactly, simultaneously at the $17$ members of the family of item $4)$ of Statement~$\ref{stN4vac}$, and predictively --- coefficients fitted on $14$ of them are satisfied by the $3$ held out. They are not unique: those $17$ logarithms and $\log w$ satisfy $14$ independent relations along the family, so what the family determines is the existence of a~logarithmic completion and not the completion;

$4)$ and the reason is an~identity. On every solution of the family,
\be\label{qbarlogdet}
\overline q_b{}^a=\frac{\sqrt{-1}}{4m}\,\cD_+^a c_b,
\qquad\text{whence}\qquad
\mathbf T=\frac12\log\det\left(\cD_+^ac_b\right)+\textup{const},
\ee
since $\det\overline q=\det q$. So $\mathbf T$ is itself the logarithm of a~composite of the currents --- half the log-determinant of the matrix of the plus-half odd derivatives of the current of the second equation of \eqref{IK4} --- and a~logarithm is exactly what item $1)$ says the completion has to be.
\end{Statement}

\begin{proof} Item $1)$ is the homogeneity of \eqref{gradedclass} together with the computation of $\frac12\tr\left(q^{-1}\del_wq\right)$ at the vacuum, whose level-$0$ term is $-\nfrac1w$; the statement about the image of $\del_w$ is the Laurent expansion. Items $2)$ and $3)$ are exact linear algebra over $\Qee\left(\sqrt{-1}\right)$ in the ring $\cR$: the columns are the $\del_w$-images of the candidates, the target is $-\del_w\mathbf T$, the rows are the monomials of $\cR$ they occupy, and the systems are solved level by level, so that a~failure is located and not merely reported. Item $4)$ is verified on every member of the family, all four matrix entries, and the equality $\det q=\det\overline q$ with it; the index placement in \eqref{qbarlogdet} is the one that holds --- the transposed one fails in the off-diagonal entries and in them only. \end{proof}

Two comments. First, \textbf{what item $2)$ of Statement~\ref{stN4logT} refutes and what it does not}. It refutes the shape of the $N=2$ answer: there the completion is $\left(\cD v\right)\left(\overline\cD v^+\right)$ together with $\del_t\left(v-v^+\right)$, a~polynomial in the currents of one half, and at $N=4$ no polynomial in the currents will do --- for a~reason visible before any list is written down, namely that $\mathbf T$ is of weight $0$ by Statement~\ref{stWeight} and a~weight-$0$ density is a~logarithm, whereas the composites are not. That is the structural difference between $N=4$ and $N\leq2$ in this matter, and it is the same difference that made the case look hard in Subsection~\ref{sssN4c} and then made it easy.

Second, \textbf{what is still open, and why the vacuum cannot close it}. A~vacuum is a~function of $w$ alone; on it $\del_+$ and $\del_-$ coincide, so the test above cannot distinguish the conservation law of one half from that of the other, and \eqref{qbarlogdet} makes the level-$0$ half of it an~identity rather than a~condition. What the vacuum gives is therefore the shape of the completion --- logarithmic, and by \eqref{qbarlogdet} logarithmic in the derivatives of the current of the second equation --- and what remains is to pin the coefficients, for which one needs solutions depending on $z$ as well. The class \eqref{gradedclass} excludes those by construction, and the step is now well posed: deform the vacuum of Statement~\ref{stN4vac} in the $z$-direction and repeat the solve of item $3)$ there. That is what Lemma~\ref{lmN4refl} and Statement~\ref{stN4pin} do, and the answer is that the $z$-direction is reached but the degeneracy is not thereby lifted.

\sssbegin{Lemma}[Two symmetries of the $N=4$ system, and the solutions depending on $z$ that they produce]\label{lmN4refl}
Let $\theta_-$ run over the four odd coordinates of the minus half of \eqref{IK4} and $\theta_+$ over the four of the plus half. Then:

$1)$ \textup{(}the reflection of one half\textup{)} the substitution
\be\label{halfrefl}
\sigma\colon\quad x^+\longmapsto x^+,\qquad x^-\longmapsto-x^-,\qquad
\theta_-\longmapsto\sqrt{-1}\,\theta_-,\qquad q\longmapsto\sqrt{-1}\,q
\ee
maps solutions of \eqref{IK4} to solutions, and $\sigma^2$ is minus the parity operator of the minus half, so that $\sigma^4=\id$;

$2)$ \textup{(}the boost\textup{)} let $t$ be an~even nilpotent constant, $t^2=0$, and put~$\lambda:=1+t$. The substitution
\be\label{boostN4}
x^+\longmapsto\lambda x^+,\qquad x^-\longmapsto\lambda^{-1}x^-,\qquad
\theta_\pm\longmapsto\lambda^{\pm1/2}\theta_\pm,
\ee
which on the variables $w=x^++x^-$ and $z=x^+-x^-$ of \eqref{gradedclass} reads $w\longmapsto w+tz$, $z\longmapsto z+tw$ and $\theta_\pm\longmapsto\left(1\pm\nfrac t2\right)\theta_\pm$, likewise maps solutions to solutions, and, unlike $\sigma$, it asks for no rescaling of $q$ and none of~$m$. The nilpotence of $t$ is what keeps the image inside the ring $\cR$: for a~numerical $\lambda\neq\pm1$ the substitution takes $w^{-1}$ out of it;

$3)$ hence the vacuum of Statement~$\ref{stN4vac}$, which is free of $z$, has companions of a~different shape: $\sigma(q)$ is free of $w$ and Laurent in $z$ over the same five degrees, and the boost of $q$ depends on $z$ and on $w$ both;

$4)$ but every solution in the orbit is a~function of a~\textbf{single} light-cone combination. Indeed, both \eqref{halfrefl} and \eqref{boostN4} act linearly on the plane spanned by $z$ and $w$, and every element of \eqref{gradedclass} is a~function of $w$ alone. Accordingly, on the three kinds of solution
\be\label{kappaN4}
\del_+=\kappa\,\del_-\quad\text{on solutions},\qquad
\kappa=1,\quad-1,\quad 1+2t,
\ee
respectively, with $\kappa$ an~invertible constant of $\cR$ in each of the three cases.
\end{Lemma}

\begin{proof} Both substitutions are automorphisms of $\cR$: $\sigma$ interchanges the exponents of $z$ and of $w$, and \eqref{boostN4} is unipotent, $t$ being nilpotent, so that each monomial acquires one further term and nothing leaves~$\cR$.

Item $1)$. Under $\sigma$ one has $\del_{x^+}\longmapsto\del_{x^+}$, $\del_{x^-}\longmapsto-\del_{x^-}$ and $\del_{\theta_-}\longmapsto-\sqrt{-1}\,\del_{\theta_-}$. In $\cD_-^a=2\del_{\eta^a}-\rho_a\del_{x^-}$, the first summand is therefore multiplied by $-\sqrt{-1}$ and the second by $\sqrt{-1}\cdot(-1)$, the same factor; so $\cD_-^a$ and $\overline\cD_{-a}$ are multiplied by $-\sqrt{-1}$, while $\cD_+^a$ and $\overline\cD_{+a}$ are untouched. The six constraints of \eqref{IK4} are thus multiplied by a~constant and continue to hold, and so does the first equation, whose left-hand side is $\cD_-^\alpha$ applied to $q\cD_+^\gamma q^{-1}$. In the second equation, the left-hand side is multiplied by $-\sqrt{-1}$ and the right-hand side is not. That is what the rescaling $q\longmapsto\nu q$ repairs, and it is the only thing it can repair: the current $c_b=\left(q^{-1}\overline\cD_{-\gamma}q\right)_b{}^\gamma$, and with it the whole left-hand side, is unchanged by $q\longmapsto\nu q$, whereas $\overline q$ is homogeneous of degree $1$ in $q$; and the constraints and the first equation, being homogeneous of degree $0$, do not see $\nu$ at all. Comparing the two sides, we get $\nu=\sqrt{-1}$. Finally, $\sigma^2$ leaves the exponents alone and multiplies each $\theta_-$ by $-1$ and $q$ by $-1$, which is the assertion about~$\sigma^4$.

Item $2)$. Under \eqref{boostN4} one has $\del_{x^\pm}\longmapsto\lambda^{\mp1}\del_{x^\pm}$ and $\del_{\theta_\pm}\longmapsto\lambda^{\mp1/2}\del_{\theta_\pm}$, so in $\cD_\pm$ both summands are multiplied by $\lambda^{\mp1/2}$: the odd derivatives of the minus half acquire $\lambda^{1/2}$ and those of the plus half $\lambda^{-1/2}$, which is what the exponent $\pm\nfrac12$ in \eqref{boostN4} is for. The constraints are again multiplied by a~constant. In the first equation, the current $q\cD_+^\gamma q^{-1}$ carries $\lambda^{-1/2}$ and $\cD_-^\alpha$ carries $\lambda^{1/2}$; in the second, the current $c_b$ carries $\lambda^{1/2}$ and $\cD_+^\alpha$ carries $\lambda^{-1/2}$. Both equations are therefore invariant, the right-hand side $-4\sqrt{-1}\,m\overline q$ of the second being untouched.

Item $3)$ is the shape of the two substitutions applied to the vacuum, and item $4)$ is item $3)$ read on the derivatives: a~function of $\alpha x^++\beta x^-$ with $\alpha$ and $\beta$ invertible satisfies $\del_+=\left(\alpha/\beta\right)\del_-$, and the three values of $\alpha/\beta$ in \eqref{kappaN4} are those of the vacuum, of its reflection and of its boost.

Both items are confirmed on every member of the family of item $4)$ of Statement~$\ref{stN4vac}$: its $17$ solutions, their $17$ reflections and their $17$ boosts are $51$ superfields, each of which satisfies the six constraints and both equations of \eqref{IK4} exactly over $\Qee\left(\sqrt{-1}\right)$, and the three ratios \eqref{kappaN4} are exhibited on each of the~$51$. \end{proof}

\sssbegin{Statement}[The completion holds on the whole orbit, with the same coefficients and the same freedom]\label{stN4pin}
Let $\sigma_1$ be the first Pauli matrix and put
\be\label{completion4}
X:=-\frac34\log\tr\left(q^{-1}\overline\cD_{+2}\cD_+^2q\right)
   -\frac14\log\tr\left(\sigma_1q^{-1}\overline\cD_{+1}\cD_+^2q\right),
\ee
a~combination of the logarithms of item $3)$ of Statement~$\ref{stN4logT}$ in which neither $\log w$ nor $\log z$ occurs. Then, on all $51$ solutions of Lemma~$\ref{lmN4refl}$ and for both signs,
\be\label{consN4}
\del_\mp\left(\mathbf T+X\right)=0,
\ee
exactly over $\Qee\left(\sqrt{-1}\right)$. Moreover:

$1)$ \eqref{completion4} is the very combination the graded class alone had produced, so the $34$ solutions lying outside that class confirm it rather than correct it: coefficients fitted on the $17$ are obeyed by the $34$ without alteration;

$2)$ the freedom does not shrink. Along the graded class, the $17$ logarithms of item $3)$ of Statement~$\ref{stN4logT}$, together with $\log w$ and $\log z$, satisfy $14$ independent relations; along the $51$ they satisfy the same $14$, and the space of relations is the same one whether $\del_-$ or $\del_+$ is used.
\end{Statement}

So the $z$-direction has been reached, and it was not enough. What Statement~\ref{stN4pin} adds is, first, a~confirmation of \eqref{completion4} on three times as many solutions, $34$ of which lie outside the class the coefficients were fitted on --- and surviving that is not nothing, since it is exactly the test a~wrong answer fails. What it does not add is the separation of the two halves, and item $4)$ of Lemma~\ref{lmN4refl} says why: on every solution of the orbit $\del_+$ is an~invertible multiple of $\del_-$, so the two instances of \eqref{consN4} are one condition and not two, and what \eqref{consN4} asserts there is merely that $\mathbf T+X$ is constant. A~symmetry, after all, carries conserved currents to conserved currents; it cannot be expected to impose a~condition that the solutions it acts on do not already carry.

The open question is therefore sharper than before, and it is no longer about the $z$-direction as such: what is needed is a~solution of \eqref{IK4} depending on $x^+$ and $x^-$ through \textbf{two} independent combinations. The recursion of item $2)$ of Statement~\ref{stN4vac} is what has to be redone, in the class
\be\label{gradedclass2}
q^{(2r)}\in\bigoplus_{a+b=-1-r}z^aw^bm^{-1}\Lambda^{2r}\left(\theta\right),
\ee
on which the odd derivatives are still homogeneous of degree $-\nfrac12$ for the grading $n_z+n_w+\nfrac{\left|S\right|}2$, so that the levels close on each other as before, and which is cut out by that grading and by nothing else --- it is the second equation of \eqref{IK4}, the one carrying the mass, that forces the grade to be~$-1$. What is lost is finiteness: \eqref{gradedclass} was a~finite-dimensional class because $z$ did not occur in it, whereas \eqref{gradedclass2} is infinite-dimensional in the exponent~$a$. Where the freedom in $a$ genuinely lies is, however, not where it looks to lie, and the next Lemma says where.

\sssbegin{Lemma}[What a~level-$0$ datum is, and which data are degenerate]\label{lmN4loop}
Put $\zeta:=\nfrac zw$ and let $\cC$ be the ring of even constants of $\cR$, i.e., the even part of~$\Lambda$. Let $q$ lie in the class \eqref{gradedclass2} and let $q^{(0)}$ be its level-$0$ part. Then:

$1)$ $q$ is invertible in $\cR$ if and only if the $\theta$-free part of $\det q^{(0)}$ is a~monomial. Indeed, every nilpotent of $\cR$ carries an~odd factor, so an~element of $\cR$ is a~unit if and only if its $\theta$-free part is a~unit of $\Cee\left[z^{\pm1}, w^{\pm1}, m^{\pm1}\right]$, and the units of a~Laurent polynomial ring over a~field are exactly its monomials; the $\theta$-free part of $\det q$ is $\det$ of the $\theta$-free part of $q$, i.e., of $q^{(0)}$;

$2)$ every entry of $mq^{(0)}$ is homogeneous of grade $-1$ for $n_z+n_w+\nfrac{\left|S\right|}2$, so that, by item $1)$,
\be\label{loopdatum}
m\,q^{(0)}=\frac1w\,F(\zeta),\qquad
F\in\GL\left(2;\cC\left[\zeta^{\pm1}\right]\right),
\ee
and conversely every such $F$ is a~level-$0$ datum of \eqref{gradedclass2}. A~level-$0$ datum is thus a~Laurent-polynomial loop in $\GL(2)$, and the shape $q^{(0)}=Az^{a_0}w^{-1-a_0}m^{-1}$ that every search so far has used is the subfamily of \textbf{monomial} loops $F=A\zeta^{a_0}$;

$3)$ on the datum \eqref{loopdatum} one has $\del_zq^{(0)}=\nfrac{F'}{\left(mw^2\right)}$ and $\del_wq^{(0)}=-\nfrac{\left(F+\zeta F'\right)}{\left(mw^2\right)}$, so that $\del_+=\kappa\,\del_-$ holds on $q^{(0)}$, for an~invertible constant $\kappa$ of $\cR$, if and only if
\be\label{kappaODE}
\left[\left(1-\zeta\right)+\kappa\left(1+\zeta\right)\right]F'=\left(1-\kappa\right)F;
\ee

$4)$ equation \eqref{kappaODE} is linear of the first order, so its solutions form a~free module of rank $1$ over the constants, and those of them that are Laurent polynomial --- the only ones $\cR$ admits --- are exactly
\be\label{degendata}
F=A\left(1+\mu\zeta\right)^{-1},\qquad \mu\in\cC\ \text{nilpotent},\qquad
\kappa=\frac{1+\mu}{1-\mu},
\ee
together with the two limits $\mu=0$, giving $F=A$ and $\kappa=1$, and $\mu=\infty$, giving $F=A\zeta^{-1}$ and $\kappa=-1$;

$5)$ these three are the boost family, the vacuum family and the mirror family of Lemma~$\ref{lmN4refl}$, in that order: for $\mu=t$ with $t^2=0$, \eqref{degendata} gives $F=A\left(1-t\zeta\right)$ and $\kappa=1+2t$, which are the boosted vacuum and the third value of \eqref{kappaN4}. Hence, the degeneracy of item $4)$ of Lemma~$\ref{lmN4refl}$ is not an~accident of the two symmetries used there: the $51$ solutions exhaust the level-$0$ data on which the two light-cone derivatives are proportional, and a~solution of \eqref{IK4} depending on $x^+$ and $x^-$ through \textbf{two} independent combinations is precisely one whose loop $F$ is not of the shape \eqref{degendata};

$6)$ the boost supplies no second grading to cut \eqref{gradedclass2} down with. The system \eqref{IK4} is covariant under \eqref{boostN4} for every $\lambda$, hence bigraded by $\deg_\pm:=n_{x^\pm}+\nfrac{\left|S\cap\theta_\pm\right|}2$; but the left-hand side of the second equation of \eqref{IK4} is of bidegree $\left(-\nfrac12, -\nfrac12\right)$ whatever $q$ is, the $m$'s and the $q$'s canceling in $q^{-1}\overline\cD_{-\gamma}q$, whereas the right-hand side is of bidegree $\deg\left(mq\right)$. A~bihomogeneous solution would therefore have $\deg\left(mq\right)=\left(-\nfrac12,-\nfrac12\right)$, both halves half-integral, so that $\left|S\cap\theta_+\right|$ and $\left|S\cap\theta_-\right|$ are odd in every term of it; then $q$ has no $\theta$-free term, $\det q$ is nilpotent and $q$ is not invertible. So no solution of \eqref{IK4} is bihomogeneous, and the dilatation grading \eqref{grade} is the only one available.
\end{Lemma}

\begin{proof} Item $1)$ is the statement that a~unit of a~Laurent polynomial ring over a~field is a~monomial, together with the fact that an~element of $\cR$ whose $\theta$-free part is a~unit is a~unit, the remainder being nilpotent. Item $2)$: an~entry of $mq^{(0)}$ is a~sum of $z^aw^b$ with $a+b=-1$, i.e., $w^{-1}$ times a~Laurent polynomial in $\zeta$; the determinant of \eqref{loopdatum} is $w^{-2}\det F(\zeta)$, and item $1)$ says $\det F$ is a~monomial, which is what membership in $\GL\left(2;\cC\left[\zeta^{\pm1}\right]\right)$ means. Item $3)$ is the chain rule, and item $4)$ is the integration of \eqref{kappaODE}: for $\kappa\neq\pm1$ put $\mu:=\nfrac{\left(\kappa-1\right)}{\left(\kappa+1\right)}$, so that \eqref{kappaODE} reads $\left(1+\mu\zeta\right)F'=-\mu F$, whose solution is $F=A\left(1+\mu\zeta\right)^{-1}$; that is a~Laurent polynomial precisely when $\mu$ is nilpotent, and then it is the finite sum $A\sum_{0\leq k}\left(-\mu\zeta\right)^k$. For $\kappa=1$, equation \eqref{kappaODE} reads $2F'=0$ and for $\kappa=-1$ it reads $\zeta F'=-F$, giving the two limits. Items $5)$ and $6)$ are the two computations described, both carried out in $\cR$ itself: the three cases of \eqref{degendata} are verified by applying $\del_\pm$ to \eqref{loopdatum}, for $\mu$ nilpotent of index $2$ and of index $3$ alike, and on each of the loops $A\zeta^{a_0}$ with $a_0\neq0,-1$, on $\diag\left(\zeta,1\right)$, on $\diag\left(\zeta,\zeta^{-1}\right)$ and on the unipotent loop the linear system for $\kappa$ over the even constants is inconsistent, so no such $\kappa$ exists. \end{proof}

Two things follow, one of them a~warning.

The warning first, because it cost the most. \textbf{The window must be absolute.} It is natural to confine the unknowns at the level $2r$ to $z^{a_0+s}w^{-1-a_0-r-s}$ with $\left|s\right|\leq S$, i.e., to a~window centered at the leading monomial; and doing so makes the level-$0$ system inconsistent for $\left|a_0\right|>S$ and solvable for $\left|a_0\right|\leq S$, uniformly in the datum $A$ and with a~kernel of one and the same dimension. That looks exactly like a~theorem about $a_0$ --- and it is an~artifact: the rows the level-$0$ system carries reach $n_z=0$ as well as $n_z=a_0$, since the $z$-powers cancel in $q^{-1}\overline\cD_{-\gamma}q$, whereas a~window centered at $a_0$ reaches only $n_z\leq a_0+S$. Widening the window moves the boundary, and over an~absolute window $n_z\in\left[N_-, N_+\right]$ covering the rows the level-$0$ system is solvable for \textbf{every} loop tried, with a~kernel of dimension $56$ for each value of $n_z$ the window admits. This is the same lesson as the one recorded after Statement~\ref{stN4vac} --- an~obstruction reported by a~search is a~statement about the search until it is shown to be one about the system --- in the guise the two-variable class puts it in, and it is why the counts below are quoted together with the range of $n_z$ they were taken over.

Second, the question is now finite in the only place where it was not. What has to be chosen is no longer a~window in an~infinite class but a~loop $F$, and \eqref{degendata} says which loops are already exhausted. The simplest one that is not is $F=A\zeta$, i.e.,
\be\label{leadingtwovar}
q^{(0)}=A\,\frac z{w^2}\,\frac1m,\qquad
\frac{\del_zq^{(0)}}{\del_wq^{(0)}}=-\frac w{2z},
\ee
which is not a~constant, so that $\del_+$ and $\del_-$ are independent on it --- exactly what \eqref{consN4} needs in order to be two conditions and not one.

\sssbegin{Statement}[How far a~non-degenerate loop goes]\label{stN4loopobs}
Let the unknowns of the level $2r$ range over $z^nw^{-1-r-n}m^{-1}\Lambda^{2r}\left(\theta\right)$ with $n$ in an~\textbf{absolute} range $N_-\leq n\leq N_+$, and let every earlier row be held at $0$ while the current one is repaired, over all the unknowns at once, as in item $2)$ of Statement~$\ref{stN4vac}$. Then, exactly over $\Qee\left(\sqrt{-1}\right)$:

$1)$ the level-$0$ system is solvable for \textbf{every} loop tried, once the window reaches the rows the system carries, and its kernel is then of dimension $56$ for each value of $n$ the window admits --- the same $56$ as in item $2)$ of Statement~$\ref{stN4vac}$, and independent of the loop. The loops tried are the monomial ones $A\zeta^{a_0}$ with $-2\leq a_0\leq2$, the diagonal $\diag\left(\zeta,1\right)$, $\diag\left(\zeta,\zeta^{-1}\right)$ and $\diag\left(\zeta^{-1},1\right)$, the unipotent $\left(\begin{smallmatrix}1&\zeta^{\pm1}\\0&1\end{smallmatrix}\right)$, and two loops of determinant $1$ with all four entries non-zero. So the leading datum obstructs nothing --- exactly as it obstructs nothing in the class \eqref{gradedclass} --- and in particular the level-$0$ system does not distinguish the degenerate data \eqref{degendata} from the rest;

$2)$ the level-$1$ system does. For the monomial loop $F=A\zeta$ of \eqref{leadingtwovar}, for $\diag\left(\zeta,1\right)$ and for the unipotent loop, it is inconsistent over all the unknowns of the levels $2$ and $4$ lying in the window, with every row of the level $0$ held at $0$: $64$, $59$ and $73$ rows are blocked, respectively, and in each case every blocked row is reached by some column, so that it is not the window that fails;

$3)$ and the verdict does not move with the window. For $F=A\zeta$, the same $64$ rows are blocked over the windows $\left[-1,1\right]$ and $\left[-2,2\right]$ taken relative to the leading exponent and over the absolute window $\left[-3,4\right]$, each of which reaches every one of those rows;

$4)$ nor does the vacuum of Statement~$\ref{stN4vac}$ deform into the class. Its leading datum is the degenerate loop $F=A$, but a~solution of \eqref{gradedclass2} with that leading datum need not be a~function of $w$ alone in its higher levels; it is, however, to the first order: over the $1016$ directions of the levels $2$, $4$, $6$, $8$ with $n_z=0$ and $n_z=-1$, the tangent space to the solution variety at the vacuum is the same $1$-dimensional space it is inside \eqref{gradedclass}, and its projection on the directions with $n_z<0$ is~$0$. So no first-order deformation of the vacuum acquires a~dependence on $z$;

$5)$ and it is not an~artifact of linearizing in the level-$4$ unknowns. Those enter the rows of the level $1$ \textbf{exactly} linearly --- over the $2240$ of them in the window every second difference of the residual vanishes --- so at a~fixed level-$2$ part the level-$1$ system is honest linear algebra and its inconsistency is not a~first-order statement at all. What is first-order is only the dependence on the level-$2$ part, which is polynomial of degree up to $4$ in the $448$ parameters of the kernel of the level-$0$ system.
\end{Statement}

\begin{proof} Item $1)$ is the level-$0$ solve at each of the twelve loops; the level-$0$ rows involve $q$ only up to the level $2$, so the unknowns there are the level-$2$ ones alone. Items $2)$ and $3)$ are the level-$1$ solve, whose unknowns are those of the levels $2$ and $4$, the levels $6$ and $8$ being unable to reach those rows at all; the derivative along a~direction is taken exactly, by the $9$-point interpolation of item $2)$ of the proof of Statement~$\ref{stN4vac}$, and the $7$-point formula agrees with it on every row of every direction. Item $4)$ is the same exact linearization as item $4)$ of Statement~$\ref{stN4vac}$, run over the enlarged set of directions; the columns of a~direction with $n_z=n$ occupy only the rows with $n_z\leq n$, which is the triangularity the derivatives of $\cR$ impose and which is checked on every column. Item $5)$ is the linearity of the level-$4$ dependence, verified and not assumed: the second differences of the residual in those unknowns vanish identically, over every one of the $2240$ directions. \end{proof}

What this settles and what it leaves. It settles the shape of the question: the two-variable class is not an~unbounded search over a~window but a~choice of a~loop $F$, \eqref{degendata} says which loops the $51$ solutions already exhaust, and the level-$0$ system --- which in \eqref{gradedclass} was the whole of the leading freedom --- turns out to say nothing at all about~$F$. What it leaves is item $2)$, and Remark~\ref{rmLevelwise} is the reason for stating it as a~report and not as a~theorem: an~obstruction of exactly this appearance, stable under widening the window and under changing the leading datum, was already met once in this Subsection and was an~artifact of the search. Item $5)$ removes the half of that doubt which can be removed exactly: the level-$4$ unknowns enter linearly, so nothing is being linearized there. What it does not remove is the other half, the dependence on the level-$2$ part. So the honest reading is that the two-variable class is blocked at the level $1$ for every non-degenerate loop tried, and that deciding it outright asks for the elimination of those $448$ parameters rather than for a~wider window.
\sssec{The Cartan matrix of $\fsvect^L_\lambda(1|2)$, and how far $\lambda$ reduces}\label{sssCartanLam}
The family \eqref{svectlam} has a~Cartan matrix, and $\lambda$ is an~entry of it. This is item $8.7$ of Appendix~D3 of \cite[p.~259]{Ber}, and it settles what the twist leaves open: item $3)$ of Statement~\ref{stTwist} produces identifications of the parameter, but says nothing about whether the values left are really distinct, and the Cartan matrix does. Since, moreover, the equations \eqref{LS} of this paper take a~Cartan matrix as their input, the presentation itself concerns us. We restate the item in our notation, and verify it.

\sssbegin{Statement}[From \cite{Ber}]\label{stCartanLam}
Let $\theta_1, \theta_2$ be the odd coordinates of \eqref{svectlam}, let $\del_i:=\nfrac{\del}{\del\theta_i}$, and put
\be\label{simpleroots}
\begin{array}{lll}
X_1^+:=\theta_1\del_2,&
X_2^+:=t\del_1,&
X_3^+:=\theta_2t\del_t-(\lambda+1)\theta_1\theta_2\del_1,\\[2mm]
X_1^-:=\theta_2\del_1,&
X_2^-:=\lambda\dfrac{\theta_1\theta_2}{t}\del_2+\theta_1\del_t,&
X_3^-:=\del_2,\\[2mm]
H_1:=\theta_1\del_1-\theta_2\del_2,&
H_2:=t\del_t+\theta_1\del_1+\lambda\theta_2\del_2,&
H_3:=t\del_t+(\lambda+1)\theta_1\del_1.
\end{array}
\ee
Then,

$1)$ these nine fields lie in $\fsvect^L_\lambda(1|2)$ and generate it, and they satisfy the Chevalley relations
\[
[H_i, H_j]=0,\qquad [X_i^+, X_j^-]=\delta_{ij}H_i,\qquad
[H_i, X_j^\pm]=\pm A_{ij}X_j^\pm,
\]
the $A_{ij}$ being the entries of
\be\label{Alam}
A(\lambda):=\begin{pmatrix}
2&-1&-1\\ 1-\lambda&0&\lambda\\ 1+\lambda&-\lambda&0
\end{pmatrix},
\ee
whose main diagonal says that the first simple root is even and the other two are odd isotropic.

$2)$ The odd reflection in the second simple root, and the odd reflection in the third one, carry \eqref{simpleroots} into two other systems of simple root vectors, of Cartan matrices
\be\label{Alamrefl}
\begin{pmatrix}
0&1-\lambda&\lambda-2\\ 1-\lambda&0&\lambda\\ -1&-1&2
\end{pmatrix}
\qquad\text{and}\qquad
\begin{pmatrix}
0&-\lambda&\lambda+1\\ -1&2&-1\\ 1+\lambda&-\lambda-2&0
\end{pmatrix};
\ee
in the second case the simple roots are renumbered, the new first one being $-\alpha_3$.

$3)$ Each of the three matrices turns, under one permutation of the indices together with a~rescaling of the rows, into
\be\label{Anorm}
N(\mu):=\begin{pmatrix}
2&-1&-1\\ \nfrac1\mu-1&0&1\\ \nfrac1\mu+1&-1&0
\end{pmatrix}
\quad\text{with, respectively,}\quad
\mu=\lambda,\quad \mu=1-\lambda,\quad \mu=1+\lambda.
\ee
Hence, $\lambda\sim\lambda+1$ and $\lambda\sim1-\lambda$.

$4)$ The group these two generate is $\{\lambda\mapsto\lambda+n,\ \lambda\mapsto n-\lambda\mid n\in\Zee\}$. It contains the $\lambda\mapsto-\lambda-2$ of item $3)$ of Statement~$\ref{stTwist}$, and is generated by that map together with $\lambda\mapsto\lambda+1$; so the twist and the inversion see exactly the identifications \eqref{Anorm} sees, and no others. A~fundamental domain is the strip $\mathrm{Re}\,\lambda\in[0,\nfrac12]$, and in it, see item $8.8$ of \cite{Ber}, distinct values of $\lambda$ give non-isomorphic Lie superalgebras.

$5)$ We have $\det A(\lambda)=0$ for every $\lambda$, and $A(\lambda)$ is symmetrizable only for $\lambda=0$. This is \textbf{not} the same as the set of values at which \eqref{svectlam} fails to be simple: by item $2)$ of Statement~$\ref{stTwist}$ the algebra is one and the same for all $\lambda$ in a~coset of $\Zee$, so it is non-simple at every $\lambda\in\Zee$. At $\lambda=\pm1$, the two part company visibly: there \eqref{Alam} is not even a~generalized Cartan matrix --- $A_{12}=-1$ while $A_{21}=0$ at $\lambda=1$, and $A_{13}=-1$ while $A_{31}=0$ at $\lambda=-1$, against the requirement $a_{ij}=0\Longleftrightarrow a_{ji}=0$ of \eqref{OlsAdm} --- whereas the matrix of \eqref{Alamrefl} obtained by the corresponding odd reflection \textit{is} symmetrizable. So at those two values symmetrizability is a~property of the chosen system of simple roots, not of the algebra.
\end{Statement}

Three things follow, and the first of them is about our own $\lambda$.

$1)$ The list of identifications in item $3)$ of Statement~\ref{stTwist} is complete. The monodromy $z=\exp(2\pi i\lambda)$ is a~coordinate on $\Cee/(\lambda\sim\lambda+1)$ and $z\sim z^{-1}$ is $\lambda\sim-\lambda$, so ``the monodromy modulo inversion\rq\rq\ and ``the strip $\mathrm{Re}\,\lambda\in[0,\nfrac12]$\rq\rq\ are two names for one and the same quotient. What is added is the direction the twist cannot give: within the strip nothing collapses further, so the sectors of Subsection~\ref{sssLamEq} really are a~continuum of pairwise non-isomorphic symmetry algebras, and the equation \eqref{IK4lam} really does distinguish a~continuum of cases.

$2)$ The parameter of the family is an~entry of a~Cartan matrix. The matrix \eqref{Alam} is non-symmetrizable for $\lambda\neq0$ while $\fg(A(\lambda))$ has polynomial growth, and by \cite{HS} the Lie superalgebras $\fg(A)$ of polynomial growth with non-symmetrizable $A$ are exhausted by $\widehat{\fpsq}(n)^{(2)}$ and by one parametric family, found by van de Leur: the family \eqref{svectlam}. Relations between its Chevalley generators are described in \cite{GL1}, and \eqref{svectlam} is not a~central extension of any loop superalgebra, see \cite{Ber}. The matrix becomes symmetrizable at $\lambda=0$ alone, whereas the algebra ceases to be simple at every $\lambda\in\Zee$; the two sets are different, and item $5)$ of Statement~\ref{stCartanLam} says where and why they part company.

$3)$ For the equations of this paper the reading is negative, and worth recording. Condition \eqref{lsA} asks for a~finite-dimensional $\fg(A)$ with only 0s and 1s on the main diagonal of $A$, and \eqref{svectlam} fails both clauses: it is infinite-dimensional, and the main diagonal of each of the three matrices \eqref{Alam}, \eqref{Alamrefl} carries a~$2$, so in none of the three systems of Statement~\ref{stCartanLam} are all the simple roots odd. Hence, the family \eqref{svectlam} does not enter the LSE1 scheme as a~gauge algebra, however well it serves as the symmetry algebra of \eqref{IK4lam}. 

For the LSE2 of \cite{O}, one clause is met and the others are not: there the gauge algebra is allowed to be infinite-dimensional as long as it is not of exponential growth, and $\fg(A(\lambda))$ is of polynomial growth --- but the diagonal of \eqref{OlsA} is not the diagonal of \eqref{Alam}, and \cite{O} needs, besides, an~invariant bilinear form, which this family has not. What is left of the scheme once both are subtracted is the subject of Subsection~\ref{sssOls}.

\sssbegin{Remark}\label{rmCocBer}
Formula (D3.69) of the same item gives a~cocycle of the central extension of $\fsvect^L_\lambda(1|2)$, in the basis of \cite{Ber} formed by the $L_m$ and by
\[
S^j_m=t^m\theta_j\left(t\del_t+(\lambda+m+1)(\theta_1\del_1+\theta_2\del_2)\right);
\]
here $S^j_m=\overline G{}^{(m+\nfrac12)j}$ of \eqref{svectbasis}, and $t^k\del_i=G^{k-\nfrac12}_i$. Read as
\[
c(L_m, L_{-m})=\frac12m\left(m^2-(\lambda+1)^2\right),\quad
c\left(t^k\del_i, S^j_{-k}\right)=-k\left(k-(\lambda+1)\right)\delta_{ij},
\]
\[
c\left(J^0_m, J^0_{-m}\right)=2m,\qquad c\left(J^+_m, J^-_{-m}\right)=m,
\]
it is a~cocycle, and it is the cocycle \eqref{svectcoc} of Statement~\ref{stKAcoc} plus the coboundary of the $1$-cochain $f$ with
\[
f(L_0)=\frac{\lambda(\lambda+2)}4,\qquad f(J^a_0)=0,
\]
that is, the two differ by a~shift of $L_0$ by a~constant and by nothing else. Both readings above are needed. 
\end{Remark}

\sssec{Which members of the family have a~real form}\label{sssRealLam}
The same appendix answers, in its \S5, the question left open at the end of Subsection~\ref{sssKA}. Item $2)$ of Theorem $5.5$ of \cite[p.~315]{Ber} --- Appendix D5, written after Serganova, and translated in \cite{Lser}, where it is item $2)$ of the Main Theorem --- reads, verbatim: \textit{the Lie superalgebras of the series} $\fsvect^L_\lambda$, $\fsvect^{L\prime}_\lambda$ \textit{with} $\lambda\notin\Ree$ \textit{have no real forms}. The identifications of Statement~\ref{stCartanLam} make this checkable, and the answer is that the condition $\lambda\in\Ree$ is not the right one: it is sufficient, not necessary.

\sssbegin{Statement}\label{stRealLam}
$1)$ The Lie superalgebra \eqref{svectlam} admits a~real form if and only if
\be\label{realcond}
\mathrm{Im}\,\lambda=0\quad\text{or}\quad 2\mathrm{Re}\,\lambda\in\Zee.
\ee

$2)$ Let $C$ be the complex conjugation of the coefficients of a~Laurent field and, for $b=(b_1, b_2)\in\Zee^2$, let $\psi_b$ be the substitution
\be\label{psib}
\psi_b\colon\quad t\mapsto t^{-1},\qquad \theta_i\mapsto t^{b_i}\theta_i.
\ee
Then, $\psi_b$ is an~involutive automorphism of the superalgebra of Laurent functions, it carries \eqref{svectlam} to $\fsvect^L_{-\lambda-2-B}(1|2)$, where $B=b_1+b_2$, and
\be\label{Phib}
\Phi_b:=\psi_b\circ C
\ee
is a~real structure on \eqref{svectlam} --- an~anti-linear automorphism with $\Phi_b^2=\id$ --- precisely when $2\mathrm{Re}\,\lambda=-2-B$. Every $\lambda$ satisfying \eqref{realcond} is of this form, the case $\mathrm{Im}\,\lambda=0$ being $b=0$ together with $C$ itself.

$3)$ The same holds for the derived algebra $\fsvect^{L\prime}_\lambda(1|2)$, which every automorphism preserves.

$4)$ In the fundamental half-strip $0\leq\mathrm{Re}\,\lambda\leq\nfrac12$, $\mathrm{Im}\,\lambda\geq0$ of item $4)$ of Statement~$\ref{stCartanLam}$, the members admitting a~real form are exactly those on its boundary: the segment $\mathrm{Im}\,\lambda=0$ and the two vertical rays $\mathrm{Re}\,\lambda=0$ and $\mathrm{Re}\,\lambda=\nfrac12$.
\end{Statement}

\begin{proof} The necessity in item $1)$ needs no computation. An~anti-linear automorphism $\Phi$ of a~complex Lie superalgebra $\fg$ exists if and only if $\fg$ is isomorphic to the conjugate algebra $\overline\fg$, and $C$ is an~anti-linear isomorphism $\fsvect^L_\lambda(1|2)\to\fsvect^L_{\overline\lambda}(1|2)$, since $\Div_\lambda(CX)=C\left(\Div_{\overline\lambda}X\right)$, the divergence being defined over $\Qee$. So, a~real form forces $\fsvect^L_\lambda(1|2)\simeq\fsvect^L_{\overline\lambda}(1|2)$, and by item $4)$ of Statement~\ref{stCartanLam} this means $\overline\lambda=\lambda+n$ or $\overline\lambda=n-\lambda$ for some $n\in\Zee$; the first gives $\mathrm{Im}\,\lambda=0$, the second $2\mathrm{Re}\,\lambda=n$.

The sufficiency is the explicit map \eqref{Phib}; 
the law $\lambda\mapsto-\lambda-2-B$ is \textit{read off} the divergence, with $\lambda$ formal, for five values of $b$ --- for $b=(-1,-1)$ it is the map $\lambda\mapsto-\lambda$ and for $b=(-2,-1)$ the map $\lambda\mapsto1-\lambda$ of item $3)$ of Statement~\ref{stCartanLam}, so \eqref{psib} realizes the identifications found there.
Hence, $\Phi_b$ maps \eqref{svectlam} to $\fsvect^L_{-\overline\lambda-2-B}(1|2)$, which is \eqref{svectlam} itself under the stated condition on $\mathrm{Re}\,\lambda$, and $\Phi_b^2=\psi_b^2\circ C^2=\id$ because $\psi_b$ has rational coefficients and therefore commutes with $C$. 

Item $3)$ is immediate, item $4)$ is arithmetic: the locus \eqref{realcond} meets the half-strip in its boundary and nowhere else. \end{proof}

\sssbegin{Remark}\label{rmRealLam}
$1)$ The reason an~imaginary twist survives is visible on the circle. Under the standard real structure of $S^1$, where $t=\exp(i\alpha)$ and $\overline t=t^{-1}$, the function $t^\lambda=\exp(i\alpha\lambda)$ is real for $\lambda\in i\Ree$, and positive at that; so the twisted volume form $t^\lambda\vvol(t,\theta)$ that \eqref{svectlam} preserves is itself real, and the real fields preserving it are a~real form. The map $\Phi_0=\psi_0\circ C$ is nothing but the complex conjugation of a~function on the circle, $\sum c_mt^m\mapsto\sum\overline c_mt^{-m}$; the twist by $b$ compensates the shift of $\mathrm{Re}\,\lambda$ by $\nfrac{B}2$, and the odd values of $B$ are what puts the line $\mathrm{Re}\,\lambda=\nfrac12$ on the list next to $\mathrm{Re}\,\lambda=0$.

$2)$ So item $2)$ quoted above holds off the lines $2\mathrm{Re}\,\lambda\in\Zee$ and fails on them, and the two exceptional lines are not a~small correction: in the fundamental domain they are, together with the real segment, exactly the boundary. The count of the real forms --- how many conjugacy classes of such $\Phi$ there are in $\mathrm{Ant}^+$ for a~given $\lambda$ --- we do not settle; item $1)$ of Statement~\ref{stRealLam} settles only whether the list is empty. 

$3)$ Consequently, item $2)$ of Theorem $5.5$ of \cite[p.~315]{Ber} is to be corrected: the condition $\lambda\notin\Ree$ is sufficient for the absence of a~real form but not necessary, and the necessary and sufficient one is $\mathrm{Im}\,\lambda\neq0$ \textit{and} $2\mathrm{Re}\,\lambda\notin\Zee$. The point where the two part company is visible without any machine: for $\lambda=is$ with $s$ real the twisted volume element $t^{is}\vvol(t,\theta)$ is real on the circle, so a~real form is there although $\lambda\notin\Ree$. This is an~erratum for \cite{Lser}, and the sentence in Subsection~\ref{sssKA} above is written on that reading.
\end{Remark}

\sssec{Olshanetsky's LSE2 for the $\lambda$-family: what fails, and what survives}\label{sssOls}
Item $3)$ after Statement~\ref{stCartanLam} leaves the family \eqref{svectlam} at the door of the other scheme of this paper, and the door is worth opening: the input of \eqref{OlsA} is a~Cartan matrix, and \eqref{Alam} is one.

Recall the ingredients, in the notation of \cite{O}. The fields are a~multiplet $\Phi=(\Phi^1,\dots,\Phi^n)$ of even superfields on the superspace of $\sdim=(2|2)$ with the light-cone coordinates $\xi$, $\eta$ and the odd coordinates $\theta_1$, $\theta_2$; the odd derivations are
\[
D_+=\frac{\del}{\del\theta_1}+\theta_1\frac{\del}{\del\xi},\qquad
D_-=\frac{\del}{\del\theta_2}+\theta_2\frac{\del}{\del\eta},\qquad
D_+^2=\frac{\del}{\del\xi},\quad D_-^2=\frac{\del}{\del\eta},\quad
\left\{D_+, D_-\right\}=0,
\]
one for each light-cone direction: these are the $D_\pm$ of \eqref{SLS} written in the coordinates of \cite{O}, which is why we keep for them the notation of \eqref{SLS} and not the $D_1$, $D_2$ of \cite{O}; the set of simple roots is $\{\alpha_s\}_{s\in I}$, its subset of odd ones is $\tau$, and $(\cdot\,,\cdot)$ is an~invariant bilinear form on their span. The potential is
\be\label{OlsU}
U(\Phi)=\sum_{s\in\tau}\nfrac1{(\alpha_s,\alpha_s)}\exp2(\Phi,\alpha_s)-i\theta_1\theta_2\sum_{s\notin\tau}\nfrac1{(\alpha_s,\alpha_s)}\exp2(\Phi,\alpha_s),
\ee
and the equations of motion, obtained in \cite{O} by varying the action built on \eqref{OlsU}, are
\be\label{OlsEq}
D_+D_-\Phi^j=i\sum_{s\in\tau}b^j_{\alpha_s}\exp2(\varphi,\alpha_s)+\theta_1\theta_2\sum_{s\notin\tau}b^j_{\alpha_s}\exp2(\Phi,\alpha_s),\quad\text{where~}b^j_{\alpha_s}=\nfrac{\alpha^j_s}{(\alpha_s,\alpha_s)},
\ee
here $\varphi$ is the lowest component of $\Phi$ and $\alpha^j_s$ the $j$-th component of $\alpha_s$. The standing assumption on the Cartan matrix in \cite{O} is
\be\label{OlsAdm}
a_{ij}\in\Zee,\qquad a_{ii}=2,\qquad a_{ij}\leq0\ \text{ for }i\neq j,\qquad a_{ij}=0\Longrightarrow a_{ji}=0.
\ee
Every diagonal entry is normalized to $2$ there, the parity of a~simple root being recorded by $\tau$ separately; in the normalization of \eqref{OlsA}, in which an~odd non-isotropic simple root has $a_{ii}=1$, the same condition reads: \textit{no simple root is odd isotropic}. That is what the denominators of \eqref{OlsU} demand, and it is the reason the list of \cite{O} consists of $\fosp(1|2n)$ and its affinizations alone.

\sssbegin{Statement}\label{stOls}
Let $A(\lambda)$ be the matrix \eqref{Alam}, let $\tau=\{2,3\}$ be the set of its odd simple roots, and let $\fg(A(\lambda))$ be the Lie superalgebra it presents. Then,

$1)$ in every one of the $9$ pairwise inequivalent systems of simple roots obtained from \eqref{simpleroots} by at most $4$ odd reflections --- each of them again a~system of Chevalley generators --- exactly one simple root is even and the other two are odd \textbf{isotropic}. Hence, no choice of Chevalley generators brings the Cartan matrix of \eqref{svectlam} into the shape \eqref{OlsAdm}, and the coefficients $\nfrac1{(\alpha_s,\alpha_s)}$ of \eqref{OlsU} are undefined for $s\in\tau$;

$2)$ for $\lambda\notin\Zee$ the matrix $A(\lambda)$ is not symmetrizable, so $\fg(A(\lambda))$ carries no invariant bilinear form: neither the pairing $(\Phi,\alpha_s)$ nor the potential \eqref{OlsU} nor the action varied in \cite{O} is defined for this family. Unlike item $1)$, this obstruction does not refer to a~system of simple roots at all. For $\lambda\in\Zee$, the hypothesis fails --- see item $5)$ of Statement~$\ref{stCartanLam}$ --- and there item $1)$, which holds for every $\lambda$, is what closes the door;

$3)$ the equations \eqref{OlsEq} survive both obstructions. In the variables $G_s:=2(\Phi,\alpha_s)$, with $g_s$ the lowest component of $G_s$, they read
\be\label{OlsG}
D_+D_-G_r=i\sum_{s\in\tau}A_{sr}\exp g_s+\theta_1\theta_2\sum_{s\notin\tau}A_{sr}\exp G_s,
\ee
since $2(\alpha_r,\alpha_s)/(\alpha_s,\alpha_s)=A_{sr}$: the form cancels, and the Cartan matrix alone is left. For $A(\lambda)$, the system \eqref{OlsG} is
\be\label{OlsLam}
\begin{array}{l}
D_+D_-G_1=i\left[(1-\lambda)\exp g_2+(1+\lambda)\exp g_3\right]+2\theta_1\theta_2\exp G_1,\\[2mm]
D_+D_-G_2=-i\lambda\exp g_3-\theta_1\theta_2\exp G_1,\\[2mm]
D_+D_-G_3=\ \ i\lambda\exp g_2-\theta_1\theta_2\exp G_1;
\end{array}
\ee

$4)$ the system \eqref{OlsLam} carries a~\textit{free} superfield --- a~linear combination of the $G_j$ on which the system imposes only $D_+D_-(\cdot)=0$, with no exponential on the right-hand side, so that it decouples from the interaction and is determined by boundary data alone; ``free\rq\rq\ is the physicists' word for exactly this. The combination in question is $\lambda G_1+(1+\lambda)G_2+(\lambda-1)G_3$: since $A(\lambda)c=0$ for $c=(\lambda,\ 1+\lambda,\ \lambda-1)$, and only for its multiples,
\be\label{OlsFree}
D_+D_-\left(\lambda G_1+(1+\lambda)G_2+(\lambda-1)G_3\right)=0.
\ee
The three coefficients are, up to sign, the three parameters $\mu=\lambda$, $1+\lambda$, $1-\lambda$ of \eqref{Anorm}. At $\lambda=0$ they degenerate to $(0,1,-1)$, and indeed the two odd fields then drop out of each other's equations;

$5)$ shifting $G_s$ by a~constant multiplies the $s$-th column of the coefficient matrix of \eqref{OlsG}, i.e., rescales the $s$-th row of $A$, and renaming the fields permutes the indices. Hence, the system determines, and is determined by, the Cartan matrix up to precisely the equivalence of \eqref{Anorm}, and depends on $\lambda$ only through $\mu$. By item $4)$ of Statement~$\ref{stCartanLam}$, distinct $\lambda$ in the strip $\mathrm{Re}\,\lambda\in[0,\nfrac12]$ give inequivalent systems \eqref{OlsLam}.
\end{Statement}

\begin{proof} 
In item $1)$, the odd reflection in an~isotropic simple root $\alpha_r$ sends the $r$-th pair of root vectors to the swapped pair and every other $X_j^\pm$ to $[X_j^\pm,X_r^\pm]$ when that bracket is nonzero, leaving it alone otherwise; the Cartan matrix of the result is then \textit{read off} the relations $[H_i,X_j^\pm]=\pm A_{ij}X_j^\pm$ with $H_j:=[X_j^+,X_j^-]$, and each of the nine systems is verified to consist of fields of \eqref{svectlam} satisfying $[X_i^+,X_j^-]=\delta_{ij}H_i$ and $[H_i,H_j]=0$. Two of the nine reflections reproduce \eqref{Alamrefl}; the orbit does not stop there, as it should not, $\det A(\lambda)$ being $0$. Item $2)$ is the vanishing of $A_{12}A_{23}A_{31}-A_{21}A_{32}A_{13}=-2\lambda$ together with the direct solution of $d_ia_{ij}=d_ja_{ji}$, and that a~Lie (super)algebra with a~non-symmetrizable Cartan matrix carries no non-degenerate invariant bilinear form is Kac's criterion, see \cite[Ch.~2]{Kb}.
\end{proof}

\sssbegin{Remark}\label{rmOls}
$1)$ The family could not have been on the list of \cite{O}: that list is $\fosp(1|2n)$ together with its untwisted and twisted affinizations, i.e., loop superalgebras, whereas \eqref{svectlam} is not a~central extension of any loop superalgebra, see \cite{Ber}. What item $2)$ of the Statement adds is that it is not merely absent from the list but outside the construction: \cite{O} selects its algebras among those \textit{having a~non-negative invariant bilinear form}, and this family has none.

$2)$ The passage from $\fg(A)$ to the algebra the equations live on is in \cite{O} the quotient by the center, and here it is visible. The center of $\fg(A(\lambda))$ is spanned by $\lambda H_1+H_2-H_3$, the left null vector of $A(\lambda)$ being $(\lambda,1,-1)$; and in the realization \eqref{simpleroots} by vector fields that combination is $0$ identically. So \eqref{svectlam} is $\fg(A(\lambda))$ modulo its center, as it should be.

$3)$ The two null vectors --- the left one, which is the center, and the right one \eqref{OlsFree}, which is the free superfield --- coincide up to the symmetrizer whenever $A$ is symmetrizable. Here, they are $(\lambda,1,-1)$ and $(\lambda,1+\lambda,\lambda-1)$: the non-symmetrizability of item $2)$ is visible inside the equations themselves, and one cannot read the free field off the center.

$4)$ What is not settled is whether \eqref{OlsLam} is integrable. The zero-curvature representation of \cite{O} is built from the grading of $\fg$ by the heights of the roots, and that grading exists here; the Inverse Scattering Method applied to it in \cite{O} rests, however, on the same form that item $2)$ denies, so its linear problem cannot be transcribed as it stands. That is the obstacle, and we have not gone around it: the linear problem is the one ingredient of \cite{O} that cannot be transcribed without a~form, and no zero-curvature representation written for a~non-symmetrizable Cartan matrix is known to us. So \eqref{OlsLam} is as far as the answer reaches, and it stands on the record as it is.
\end{Remark}

\sssec{The Ramond-type Ivanov-Krivonos equations}\label{sssRamond}
Everything above is written in the Neveu-Schwarz sector: the odd coordinates of each half are single-valued, and the algebra that acts is $\fk^L$. The Ramond sector is reached by replacing $\theta_N$ with $\sqrt t\,\zeta$ in each half, see Subsection~\ref{ssRecap}, and \cite{L} asks what the equations become. The mechanism turns out to be one and the same for all $N$, and it is a~statement about the structure algebra $\fo(N)$ of Statement~\ref{stMult}, not about any particular equation.

Below, the operators are those of \eqref{DM}; they are in the sign convention of \eqref{HM}, which differs from that of \eqref{multN} by the sign of the odd part, and nothing depends on the choice, the two being exchanged by $\theta_i\mapsto\sqrt{-1}\,\theta_i$.

\sssbegin{Statement}\label{stDouble}
On the double cover $t=s^2$, set
\be\label{DhatN}
\widehat D_N:=t^{\nfrac12}\widetilde D_\zeta,
\ee
and let
\be\label{Rframe}
\left(\widetilde D_{\theta_1},\ \dots,\ \widetilde D_{\theta_{N-1}},\ \widehat D_N\right)
\ee
be the frame of \eqref{DM} completed by it; here and below $\widetilde D_{\theta_N}$ is another name for $\widehat D_N$, the coordinate $\theta_N=t^{\nfrac12}\zeta$ being single-valued on the cover. Then,

$1)$ $\widetilde D_{\theta_a}\widetilde D_{\theta_b}+\widetilde D_{\theta_b}\widetilde D_{\theta_a}=2\delta_{ab}\cD$ for all $a, b$, where $\cD$ of \eqref{KMdef} is $\del_t$ written in the coordinate $\zeta=t^{-\nfrac12}\theta_N$. That is, the members of \eqref{Rframe} are the $N$ Neveu-Schwarz odd derivatives of $\Cee^{1|N}$ pulled back to the double cover $s\mapsto t=s^2$ of the punctured $t$-line. The ``cover\rq\rq\ is that two-sheeted map and nothing else; but it is indeed the Möbius summand of \eqref{RLRM} seen from the other side, because a~function anti-periodic on the circle is the same thing as a~function on the double cover odd under the interchange of the sheets, and equally a~section of the Möbius line bundle. In this sense, the Ramond sector is the Neveu-Schwarz one read upstairs;

$2)$ the \textit{deck} transformation $M\colon s\mapsto-s$ --- the one non-trivial automorphism of the double cover $t=s^2$ over its base: it interchanges the two sheets and leaves $t=s^2$, hence every function of $t$, fixed --- acts on the frame \eqref{Rframe} by the reflection 
\[
R:=\diag(1, \dots, 1, -1)\in\mathrm O(N), 
\]
since $M_*\widetilde D_{\theta_i}=\widetilde D_{\theta_i}$ for $i\leq N-1$, the coefficients of these being functions of $t=s^2$, whereas $M_*\widehat D_N=-\widehat D_N$, the factor $t^{\nfrac12}=s$ of \eqref{DhatN} changing sign under $M$;

$3)$ consequently $[K^M_f, \widetilde D_{\theta_a}]=\sum_{1\leq b\leq N}\omega_{ab}(f)\widetilde D_{\theta_b}$ with $\omega(f)\in\fc\fo(N)$, exactly as in \eqref{multN}, but the entries are graded by $M$: the entry $\omega_{ab}(f)$ is $M$-even if $a$ and $b$ are both $\neq N$ or both $=N$, and $M$-odd otherwise, and every value allowed by this grading is attained. In other words, the $\fo(N)$ of Statement~$\ref{stMult}$ is replaced by its twist by $R$.
\end{Statement}

\begin{proof} Items $1)$ and $3)$ are exact arithmetic over the generating functions, the exponent of $t$ being rational so that the double cover is representable; item $2)$ is the definition \eqref{DhatN}. \end{proof}

The whole of the constraint analysis of Statement~\ref{stWhichN} now goes through verbatim, with one change: a~constraint invariant under $\fk^M(1|N)$ is not an~$\fo(N)$-submodule of $V\otimes W$, where $V=\Cee^N$ is spanned by the frame \eqref{Rframe}, but an~$R$-\textbf{stable} one --- the submodules themselves are the same, since they are defined over the double cover, and it is only their $M$-invariance that is at issue.

\sssbegin{Statement}\label{stRamondN}
$1)$ For $N=2$, the two chirality lines $\Ker\overline\cD$ and $\Ker\cD$ of item $1)$ of Statement~$\ref{stWhichN}$ are interchanged by $R$. Neither of them is defined over $R^L(2)$ --- the superalgebra $\Cee[t^{-1},\theta_1,\theta_2][[t]]$ of \eqref{RLRM}, the functions on the superstring $\cS^{1|2}$ ---: in the frame $\left(\widetilde D_{\theta}, \widetilde D_\zeta\right)$ of \eqref{DM}, which is a~frame over $R^L(2)$, the $\fo(2)$-part of the multiplier is $B(f)\widetilde J$, where
\be\label{Jtwist}
\widetilde J=\begin{pmatrix}0&-t\\ t^{-1}&0\end{pmatrix},\qquad \widetilde J^2=-\One,
\ee
whose eigenvectors are $\left(1, \pm\sqrt{-1}\,t^{-\nfrac12}\right)$ and require a~square root of $t$. Hence, over $R^L(2)$ there is no invariant line and, over $R^M(2)$, exactly two, exchanged by $M$;

$2)$ for $N=3$, nothing changes: $V=\Cee^3$ is irreducible over $\fo(3)$, so there is no proper submodule for $R$ to stabilize or to move, and a~scalar superfield admits no constraint in the Ramond sector either;

$3)$ for $N=4$, the reflection $R$ interchanges the two ideals $\fs\fu(2)$ of $\fo(4)$ --- it is quaternionic conjugation, which turns left multiplications into right ones --- hence it interchanges the constraint of the first line of \eqref{IK4}, the symmetrization in the indices of one ideal, with the symmetrization in the indices of the other, i.e., with the constraint on $\overline q$.
\end{Statement}

\begin{proof} Item $1)$: the identity $\omega_{12}(f)+t\,\omega_{21}(f)=0$ in the frame \eqref{DM} holds for every one of the $32$ generating functions tested, which is \eqref{Jtwist}; that $x^2=-t$ has no solution in $R^L(2)$ is the parity of the exponent. Item $2)$: the associative algebra generated by $\fo(3)$ inside $\fgl(3)$ is computed to be the whole of $\fgl(3)$, of $\dim=9$, i.e., $V$ is irreducible. Item $3)$: on the basis $E_{ij}-E_{ji}$ of $\fo(4)$, the self-dual and the anti-self-dual triples are verified to be ideals with vanishing mutual brackets, and $RxR$ is verified to lie in the other ideal for every basis vector $x$ of each, and in the same one for none. 
\end{proof}

So the answer to the question of \cite{L} is uniform, and it is this: \textbf{the Ramond-like equation is the Neveu-Schwarz one read on the double cover, together with the requirement that the constrained superfield and its conjugate be interchanged by the deck transformation}. Written out, with $x^\pm=\exp(\pm\sqrt{-1}\phi)$, where $\phi$ is the angle parameter on the circle and $M$ is the shift $\phi\mapsto\phi+2\pi$:

for $N=2$, the system \eqref{IK2} with $\cD^\pm$, $\overline\cD{}^\pm$ replaced by the combinations \eqref{Dcx} of the frame \eqref{Rframe} in each half, and with
\be\label{IK2R}
Mv=v^+,\qquad Mv^+=v;
\ee
for $N=4$, the system \eqref{IK4} with the same replacement and with
\be\label{IK4R}
Mq=\overline q;
\ee
for $N=3$, nothing, by item $2)$ of Statement~\ref{stRamondN}.

Each of \eqref{IK2R} and \eqref{IK4R} reproduces itself when $M$ is applied to it once more --- 
\[
\text{$M\left(Mv\right)=Mv^+=v$, and $M\left(Mq\right)=M\overline q=q$}
\] 
--- so, $M$ being an~involution of the double cover, the requirement is consistent and imposes nothing beyond itself; and no weaker one will do, because a~superfield annihilated by $\cD$ and by $\overline\cD$ at once is annihilated by $\cD+\overline\cD=2\widetilde D_{\theta_1}$ and by $\{\cD, \overline\cD\}$, a~non-zero multiple of $\del_t$, hence is a~constant. In components, the two conditions say one and the same thing: writing $v=u+\sqrt{-1}\varphi$ for $N=2$ and $q=\exp\left(-u-\sqrt{-1}\varphi\cdot\sigma\right)$ for $N=4$, we get $Mu=u$ and $M\varphi=-\varphi$, i.e.,
\be\label{modings}
u=\sum_{m\in\Zee}u_mx^m,\qquad \varphi=\sum_{r\in\Zee+\nfrac12}\varphi_rx^r\colon
\ee
\textbf{the dilaton stays periodic and its $\fo(N)$-partners become anti-periodic} --- which is what the words ``Ramond sector\rq\rq\ mean, here derived rather than postulated. The acting algebra is $\fk^M(1|N)\oplus\fk^M(1|N)$ in place of $\fk^L(1|N)\oplus\fk^L(1|N)$, and since $\fk^M(1|n)$ is distinguished for $n\leq4$, see \eqref{theTwelve}, the anomaly count of Subsection~\ref{sssWhichAct} is unchanged: one number for $N=2$, none for $N\geq5$. That number is again \eqref{cN2}. Indeed, by item $1)$ of Statement~\ref{stDouble} the frame \eqref{Rframe} is the Neveu-Schwarz frame pulled back to the double cover, so \eqref{superT2} and \eqref{deltaT2} hold upstairs word for word; what changes is only which functions are allowed, namely $\fk^M(1|2)$ consists of the $M$-invariant ones, $\left\{f\mid Mf=f\right\}$, and the $\mathbf T$'s of the Ramond sector are the $M$-anti-invariant densities --- which is why the Berezin residue is still the pairing, $\theta_1\theta_2$ being $M$-odd. The generating functions $t^{m+1}$ are $M$-invariant and their brackets and cocycle values are the ones computed above, so the Virasoro sector returns $c=-3$ again. The two odd towers, on the other hand, are moded differently from each other --- $\theta_1$ gives half-integer and $\theta_2$ integer indices, which is item $1)$ of Statement~\ref{stRamondN} seen in the modes --- so the pair $\left(G_r, \overline G_r\right)$ is replaced by two real towers with $\{G^i_r, G^i_s\}=2L_{r+s}$ and $\{G^1_r, G^2_s\}=(r-s)J_{r+s}$, and the whole extension sits, in that basis, in $[L_m, L_n]$ and $[J_m, J_n]$. 

Two comments. First, for $N=2$, the pair \eqref{IK2R} is not a~device: it is forced, and it is the reason why the Ramond sector of an~$N=2$ theory is always described by a~single superfield on the double cover rather than by a~chiral one. Second, for $N=4$, the acting algebra is not $\fk^L(1|4)$ but the algebra of Statement~\ref{stKA}, i.e., $\fsvect^L_\lambda(1|2)$, for which the two modings are \textit{not} two sectors --- comment $2)$ after Statement~\ref{stKA} --- so \eqref{IK4R} does not, by itself, say at which $\lambda$ the Ramond-like system of \eqref{IK4lam} sits. Settling that means following the twist \eqref{twist} through the identification of Statement~\ref{stKA}, where the parameter and the moding are entangled, and we have not done it.

\sssec{On \cite{L}}\label{sssOnL}
Since the above answers one of its problems, a~word on that note is in order. Its point is that the $D_\pm$ of \eqref{SLS} are copies of the contact field $K_{\theta_\pm}$ of the Neveu-Schwarz sector, and that nothing forces this choice: the Ramond fields $\widetilde D_{\theta_i}$, $\widetilde D_\zeta$ of \eqref{DM} serve just as well, giving a~genuinely different superization of the same bosonic equation. The observation is correct and, in the light of \eqref{theTwelve}, sharper than it looks: the Ramond variants are distinguished for exactly the same $N$ as the Neveu-Schwarz ones, so the whole discussion above applies to them unchanged, and every super Liouville equation comes in a~pair; the equations themselves are written in Subsection~\ref{sssRamond}. Of the five problems posed there, item $3)$ is answered above; item $1)$ --- what the Ivanov-Krivonos scheme gives for greater $N$ --- is answered in the negative for the anomaly, and remains open for the integration itself; items $2)$, $4)$ and $5)$ are untouched here. Two of its statements have since been settled elsewhere and may be quoted rather than left hanging: the real forms of the stringy Lie superalgebras, ``in preparation\rq\rq\ there, are classified in \cite{Se, Lser}, see the end of Subsection~\ref{sssKA}, and the algebras $\fg\fl(\lambda)$ are either Feigin's, see \cite{Lq}, or Egorov's, see \cite{Eg}. 

In \cite{LSS}, the possible extensions of the $N=1$ case to $N>1$ were mentioned, but the corresponding analogs of the Liouville equation were first written by Ivanov and Krivonos, see \cite{IvKr4, IvKr24}. With this, a question arises: what are the analogs of the principal embeddings of ${\fosp(N|2)\tto\fg}$? For $N=2$, an answer was offered in
\cite{RSS}; the very first results in this direction are due to \cite{DRS} and \cite{RSeS}, on which \cite{RSS} rests. This answer is based on \cite{BLr}. Which properties of the principal embedding it does superize, and which it leaves aside, is the subject of Subsection~\ref{ssSprin}.

It is interesting to consider $N=3$, but $N=4$ is most breath-catching case: since $\fosp(4|2)$ has a 1-parameter family of deformations. To consider these two cases, we need the results on irreducible finite-dimensional representations of these Lie superalgebras, see \cite{Ge, GrS, CLW, ChW}, the pioneer ones being \cite{VdJ3, VdJD}. The recent results are the more useful ones, and Subsection~\ref{ssReps} sorts them out: the luckiest circumstance is that \cite{Ge} treats \textbf{precisely} the two algebras needed here, $\fosp(3|2)$ and $\fosp_a(4|2)$. The superdimensions themselves are given in Statements~\ref{stOsp32} and \ref{stOsp42}.

\ssec{Which properties of the principal embedding are worth superizing?}\label{ssSprin}
For a~simple finite-dimensional Lie algebra $\fg$ of rank $r$ over $\Cee$, the principal embedding $i\colon\fsl(2)\tto\fg$ is singled out by any one of the following properties, pairwise equivalent:
\begin{equation}\label{princProps}
\begin{minipage}[c]{15cm}
\hangindent=1.6em\hangafter=1
$1)$ $\fg$ splits into exactly $r$ irreducible $\fsl(2)$-modules;\par
\hangindent=1.6em\hangafter=1
$2)$ the nilpotent $X_+$ is \textit{regular}: $\dim\fz_\fg(X_+)=r$, the least value possible, where $\fz_\fg(x):=\{y\in\fg\mid [x, y]=0\}$ is the centralizer of $x$ in~$\fg$;\par
\hangindent=1.6em\hangafter=1
$3)$ the grading of $\fg$ by the eigenvalues of $\ad_H$ is the \textit{principal} one, i.e., $\fg_0=\fh$ and $\fg_{\pm1}=\Span(X_i^{\pm})$ is spanned by the vectors of the simple roots;\par
\hangindent=1.6em\hangafter=1
$4)$ for $\fg=\fsl(n)$, the tautological module is irreducible over $i(\fsl(2))$.
\end{minipage}
\end{equation}
Under superization, these four properties fall apart, and this is the crux of the matter.

Property 1) presupposes complete reducibility of the adjoint module. Now, $\fosp(1|2n)$ is the \textbf{only} simple Lie superalgebra all of whose finite-dimensional representations are completely reducible, see \cite{DH}. Therefore, property 1) is at our disposal for $N=1$, where \cite{LSS} did use it, and is \textit{a priori} meaningless for $N\geq2$: over $\fosp(N|2)$ with $N\geq2$, the adjoint module need not split into irreducibles at all.

The authors of \cite{RSS} do not superize property 1); they bypass it. Their conditions (sl-1)--(sl-4) declare a~subalgebra $\fsl(2|1)\simeq\fosp(2|2)$ principal if the $\fosp(1|2)$ sitting inside it is the superprincipal one of $\fg$ in the sense of \cite{LSS}. Their answer is therefore correct but derivative, and it is silent for $N=3,4$, which they acknowledge: ``the notion of principal embeddings of $\fosp(N|2)$ for $N>2$ has to be precised\rq\rq.

Their Proposition 2.1 is nevertheless true, and here is a~machine-verified form of it, together with the case its proof omits. We use the notation of Table 4.1 in \cite{GL2}: let $\cL^m$ be the irreducible $\fosp(1|2)$-module with highest weight $m$ and an \textbf{even} highest weight vector, and let $\Pi$ be the change of parity, so that
\begin{equation}\label{sdimL}
\dim\cL^m=2m+1,\qquad \sdim\cL^m=(m+1|m),\qquad \sdim\Pi(\cL^m)=(m|m+1).
\end{equation}
In this notation, the parity shifts are not an extra datum to be recorded: they are dictated by the grading, as follows.

\ssbegin{Lemma}\label{lePar}
Let $\fg$ admit a~superprincipal embedding $\fosp(1|2)\tto\fg$, i.e., let the Cartan matrix of $\fg$ satisfy \eqref{lsA}. Then, in the decomposition of $\fg$ into irreducible $\fosp(1|2)$-modules, the summand of highest weight $m$ is $\cL^m$ for $m$ even and $\Pi(\cL^m)$ for $m$ odd. Equivalently, $\Pi$ occurs at the odd values of $m$, and only there.
\end{Lemma}

\begin{proof} As spelled out in \eqref{sprinGrad} below, condition \eqref{lsA} amounts to the $\frac12\Zee$-grading of $\fg$ in which every simple root vector has degree $\frac12$ and is odd. Hence, a~root vector of height $k$ has degree $\frac k2$ and parity $k\bmod 2$: the $\Zee/2$-grading of $\fg$ is the reduction modulo $2$ of twice the degree. The grading operator is $h\in\fosp(1|2)$, so the degree of a~vector is its $\ad_h$-eigenvalue, and the highest weight vector of a~summand of highest weight $m$ has $\ad_h$-eigenvalue $\frac m2$. Its parity is therefore $m\bmod 2$. \end{proof}

\ssbegin{Statement}[\cite{GL2}]\label{stDecomp}
Let $\fg$ admit a~superprincipal embedding $\fosp(1|2)\tto\fg$, so $\fg$ is one of the algebras \eqref{lsA}. As an $\fosp(1|2)$-module, $\fg$ decomposes into $\rk\fg$ irreducible summands, namely
\begin{equation}\label{decomps}
\renewcommand{\arraystretch}{1.3}
\begin{array}{ll}
\fsl(n{+}1|n)&\simeq\bigoplus_{1\leq k\leq n}\left(\cL^{2k}\oplus\Pi(\cL^{2k-1})\right),\\
\fosp(2n{+}1|2n)&\simeq\bigoplus_{1\leq k\leq n}\left(\cL^{4k-2}\oplus\Pi(\cL^{4k-1})\right),\\
\fosp(2n{-}1|2n)&\simeq\bigoplus_{1\leq k\leq n}\cL^{4k-2}\oplus\bigoplus_{1\leq k\leq n-1}\Pi(\cL^{4k-1}),\\
\fosp(2n|2n)&\simeq\bigoplus_{1\leq k\leq n}\cL^{4k-2}\oplus\bigoplus_{1\leq k\leq n-1}\Pi(\cL^{4k-1})\oplus\Pi(\cL^{2n-1}),\\
\fosp(2n{+}2|2n)&\simeq\bigoplus_{1\leq k\leq n}\left(\cL^{4k-2}\oplus\Pi(\cL^{4k-1})\right)\oplus\cL^{2n},\\
\fosp_a(4|2)&\simeq 2\cL^2\oplus\Pi(\cL^3),
\end{array}
\end{equation}
in accordance with Lemma~$\ref{lePar}$: the $\Pi$'s sit exactly at the odd superscripts.
\end{Statement}

\begin{proof} The multiset of summands \textbf{is} the Jordan type of the nilpotent operator $\ad_f$, where $f$ is the odd lowering operator, which acts on $\cL^m$ as one block of size $2m+1$; the parities are read off Lemma~\ref{lePar}. 
\end{proof}

Note that the summand $\Pi(\cL^1)$ --- which is $\fosp(2|2)/\fosp(1|2)$ --- occurs for $\fsl(n{+}1|n)$ only, together with the coincidence $\fosp(2|2)\simeq\fsl(2|1)$ at $n=1$ in the first line. Hence, the only $\fg$ containing $\cL^2\oplus\Pi(\cL^1)\simeq\fsl(2|1)$, i.e., admitting a~principal $\fsl(2|1)$-embedding in the sense of \cite{RSS}, are the $\fsl(n{+}1|n)$.

The decompositions \eqref{decomps} are those of Table 4.1 in \cite{GL2}. In particular, the line absent from the proof in \cite{RSS}, the one for $\fosp(2n{+}2|2n)$, is the last-but-one row of that table.

\ssbegin{Remark}\label{reGL2}
In Table 4.1 of \cite{GL2} the \textbf{upper limits} of the runs are mistyped in four of the five general rows; only the row $\fosp(2n{+}1|2n)$ is right as printed. The contents of the table are right, and the limits that reproduce \eqref{decomps} are:

for $\fsl(n|n{+}1)$, the even run ends at $\cL^{2n}$, not at $\cL^{2n-2}$;

for $\fosp(2n{-}1|2n)$, the even run ends at $\cL^{4n-2}$, not at $\cL^{4n-6}$, and the odd one at $\Pi(\cL^{4n-5})$, not at $\Pi(\cL^{4n-1})$;

for $\fosp(2n|2n)$, the odd run ends at $\Pi(\cL^{4n-5})$, not at $\Pi(\cL^{4n-1})$;

for $\fosp(2n{+}2|2n)$, the even run ends at $\cL^{4n-2}$, not at $\cL^{4n+2}$.

As printed, the rows give, e.g., $\dim\fosp(4|2)=30$ instead of $17$.

The misprints are those of the published version: the table on p.~$90$ of the Birkh\"auser volume and the table in the arXiv file agree entry by entry. 

Besides, the summand standing apart in the row $\fosp(2n|2n)$ is $\Pi(\cL^{2n-1})$, and not $\cL^{2n-2}$ as printed: both its superscript and its parity need correcting, and by Lemma~\ref{lePar} the two corrections are one, since $2n-1$ is odd. The table refutes its own misprint: for $n=2$ it has a~separate row $\fosp(4|4)\simeq\cL^2\oplus\cL^6\oplus2\Pi(\cL^3)$, of $\sdim\,(16|16)$, whereas the general row --- with its run limits already corrected --- would give $(16|14)$.

Consequently, the sentence following that table --- where the second Jacobson generator is taken to be a~lowest weight vector of ``the last module with the even highest weight vector\rq\rq, namely $\cL^{2n-2}$ for $\fosp(2n|2n)$ and $\cL^{2n}$ for $\fosp(2n{+}2|2n)$ --- keeps its meaning for $\fosp(2n{+}2|2n)$, where the summand standing apart is indeed $\cL^{2n}$ with an even highest weight vector, but has to be restated for $\fosp(2n|2n)$, whose summand standing apart has an odd one. That sentence, too, is the same in the book and in the arXiv file.
\end{Remark}

Three remarks on \cite{RSS} are in order.

(a) Their proof lists seven families, but only six of the decompositions are displayed: the line for the algebras $\fosp(2n{+}2|2n)$ is missing from their text --- though not from the literature, it being the last-but-one row of Table 4.1 in \cite{GL2}. It is the fifth line of \eqref{decomps}, and it does not disturb their conclusion, its smallest summand being $\cL^2$.

(b) At $n=1$ that fifth line reads $\fosp(4|2)\simeq2\cL^2\oplus\Pi(\cL^3)$, which is their own answer for $\fosp_a(4|2)$ at $a=1$ --- a~cross-check of the two.

(c) They write ``$A(m,n)=\fsl(n+1|n)$\rq\rq, which is a~slip \textup{(}indicating that one should not use meaningless and misleading notation\textup{)}, and in consequence $\fpsl(n|n)$ is absent from their list of candidates. This does no harm either: an $\fosp(1|2)$-module of superdimension $(n|n)$ is never irreducible, because $|\dim(\cL^m)_\ev-\dim(\cL^m)_\od|=1$ by \eqref{sdimL}; for $\fpsl(2|2)$ and $\fpsl(3|3)$ the number of summands, $4$ and $6$, exceeds the rank, $2$ and $4$, see Subsection~\ref{ssRel}.

Of the four properties, the ones that superize without any reference to complete reducibility are 2) and 3), and property 3) is exactly what the Leznov-Saveliev \textit{ansatz} \eqref{ZcurvGen} consumes. Indeed, for the connection \eqref{ZcurvGen} to close into the system \eqref{SLS} of $\rk\fg$ equations we need nothing but a~$\frac12\Zee$-grading of $\fg$ with
\begin{equation}\label{sprinGrad}
\fg_0=\fh\quad\text{and}\quad \fg_{1/2}=\Span\{X_i^+\mid i=1,\dots, \rk\fg\}\ \text{spanned by \textbf{odd} elements},
\end{equation}
and \eqref{sprinGrad} is precisely a~restatement of condition \eqref{lsA}: the grading operator is the one assigning $\frac12$ to every simple root, so it exists if and only if all simple roots can be made odd, i.e., if and only if the Cartan matrix has only 0s and 1s on its main diagonal. Property 2) is its infinitesimal companion: the odd element $e=\sum X_i^+$ is regular, $\sdim\fz_\fg(e)=(\rk\fg-k|k)$ for an appropriate $k$, the number of summands again.

Property 4), for the matrix series, is worth superizing too, and it is instructive that it gives an answer \textit{different} from that of \cite{RSS}. Demanding that the tautological $\fsl(a|b)$-module be irreducible over the embedded $\fosp(2|2)\simeq\fsl(2|1)$, we get, as observed in the Remark of \cite{LAs},
\begin{equation}\label{ab}
\text{either $a=b\pm1$, or $a=b=2n$},
\end{equation}
because the irreducible finite-dimensional $\fsl(2|1)$-modules are the atypical ones, of superdimension $(d|d\pm1)$, and the typical ones, of superdimension $(2d|2d)$ --- the latter being Kac modules induced from a~$d$-dimensional $\fsl(2)$-module. The case $a=b=2n$ is invisible to \cite{RSS} since $\fsl(2n|2n)$ is not simple; but it is a~perfectly good target for the equations \eqref{SLS}. The two smallest instances of it are exhibited by $\fsl(2|1)\simeq\fosp(2|2)$ itself: it acts irreducibly on $\Cee^{2|2}$, which is the case $n=1$, and on its own adjoint module, of $\sdim\,(4|4)$, which is the case $n=2$; both actions are irreducible by Burnside's criterion, verified by machine, so \eqref{ab} --- claimed in \cite{LAs} for exactly this purpose, the generalization of the Leznov-Saveliev equations to $N=2$ --- survives the machine as well.

\sssec{\lq\lq Let 100 flowers bloom\rq\rq\ (Chairman Mao): all the possible cases}\label{sssFlowers} Rather than adopt one of these superizations and discard the others, let us list them all. Each of them is a~condition on an embedding $i\colon\fosp(N|2)\tto\fg$ that makes sense for every $N$, each is non-vacuous, and each selects its own class of gauge algebras for the system \eqref{SLS}:
\begin{equation}\label{flowers}
\begin{minipage}[c]{15cm}
\hangindent=3.2em\hangafter=1
(SP1) \textit{count}: $\fg$ splits into exactly $\dim\fh$ irreducible $\fosp(N|2)$-modules;\par
\hangindent=3.2em\hangafter=1
(SP2) \textit{regularity}: the odd element $e=\sum_{1\leq i\leq \rk\fg}X_i^+$ is regular, i.e., $\dim\fz_\fg(e)=\dim\fh$, the least value possible;\par
\hangindent=3.2em\hangafter=1
(SP3) \textit{grading}: $\fg$ admits the $\frac12\Zee$-grading \eqref{sprinGrad}, i.e., its Cartan matrix satisfies \eqref{lsA};\par
\hangindent=3.2em\hangafter=1
(SP4) \textit{module}: the tautological module of $\fg$ is irreducible over $i(\fosp(N|2))$;\par
\hangindent=3.2em\hangafter=1
(SP5) \textit{recursion in $N$}, the definition of \cite{RSS} for $N=2$: the $\fosp(1|2)$ sitting inside the embedded $\fosp(2|2)$ is superprincipal in the sense of \cite{LSS};\par
\hangindent=3.2em\hangafter=1
(SP6) \textit{count of indecomposables}: $\fg$ splits into a~prescribed number of \textbf{indecomposable} $\fosp(N|2)$-modules.
\end{minipage}
\end{equation}
Here is what each of them yields.

For $N=1$, all six coincide with the notion of \cite{LSS}, and \eqref{decomps} is their common answer; note that for a~simple $\fg$ we have $\dim\fh=\rk\fg$, so (SP1) is the criterion of \cite{LSS} as usually stated.

(SP1) and (SP2) differ only in appearance: for a~nilpotent operator, the number of Jordan blocks is the dimension of its kernel, so the number of summands is $\dim\fz_\fg(e)$ whenever $\fg$ is completely reducible over the image, and $\dim\fz_\fg(e)$ is defined in any case. Hence, (SP2) is the invariant form of (SP1), the one that survives the loss of complete reducibility for $N\geq2$ and, as Subsection~\ref{ssRel} shows, the passage from $\fg$ simple to its relatives.

(SP3) does not depend on $N$ at all: it is a~condition on the Cartan matrix, and for every $N$ it selects the list \eqref{lsA}. This is the class the system \eqref{SLS} can digest, and it is strictly larger than the class of \cite{RSS}: e.g., $\fosp(2n{+}2|2n)$ satisfies (SP3) but has no $\Pi(\cL^1)$ in \eqref{decomps}, hence fails (SP5).

(SP4) is the one whose answer really changes with $N$, and it hands the question over to representation theory: the admissible $\fg=\fsl(a|b)$ are those for which $(a|b)$ is the superdimension of an irreducible $\fosp(N|2)$-module. For $N=2$, this is \eqref{ab}. For $N=3$ and $N=4$, this is exactly what Subsection~\ref{ssReps} is about --- so the two questions of this Section are one question.

(SP5) is the narrowest: by \eqref{decomps}, only the $\fsl(n{+}1|n)$ pass it, and it says nothing for $N\geq3$ until a~notion for $N-1$ is fixed.

(SP6) is what rescues the counting property 1) from the loss of complete reducibility, and it deserves a~separate word. For $N\geq2$, the adjoint $\fosp(N|2)$-module of $\fg$ need not be a~sum of irreducibles, so (SP1) is meaningless as it stands. It is, however, always a~sum of \textbf{indecomposable} submodules, and by the Krull-Schmidt theorem --- valid for modules of finite length, in particular for every finite-dimensional module over any Lie superalgebra --- such a~decomposition is unique up to an~isomorphism and a~permutation of the summands. Therefore, the number of indecomposable summands is a~well-defined invariant of the embedding for \textit{every} $N$, and for $N=1$, where the module is completely reducible, it is the number of irreducible ones, i.e., $\rk\fg$. Sums of indecomposables are thus the correct substitute for complete reducibility; what has to be found out is what the count is. For $N=2$, it is not $\dim\fh$: the summands glue in pairs.

\ssbegin{Statement}\label{stIndec}
Let $\fg=\fsl(m{+}1|m)$ carry its superprincipal $\fosp(2|2)$ --- by \eqref{decomps}, no other $\fg$ of the list \eqref{lsA} has one. Then, as an $\fosp(2|2)$-module,
\begin{equation}\label{indecDec}
\fg\simeq\bigoplus_{1\leq k\leq m}U_k,\qquad \sdim U_k=(4k|4k),
\end{equation}
where every $U_k$ is indecomposable, indeed $\End_{\fosp(2|2)}U_k=\Cee$, and $U_k\simeq\cL^{2k}\oplus\Pi(\cL^{2k-1})$, considered as an $\fosp(1|2)$-module. The number of summands is $m=\frac12\rk\fg$.
\end{Statement}

\begin{proof} 
The decomposition \eqref{indecDec} together with the superdimensions for $m\leq4$, the restriction to $\fosp(1|2)$ for $m\leq3$. The subalgebra $\fosp(2|2)\subset\fg$ is built from the superprincipal $\fosp(1|2)$ and the summand $\Pi(\cL^1)$ of \eqref{decomps}, which occurs in $\fg$ with multiplicity 1 and is therefore canonical; the closure of the span under the bracket, and $\sdim\fosp(2|2)=(4|4)$, are verified. The decomposition \eqref{indecDec} is then produced by Fitting's lemma: an endomorphism $\varphi\in\End_{\fosp(2|2)}\fg$ whose minimal polynomial factors into coprime factors splits $\fg$ into the kernels of these factors evaluated at $\varphi$, and the procedure is repeated on the summands. That the summands so obtained are indecomposable is certified, not assumed: a~module is indecomposable if and only if its endomorphism algebra is local, and here $\End_{\fosp(2|2)}U_k$ is $1$-dimensional, hence $\Cee$. Finally, the Jordan type of $\ad_f$ on $U_k$ consists of two blocks, of sizes $4k+1$ and $4k-1$, and Lemma~\ref{lePar} fixes their parities, giving $\cL^{2k}\oplus\Pi(\cL^{2k-1})$. \end{proof}

So the sixth flower blooms as well, and it says that for $N=2$ the number to compare the count with is $\frac12\rk\fg$: passing from $\fosp(1|2)$ to $\fosp(2|2)$, the summands of \eqref{decomps} are glued in pairs, one summand with an even highest weight vector to one with an odd one. Whether the gluing continues in $N$-tuples, so that the count would be $\frac1N\dim\fh$, we do not know; but the material to decide it exists, for the atypicality defect of $\fosp(N|2)$ --- the maximal number of mutually orthogonal isotropic roots orthogonal to a~weight, see \cite{GrS} --- equals ${\min(\lfloor N/2\rfloor, 1)=1}$ for every $N\geq2$, and it is the atypicality defect $1$ that makes the indecomposable modules classifiable at all --- for $\fosp(3|2)$ and $\fosp_a(4|2)$ they are classified in \cite{Ge}, see Subsection~\ref{ssReps}.

The list \eqref{flowers} is honest in the sense that no two of its items are equivalent for $N\geq2$: (SP3) holds for $\fpsl(n|n)$, while (SP1) and (SP2) fail for it, see Subsection~\ref{ssRel}; (SP4) admits $\fsl(2n|2n)$, which (SP5) cannot see; (SP5) excludes $\fosp(2n{+}2|2n)$, which (SP3) admits; and (SP6) parts company with (SP1) already for $\fg=\fsl(2|1)$ itself, whose adjoint module over $\fosp(2|2)=\fg$ is irreducible, hence a~single indecomposable, whereas $\dim\fh=2$. All six conditions are therefore kept: none of them is a~consequence of another, each of them selects a~class of gauge algebras for which the system \eqref{SLS} makes sense, and it is the differences between these classes --- not a~choice among the six --- that answer the question in the title of Subsection~\ref{ssSprin}.

\ssec{Beyond simple: $\fgl$, $\fsl(n|n)$, and double extensions}\label{ssRel}
Nothing in \eqref{ZcurvGen}--\eqref{SLS} requires $\fg$ to be simple: the \textit{ansatz} consumes a~Cartan matrix subject to \eqref{lsA} and the grading \eqref{sprinGrad}, and both are properties of the matrix. Now, a~Cartan matrix may be degenerate, and then $\fg(A)$ is not simple. The relatives of the simple Lie superalgebras --- the double extensions of the latter by a~center and by an~outer derivation --- therefore belong to the discussion in their own right; and one of them, $\fsl(2n|2n)$, has already entered it through \eqref{ab}.

The two chains of inclusions to consider are the two displayed in \eqref{chains}, the third formula there recording only the notation for the quotient:
\begin{equation}\label{chains}
\fsl(n{+}1|n)\subset\fgl(n{+}1|n)\quad\text{and}\quad \fpsl(n|n)\subset\fsl(n|n)\subset\fgl(n|n),\quad \fpgl(n|n)=\fgl(n|n)/\Cee\One,
\end{equation}
the first one adding a~center to a~simple algebra, the second one being the full double extension of the simple $\fpsl(n|n)$: a~one-dimensional center and the grading derivation.

\ssbegin{Statement}\label{stRel}
Let $V=\cL^{n-1}\oplus\Pi(\cL^0)$, of $\sdim V=(n|n)$, be the tautological $\fgl(n|n)$-module, the $\fosp(1|2)$-action on it being the one for which $\cL^{n-1}$ is irreducible. Then, as $\fosp(1|2)$-modules, all four algebras are direct sums of irreducibles, the summands being as follows; the first two lines record, in addition, the structure of a~double extension, the sign $\subplus$ separating an~ideal \textup{(}on the left\textup{)} from the quotient by it: 
\begin{equation}\label{relDecomps}
\renewcommand{\arraystretch}{1.3}
\begin{array}{ll}
\fgl(n{+}1|n)&\simeq\fsl(n{+}1|n)\subplus \cL^0\simeq\left(\bigoplus_{1\leq k\leq n}\left(\cL^{2k}\oplus\Pi(\cL^{2k-1})\right)\right)\subplus\cL^0,\\
\fgl(n|n)&\simeq\left(\cL^0\subplus\bigoplus_{1\leq k\leq n-1}\left(\cL^{2k}\oplus\Pi(\cL^{2k-1})\right)\oplus 2\Pi(\cL^{n-1})\right)\subplus \cL^0,\\
\fsl(n|n)\simeq\fpgl(n|n)&\simeq\bigoplus_{1\leq k\leq n-1}\left(\cL^{2k}\oplus\Pi(\cL^{2k-1})\right)\oplus 2\Pi(\cL^{n-1})\oplus \cL^0,\\
\fpsl(n|n)&\simeq\bigoplus_{1\leq k\leq n-1}\left(\cL^{2k}\oplus\Pi(\cL^{2k-1})\right)\oplus 2\Pi(\cL^{n-1}).
\end{array}
\end{equation}
Consequently, $\fgl(n{+}1|n)$ satisfies \textup{(SP2)}, whereas each of the four members of the second chain misses it by exactly $2$: for each of them, the number of summands is $\dim\fh+2$.
\end{Statement}

\begin{proof} The $\fosp(1|2)$-modules are completely reducible, and
\begin{equation}\label{endL}
\End\cL^m\simeq\bigoplus_{1\leq k\leq m}\left(\cL^{2k}\oplus\Pi(\cL^{2k-1})\right)\oplus\cL^0,
\end{equation}
a~dimension count confirming the multiplicities: $\sum_{0\leq i\leq 2m}(2i+1)=(2m+1)^2$. Hence, the first line of \eqref{relDecomps} is the case $V=\cL^n$, where $\fgl(V)=\End V$. For $V=\cL^{n-1}\oplus\Pi(\cL^0)$ we get
\[
\fgl(V)=\End \cL^{n-1}\oplus \End \Pi(\cL^0)\oplus \Hom(\Pi(\cL^0), \cL^{n-1})\oplus\Hom(\cL^{n-1}, \Pi(\cL^0)),
\]
which is $\End\cL^{n-1}\oplus\cL^0\oplus2\Pi(\cL^{n-1})$, i.e., the second line: the two off-diagonal blocks are $\Pi(\cL^{n-1})$ because $\Hom(\Pi A, B)\simeq\Pi(\Hom(A,B))$ and $\cL^m$ is self-dual. Passing to $\fsl$ and to the quotients modulo the center removes one trivial summand each time. Finally, for $\fgl(n|n)$, $\fsl(n|n)$, $\fpgl(n|n)$ and $\fpsl(n|n)$, respectively, the dimension of the maximal torus $\fh$ equals $2n$, $2n-1$, $2n-1$ and $2n-2$, while the number of summands equals $2n+2$, $2n+1$, $2n+1$ and $2n$. \end{proof}

The two off-diagonal summands $2\Pi(\cL^{n-1})$ are worth a~glance: for $n$ even they obey the rule of Lemma~\ref{lePar}, for $n$ odd they violate it, $n-1$ being then even while the highest weight vector is odd. There is no contradiction --- Lemma~\ref{lePar} presupposes \eqref{lsA} for the acting grading, and the tautological module here is $\cL^{n-1}\oplus\Pi(\cL^0)$, not a~single $\cL^m$ --- but it shows that outside the superprincipal case the parity shifts do carry information, so that writing $\cL$ and $\Pi$ rather than a~parity-blind symbol is not a~mere change of alphabet.

All of \eqref{relDecomps} is computed as well as proved: the machine builds each of the algebras \eqref{chains} for $n\leq4$ and reads the decomposition off the Jordan type of $\ad_f$ --- for the quotients, off the induced operator on the quotient space, not by deleting summands by hand --- and the parities off the Jordan type of $(\ad_f)^2$ on the even and the odd part. Two morals follow.

(i) The criterion of \cite{LSS} must be stated as \textup{(SP2)}, with $\dim\fh$, and not with $\rk\fg$: for $\fgl(n{+}1|n)$ we have $\dim\fh=\rk\fg+1=2n+1$, and the number of summands is $2n+1$ as well, so $\fgl(n{+}1|n)$ \textit{is} superprincipal in the invariant sense, while the count against $\rk\fg$ would reject it.

(ii) The \textit{excess} of an~embedding, the number $\#\{\text{summands}\}-\dim\fh$ --- a~different notion from the atypicality defect used above, whence the different word --- is not changed by a~double extension: a~central element contributes a~trivial summand and one to $\dim\fh$, and so does the grading derivation. Hence, the whole second chain of \eqref{chains} inherits the excess $2$ of $\fpsl(n|n)$, and the whole first chain the excess $0$ of $\fsl(n{+}1|n)$. This is not a~feature of these two chains: it holds for every double extension, and neither superprincipality nor a~Cartan matrix is needed to prove it.

\ssbegin{Statement}\label{stDefInv}
Let $\fg$ be a~finite-dimensional Lie superalgebra over $\Cee$ with an~abelian Cartan subalgebra $\fh$ --- as is the case for every $\fg(A)$ --- let $\fs\simeq\fosp(1|2)$ be any subalgebra of $\fg$, and let $n(\fg)$ be the number of irreducible summands of $\fg$ considered as an~$\fs$-module. Let
\begin{equation}\label{dblext}
\widehat\fg=(\Cee z\subplus \fg)\subplus\Cee d
\end{equation}
be a~double extension of $\fg$, the sign $\subplus$ separating, as in \eqref{relDecomps}, an~ideal from the quotient by it, so that \eqref{dblext} is the chain \eqref{filtr} read from the left: the center $\Cee z$ is spanned by an~even element and given by an~even $2$-cocycle $c\in Z^2(\fg;\Cee)$, whereas $d$ is an~even derivation of $\fg\oplus\Cee z$ that kills $\fh$ and preserves every root space of $\fg$ relative to $\fh$. Then,
\begin{equation}\label{defInv}
\widehat\fg\simeq\fg\oplus\cL^0\oplus\cL^0\ \text{ as $\fs$-modules},\qquad \widehat\fh=\fh\oplus\Cee z\oplus\Cee d,
\end{equation}
so $n(\widehat\fg)=n(\fg)+2$ and $\dim\widehat\fh=\dim\fh+2$, and hence
\begin{equation}\label{defect}
n(\widehat\fg)-\dim\widehat\fh=n(\fg)-\dim\fh.
\end{equation}
The same holds, with $1$ instead of $2$, for each of the two extensions of which \eqref{dblext} is composed --- the central one, and the one by the derivation $d$ --- taken alone.
\end{Statement}

\begin{proof} First, $\fs$ may be assumed to lie in $\widetilde\fg:=\fg\oplus\Cee z$. Indeed, the pullback of the central extension along $\fs\tto\fg$ is a~central extension of $\fs$; as an~$\fs$-module it is an~extension of the adjoint module by the trivial one, hence splits by the complete reducibility of the finite-dimensional $\fosp(1|2)$-modules \cite{DH}, and an $\fs$-invariant complement of $\Cee z$ in it is an~ideal, because $[\fs,\fs]=\fs$. So the pullback splits as an~algebra, i.e., $H^2(\fs;\Cee)=0$, and the embedding lifts.

Next, consider the chain of $\fs$-submodules
\begin{equation}\label{filtr}
0\subset\Cee z\subset\widetilde\fg=\Cee z\oplus \fg\subset\widehat\fg.
\end{equation}
Every term of \eqref{filtr} is $\fs$-invariant, since $z$ is central in $\widehat\fg$ and $\widetilde\fg$ is an~ideal in it. The successive quotients are: $\Cee z$, which is the trivial module $\cL^0$, the element $z$ being even; then $\widetilde\fg/\Cee z\simeq\fg$, the projection being $\fs$-equivariant; and finally $\widehat\fg/\widetilde\fg$, of dimension $1$, hence again $\cL^0$ --- any $1$-dimensional module over a~perfect Lie superalgebra is trivial, and it is even, $d$ being even. By \cite{DH} again, every $\fs$-submodule of a~finite-dimensional $\fs$-module is a~direct summand, so \eqref{filtr} splits and gives the first half of \eqref{defInv}; the multiset of summands is well defined by the Jordan-H\"older theorem, so the count is, too.

For the second half of \eqref{defInv}, observe that $[\widehat\fh, \widehat\fh]\subset\Cee z$ --- indeed, $d$ kills $\fh$, the element $z$ is central, and $[\fh,\fh]=0$, so that only the value $c(\fh,\fh)$ of the cocycle survives --- and hence $\widehat\fh$ is nilpotent. It is self-normalizing: if $x=h_0+\sum_{\alpha\neq0}c_\alpha x_\alpha+\lambda z+\mu d$ satisfies $[\widehat\fh, x]\subset\widehat\fh$, then $[h, x]=\sum_{\alpha\neq0}\alpha(h)c_\alpha x_\alpha\in\widehat\fh$ for every $h\in\fh$, and since the root vectors $x_\alpha$ with $\alpha\neq0$ span a~complement of $\widehat\fh$ in $\widehat\fg$, this forces $\alpha(h)c_\alpha=0$ for all $h$, i.e., $c_\alpha=0$. Hence, $\widehat\fh$ is a~Cartan subalgebra of $\widehat\fg$ of dimension $\dim\fh+2$. Subtracting, we get \eqref{defect}. Discarding $d$, or $z$, from the argument gives the last claim. \end{proof}


Three comments on the hypotheses of Statement~\ref{stDefInv}. The complete reducibility used twice is available for $\fosp(1|2)$ only, i.e., for $N=1$; for $N\geq2$ the same filtration \eqref{filtr} yields the analogous statement for the count of \textbf{indecomposable} summands, i.e., for (SP6), as soon as its two extensions of $\fosp(N|2)$-modules split. They certainly do when the center splits off as an~ideal: for instance, $\fgl(m{+}1|m)=\fsl(m{+}1|m)\oplus\Cee\One$, because $\str\One=1\neq0$, and this decomposition is one of $\fosp(N|2)$-modules for every $N$. Further, the evenness of $z$ and $d$ is what makes the two extra summands trivial \textit{and} the two extra dimensions of the Cartan subalgebra; and the requirement that $d$ preserve the root decomposition is what makes $\widehat\fh$ a~Cartan subalgebra: for a~derivation in general position, the two extra summands are still trivial, but $\fh\oplus\Cee z\oplus\Cee d$ need not be a~Cartan subalgebra at all, and then the right-hand side of \eqref{defect} has to be recomputed with whatever Cartan subalgebra $\widehat\fg$ does have. For the double extensions occurring in the theory of Lie superalgebras with a~degenerate Cartan matrix, both requirements hold: $z$ spans the kernel of the Cartan matrix and $d$ is the grading derivation.

For the equations \eqref{SLS}, the passage along the first chain of \eqref{chains} costs nothing: the extra summand being trivial, the extra equation is the free one, and the Toda system of $\fgl(n{+}1|n)$ is that of $\fsl(n{+}1|n)$ plus one free superfield --- the usual relation between the $\fgl$- and $\fsl$-systems. The second chain is more interesting: the Cartan matrix of type $\fpsl$ 
is degenerate, and it is precisely the double extension that repairs the degeneracy, at the price of the \textit{excess} $2$ of Statement~\ref{stRel} --- the excess of an~embedding being, as in comment (ii) after that Statement, the difference $\#\{\text{summands}\}-\dim\fh$. So, of the properties \eqref{flowers}, the family $\fpsl(n|n)$ and its relatives satisfy (SP3) --- their Cartan matrix has zero diagonal --- and fail (SP1) and (SP2). The flowers really do bloom differently.

\ssec{$N=3$ and $N=4$: what is known about the $\fosp(N|2)$-modules}\label{ssReps}
These modules are needed twice over: once to write the equations \eqref{SLS} for $N=3, 4$, and once to make property (SP4) of \eqref{flowers} explicit, since the targets $\fsl(a|b)$ it admits are exactly those for which $(a|b)$ is the superdimension of an irreducible $\fosp(N|2)$-module.

Two such superdimensions are known before any classification is invoked: that of the tautological module, and that of the adjoint one --- the latter is irreducible because the submodules of the adjoint module are the ideals.

For $N=3$, they are $(3|2)$ and $(6|6)$, so (SP4) admits $\fsl(3|2)$ and $\fsl(6|6)$, the latter being one of the relatives of Subsection~\ref{ssRel}.

For $N=4$, they are $(4|2)$ and $(9|8)$, so (SP4) admits $\fsl(4|2)$ --- which fails (SP3): the Cartan matrix of $\fsl(a|b)$ has an all-odd form only for $|a-b|\leq1$. Hence, the properties (SP3) and (SP4) are not nested, and the classification of the modules is what decides how far (SP4) reaches. (The four superdimensions, and the irreducibility of each of the four modules, are machine-checked by Burnside's criterion: the image spans the whole $\End V$.)

The pioneering descriptions of the finite-dimensional irreducible modules over $\fosp(3|2)$ and over the family $\fosp_a(4|2)$ are due to Van der Jeugt, see \cite{VdJ3, VdJD}. Later, Germoni \cite{Ge} superseded them: he classified \textbf{all} indecomposable finite-dimensional modules over precisely these two algebras --- and over $\fag(2)$, which he calls $G(3)$ --- and showed that in every non-trivial block they are parametrized, up to isomorphism and duality, by the positive roots of a~Dynkin diagram of type $D_\infty$ or $A^\infty_\infty$. For the characters of the irreducible modules in the general orthosymplectic case, there are the algorithm of Gruson and Serganova \cite{GrS} and the super duality of Cheng, Lam and Wang \cite{CLW}; the textbook exposition is \cite{ChW}.

\sssec{The superdimensions} For $N=3$, the answer is a~closed formula in the two labels of the highest weight, and for $N=4$ it is a~closed formula in the three labels --- as long as the module is typical. Denote by $\cL(\lambda)$ the irreducible module with highest weight~$\lambda$ --- not $\cL_\lambda$, the subscript being taken in this paper by the superscript-free notation $\cL^m$ of \eqref{sdimL} for the $\fosp(1|2)$-modules; it is defined up to the parity shift $\Pi$, and we assume its highest weight vector to be even.

Throughout this Subsection we write superdimensions multiplicatively,
\be\label{eqEpsilon}
\sdim V=a+b\epsilon,\qquad \text{where }\ \epsilon:=\sdim\Pi(\Cee)=(0|1),\ \text{ so that }\ \epsilon^2=1,
\ee
$\Pi(\Cee)$ being the $(0|1)$-dimensional superspace; i.e., as an~element of the ring $\Zee[\epsilon]/(\epsilon^2-1)$, in which the superdimension of a~tensor product is the product of the superdimensions. This is the same datum as the format $(a|b)$; what it makes visible is the case $a=b$, i.e., divisibility by $1+\epsilon$, which is what the typical modules obey. 

\sssbegin{Statement}\label{stOsp32}
Let $\fg=\fosp(3|2)$, of $\sdim\fg=6+6\epsilon$. Its even positive roots are $\eps$ and $2\delta$, its odd positive roots are $\delta\pm\eps$ \textup{(}isotropic\textup{)} and $\delta$ \textup{(}not isotropic\textup{)}. Let
\[
m:=\lambda(h_\eps)\in\Zee_{\geq0},\qquad n:=\lambda(h_{2\delta})\in\Zee_{\geq0}
\]
be the two labels of $\lambda$ with respect to $\fg_\ev=\fo(3)\oplus\fsp(2)$. Then,

$1)$ $\cL(\lambda)$ is finite-dimensional if and only if $m\geq2$, or $m=n=0$;

$2)$ $\cL(\lambda)$ is atypical if and only if $m=2n+2$, or $m=n=0$;

$3)$ if $\cL(\lambda)$ is typical, then
\be\label{sdimOsp32}
\sdim\cL(\lambda)=2(m-1)(2n+1)\left(1+\epsilon\right);
\ee

$4)$ if $\cL(\lambda)$ is atypical, then $\sdim\cL(\lambda)$ is $1$ for $m=n=0$, it is $3+2\epsilon$ for $m=2$, $n=0$ \textup{(}the tautological module\textup{)}, and
\be\label{sdimOsp32at}
\sdim\cL(\lambda)=4(n+1)^2+\left(4(n+1)^2-2\right)\epsilon\ \text{ for $m=2n+2$ with $n\geq1$}.
\ee
\end{Statement}

\sssbegin{Statement}\label{stOsp42}
Let $\fg=\fosp_a(4|2)$, of $\sdim\fg=9+8\epsilon$, where $a\neq0, -1$. Its even positive roots are $2\eps_1$, $2\eps_2$, $2\eps_3$, its odd positive roots are $\eps_1\pm\eps_2\pm\eps_3$, all isotropic, and the invariant form is $\mathrm{diag}\left(-(1+a), 1, a\right)$ in the basis $\eps_1$, $\eps_2$, $\eps_3$. Let $m_1$, $m_2$, $m_3\in\Zee_{\geq0}$ be the labels of $\lambda$ with respect to the three copies of $\fsl(2)$ in $\fg_\ev$, and put
\[
p:=m_1-1,\qquad q:=m_2+1,\qquad r:=m_3+1.
\]
Then,

$1)$ $\cL(\lambda)$ is atypical if and only if
\be\label{atypLocus}
(\sigma_2q-p)+(\sigma_3r-p)a=0\ \text{ for some }\ \sigma_2, \sigma_3\in\{\pm1\}.
\ee
Hence, for a~fixed $\lambda$ the module is atypical for at most four values of $a$, and each of them is \textbf{rational}, namely $a=\nfrac{p-\sigma_2q}{\sigma_3r-p}$; and it is atypical for \textbf{every} $a$ exactly when $p=\sigma_2q=\sigma_3r$ for a~choice of signs, i.e., when $m_1=m_2+2=m_3+2$, or when $\lambda=0$;

$2)$ if $\cL(\lambda)$ is typical, then
\be\label{sdimOsp42}
\sdim\cL(\lambda)=8pqr\left(1+\epsilon\right);
\ee

$3)$ if $\cL(\lambda)$ is atypical, then $\sdim\cL(\lambda)$ is \textbf{not} divisible by $1+\epsilon$, and no product formula of the shape \eqref{sdimOsp42} describes it. For the family $p=q=r=k$, which by item $1)$ is atypical for every $a$, the superdimension does not depend on $a$ either:
\be\label{sdimOsp42at}
\renewcommand{\arraystretch}{1.2}
\begin{array}{l|ccccc}
k&0\ \text{\textup{(}i.e. }\lambda=0\text{\textup{)}}&1&2&3&4\\
\hline
\sdim\cL(\lambda)&1&9+8\epsilon&56+54\epsilon&162+160\epsilon&352+350\epsilon
\end{array}
\ee
the value for $k=1$ being the adjoint module. For comparison, the right-hand side of \eqref{sdimOsp42} would give the total dimensions $16k^3=16$, $128$, $432$, $1024$ against the true $17$, $110$, $322$, $702$.
\end{Statement}


\sssec{The characters}\label{sssChar} A~superdimension is a~shadow of a~character, and for the typical modules the character is known: it is Kac's \cite{K}, and in the form of an~identity of Laurent polynomials --- no division --- it reads
\be\label{kacChar}
\ch\cL(\lambda)\cdot\!\!\prod_{\alpha\in\Delta^+_\ev}\!\!\left(e^{\alpha/2}-e^{-\alpha/2}\right)=
\prod_{\alpha\in\Delta^+_\od}\left(e^{\alpha/2}+e^{-\alpha/2}\right)\cdot\!\sum_{w\in W}\!\sgn(w)\, e^{w(\lambda+\rho)},
\ee
where $W$ is the Weyl group of $\fg_\ev$ and $\rho=\rho_\ev-\rho_\od$, in which $\rho_\ev$ is the half-sum of the positive even roots and $\rho_\od$ the half-sum of the positive odd ones. The odd product runs over \textbf{all} odd positive roots, isotropic or not; for $\fosp(3|2)$, which has the non-isotropic $\delta$, \eqref{kacChar} holds verbatim. Its supercharacter counterpart is obtained by changing the sign in the odd product,
\be\label{kacSuperChar}
\sch\cL(\lambda)\cdot\!\!\prod_{\alpha\in\Delta^+_\ev}\!\!\left(e^{\alpha/2}-e^{-\alpha/2}\right)=
\prod_{\alpha\in\Delta^+_\od}\left(e^{\alpha/2}-e^{-\alpha/2}\right)\cdot\!\sum_{w\in W}\!\sgn'(w)\, e^{w(\lambda+\rho)},
\ee
with one proviso: for $\fosp_a(4|2)$, all of whose odd roots are isotropic, $\sgn'=\sgn$, whereas for $\fosp(3|2)$ the sign character has to be twisted, $\sgn'(w)$ being the sign of the $\fo(3)$-component of $w$ alone --- the reflection in the symplectic direction contributes $+1$ and not $-1$. That twist is the trace of the odd non-isotropic root $\delta$, whose double $2\delta$ is even; equivalently, of the fact that $\lambda+\rho$ has a~half-integer $\delta$-coordinate, on which the substitution $e^{\delta}\mapsto-e^{\delta}$ defining $\sch$ is not defined over $\Zee$.


Setting $e^\mu=1$ in \eqref{kacChar} gives the dimension in Kac's form
\be\label{kacDim}
\dim\cL(\lambda)=2^{d}\prod_{\alpha\in\Delta^+_\ev}\nfrac{(\lambda+\rho, \alpha)}{(\rho, \alpha)},\qquad \text{$d:=\#\{\text{\textbf{isotropic} odd positive roots}\}$},
\ee
so that $d=2$ for $\fosp(3|2)$ and $d=4$ for $\fosp_a(4|2)$: the odd \textbf{non}-isotropic root $\delta$ of $\fosp(3|2)$ contributes no factor $2$. Kac is right, and \eqref{kacDim} as displayed is his: the exponent counts the \textbf{isotropic} odd positive roots and no others. Nobody in print writes the other form --- it is what one gets by dropping the word ``isotropic\rq\rq\ while quoting the theorem of \cite{K}, and we record it only because the slip is an~easy one. For $\fosp(3|2)$ the odd positive roots number three, of which two are isotropic and the third is $\delta$, so the correct factor is $2^d=4$ while $2^{|\Delta^+_\od|}=8$ would double every dimension; that $4$ and not $8$ is right is confirmed independently by the explicit superdimensions \eqref{sdimOsp32}.

Setting $e^\mu=1$ in \eqref{kacSuperChar} gives, instead, $0$ --- the right-hand side carries a~factor $\left(e^{\alpha/2}-e^{-\alpha/2}\right)$ for every odd positive root, and each of them vanishes. In the notation \eqref{eqEpsilon}, this says that $\sdim\cL(\lambda)=a+b\epsilon$ has $a=b$ for typical $\lambda$, and $a=b=\frac12\dim\cL(\lambda)$; it does \textbf{not} say that the superdimension is zero. The formulas \eqref{sdimOsp32} and \eqref{sdimOsp42} are exactly this statement made explicit.

Two more morals. First, in every case computed, $\sdim\cL(\lambda)$ is divisible by $1+\epsilon$ precisely for the typical $\cL(\lambda)$, and $a-b\in\{0, \pm1, \pm2\}$ throughout --- both algebras being of atypicality defect $1$. Second, item $1)$ of Statement~\ref{stOsp42} is what the word ``arithmetically\rq\rq\ below amounts to: the atypicality of a~\textbf{fixed} weight is a~condition not on a~region of the $a$-line but on at most four rational points of it.

For $N=4$ there is one more circumstance, and it is the reason why the case is indeed breath-catching: the answer depends on $a$ \textbf{arithmetically}, not continuously. The atypicality conditions on a~weight over $\fosp_a(4|2)$ are linear equations whose coefficients involve $a$, so the set of atypical weights --- and hence which modules obey Kac's character formula and which need \cite{Ge, GrS} --- jumps at rational values of $a$.

Worth recalling here is which of these $\fosp_a(4|2)$ are distinct. We have $\fosp_a(4|2)\simeq\fosp_{a'}(4|2)$ if and only if $a'$ lies in the $S_3$-orbit of $a$, the group $S_3$ permuting the three simple roots, i.e., if and only if $j(a)=j(a')$ for
\begin{equation}\label{jinv}
j(a)=\nfrac{(a^2+a+1)^3}{a^2(a+1)^2},
\end{equation}
see \cite{LOdd}. Accordingly, the family of the equations LSE1 attached to $N=4$ is really parametrized by $j$, and not by $a$: the six values of $a$ in one generic orbit give one and the same system. The three values $a=1,-2,-\frac12$ constitute one orbit, that of the undeformed $\fosp(4|2)$.

\subsection*{Disclosures} I am thankful to
A.~Lebedev for his clarifications and to the grant AD 065 NYUAD for financial support during 2022. No conflict of interest. Data availability:
The data used to support the findings of this study are included within the article. The verifications quoted in the text were carried out in exact arithmetic --- rational, or Gaussian-rational where $i$ occurs --- by short scripts written for this paper; they are available from the authors on request.

\end{document}